Leonard M. Adleman

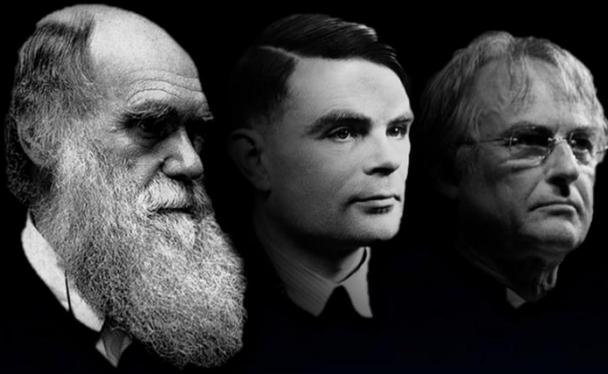

*DARWIN*
*TURING*
*DAWKINS*

Building a General Theory of Evolution

# <u>Darwin Turing Dawkins</u>

Building a General Theory of Evolution

Leonard M. Adleman

# Preface

This is not the book I wanted to write. Inspired by the works of Charles Darwin, Alan Turing, and Richard Dawkins, I have been thinking about a general theory of evolution, what I have come to call prene-theory, for over 40 years. I had hoped to reach a level of understanding sufficient to present the theory in a completed form. I did not succeed. What is presented here is, at best, a first approximation to prene-theory. Hopefully, it is sufficient to convince the reader that further investigation is warranted.

As it is, this book will let you see the world from my prene-theoretic point of view. The next time you run for president, fight a war, or just deal with the ordinary problems that humans are heir to, perhaps the prene-theoretic perspective will be of use. If you want to understand why and when you are likely to die, or if you want to achieve greatness, prenes may help here as well. If you are confused about where the "computer revolution" is headed, then prenes may provide some answers.

I will apply prene-theory to bees, cells, computers, cultures, history, humans, literature, music, politics, religion, science, viruses (both biological and computer), and other things. In many cases my expertise is limited. No doubt this and other factors have led to mistakes, oversights, over-simplifications, and speculations that the future will determine to be unwarranted.

# Table of Contents

# I: Prenes

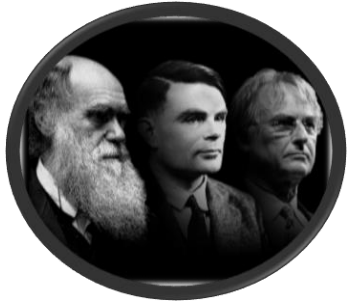

Figure 1

I was a young mathematics professor at MIT when I began reading Richard Dawkins' 1976 book "The Selfish Gene" (Dawkins, 1976) about the primacy of genes in Darwinian evolution. In the last chapter, Dawkins introduced what he called "memes". Memes were things like ideas and beliefs that were analogs of genes but resided in peoples' brains rather than in molecules of DNA.

I had long been aware of the work of Alan Turing (the father of the computer) and other logicians. It had been apparent to me that their results implied that the things stored in computers (which I later called Turenes, in honor of Turing) were also analogs of genes.

Had Turing and Dawkins set the table for a broad generalization of Darwin's theory of evolution? Were genes, memes, and Turenes simply special cases of a single more fundamental thing?

It turned out they were. I called these fundamental things "prenes" (portmanteau from primary and genes)[1]. Roughly speaking, prenes are units of information. They are the main characters of this book and are surprisingly powerful. As you will see, they have a profound impact on everything you think, feel, and do.

Genes have become central to the study of biological evolution. I believe that memes, Turenes, and other prenes have the potential to occupy a similar position with respect to the study of societal and computer evolution.

Ideally, the theory of prenes will provide a more complete understanding of Darwin's great insight.

───────────────────────

[1] *Genes, Turenes, and prenes rhyme.*

# What is a prene?

Unfortunately, I don't exactly know. You may see that as a bad way to start this chapter but let me explain why it's not my fault.

Consider the question: What is a gene? The concept of a gene has been with us for over a century and is one of the most important in biology. You probably think that biologists know exactly what a gene is, but you would be wrong. If you look up gene in The Oxford English Dictionary (OED, 2022), you'll find this:

> *1. Biology.*
> *a. The basic unit of heredity in living organisms, originally recognized as a discrete physical factor associated with the inheritance of a particular morphological or physiological trait, and later shown to be located at a specific site on a chromosome and to consist of a sequence of DNA (or RNA in certain viruses) containing a code for a protein or RNA molecule, together with any associated sequences necessary for transcription and translation.*

Nice words, but what do they mean? "The basic unit of heredity", what does that mean? "located at a specific site on a chromosome", what if it is located on a synthesized DNA molecule in a lab? "and to consist of a sequence of DNA (or RNA in certain viruses)", what if it is a sequence of mRNA in a cat? I could go on.

In fairness to the Oxford English Dictionary, some amount of imprecision in their definition was inevitable. For deep philosophical and mathematical reasons, we cannot (and likely can never) provide an exact definition of gene. We will return to this in *Socrates' bed*.

For the same reason, I cannot provide an exact definition of prene. So, I'll content myself with the following Oxford English Dictionary-like "definition":

> *A basic unit of information that can be stored in physical things.[2]*

Let's consider some examples.

> *To be, or not to be, that is the question …*

Hamlet's soliloquy. It is stored in books as a sequence of letters. It is stored in computers as a sequence of zeros and ones. It is stored in people's brains in a manner yet to be elucidated by science.

So, what is Hamlet's soliloquy? It is a prene, which we will call the "Hamlet's-soliloquy-prene", and it is currently stored in many physical things including books, computers, and brains.

───────────────────────

[2] *If you are interested in a definition better suited for research, then see the chapter What is a prene - really? One advantage of the vague definition just given is that it allows us to begin this book with an informal exposition where the reader's intuition about the relevant concepts should be sufficient.*

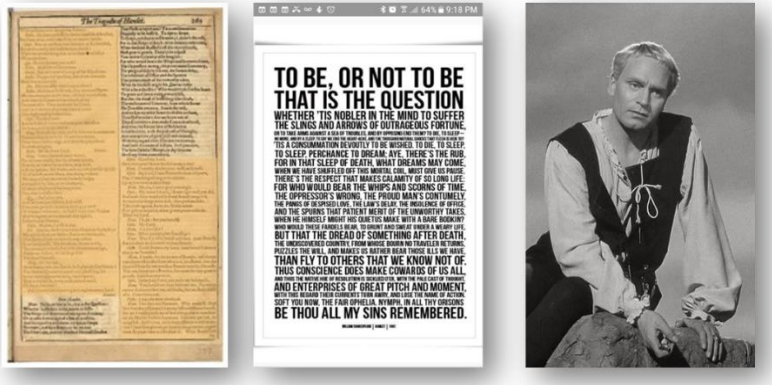

Like Hamlet's soliloquy, the smallpox genome is a prene. The smallpox genome is not a molecule of DNA, nor is it a sequence of A's, T's, C's, and G's in a computer; rather, each molecule of DNA in a smallpox virus is storing the smallpox-genome-prene, and each computer with the appropriate sequence of A's, T's, C's, and G's is also storing it.

A "copy" of a prene is a physical object that stores it. The most important thing to know about a prene is its current copy number – the number of distinct physical things in which it is stored at this moment. It is likely that the Hamlet's-soliloquy-prene is currently stored in millions of books, millions of computers, and tens of thousands of brains. Perhaps, it's current copy number exceeds a billion.

If the current copy number of a prene drops to zero, the prene has gone extinct.

For example, it is known that some of Shakespeare's plays have been lost; so, barring a miraculous find, the prenes once stored in written copies of those plays have gone extinct. The same can be said for the genome prenes of many ancient creatures.

## Proposition 1

### All prenes struggle to avoid extinction.

In light of Darwin, it is easy to see how the smallpox virus, and hence the smallpox-genome-prene it stores, has struggled. But can it really be that the Hamlet's-soliloquy-prene has also struggled? The answer is yes. The next several chapters will provide the foundation necessary to understand this.

As you read further, you may see ideas that seem familiar. For example, perhaps you are aware that cultures evolve. But is this just an analogy or is something deeper at play? One of the goals of this book is to demonstrate that this and numerous other important observations have a common source: *Proposition 1*.

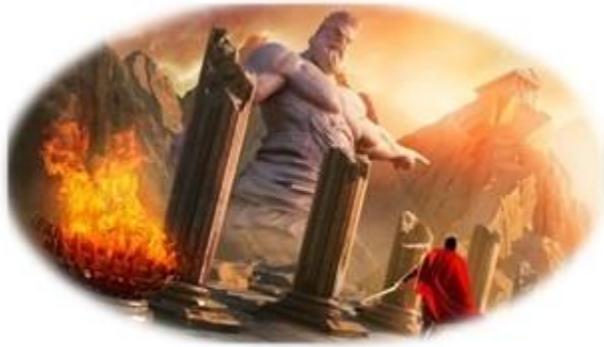

Figure 3: Destiny

Like the gods of mythology, prenes are immaterial but carry out their activities through humans and other physical things. They began their journey long ago and will continue it long into the future. We are merely evanescent creatures swirling in their wake.

# Special-prenes

In what follows, certain sets of prenes will be of particular interest. Accordingly, we define a gene to be a prene for which there is at least one nucleic acid molecule that stores it[3]. This definition is more general than many commonly used in biology where, for example, a gene may be required to encode a protein.

Analogously, we define a meme to be a prene for which there is at least one brain that stores it, and a Turene to be a prene for which there is at least one computer that stores it.

So, since the Hamlet's-soliloquy-prene is currently stored in computers and brains, it is both a Turene and a meme. As we will see in *The resurrection of smallpox*, the smallpox-genome-prene is both a Turene and a gene. What we call "Internet memes" are typically both memes and Turenes. Soon we will see an example of a prene that is a gene, a meme, and a Turene.

Now, consider the extinct Maya society of the Yucatán Peninsula. Presumably, the Maya had prenes concerning law, religion, war, food, relationships, and lots of other things. Some of these prenes would probably have been stored in brains, for example as stories or songs. The Maya developed writing and stored some in stone, skin, and paper. Taken together, these Maya prenes correspond to what social scientists might call the Maya culture. We will refer to prenes in this set as cultural-prenes of the Maya culture and the set itself as the Maya cultural-prene-set.

---

[3] *More precisely, there is a molecule that contains a linear sequence of nucleotides that store it.*

All societies, including religions, companies, nations, political parties, family units, book clubs, etc. have cultural-prene-sets.

We will have much to say about cultural-prenes in what follows. We will see that, roughly speaking, cultural-prenes are to social science what genes are to biology (see *Cultural-prenes in the social sciences*).

While it is common to say that societies are "all about money" or "all about power", they are not; they are all about the survival of their cultural-prenes.

## Why do bees kill themselves?

We will use prenes to investigate the shocking number of suicides among honeybees. We will see that applying evolution to prenes rather than just genes can provide insights into our own behavior.

If a honeybee stings you, you'll probably be angry because it really hurts. You may find a bit of solace knowing that the bee that did this has just signed her own death warrant. Her entrails will be ripped from her body. You probably hope that really hurts too. It probably doesn't, but that's another story (see *Perceptions*).

So, why do bees sacrifice themselves to attack you? Suicide is not usually considered a good evolutionary strategy.

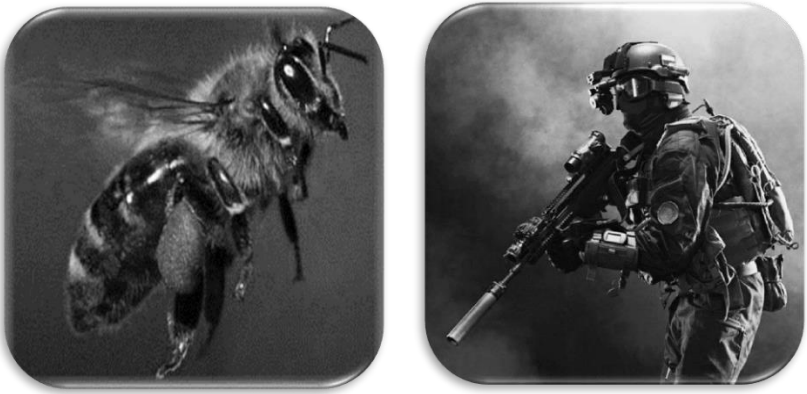

Figure 4: Brave prene-warriors.

Dawkins and others have championed a gene-centric view of biology. From this view, the struggle of living things to survive is a struggle between genes. Living things are built by their genes, their behavior is determined by their genes, and their purpose for living is to preserve and reproduce their genes.

The gene-centric view has been the starting point for many proposed explanations of sacrifice among living things.

I'll use a gene-centric perspective to provide an explanation of the honeybee's sacrifice.

Though it may seem paradoxical, the honeybee has sacrificed her (only worker honeybees sting, and all workers are female) life to reproduce some of her and her colony mate's genes, that is, to make new copies of genes the bees in her colony store in their DNA molecules.

While her death subtracts 1 from the copy number of each gene she stored, it may also protect the colony, and in particular the queen. The queen is the reproductive organ of the colony. She, and, with a few exceptions which for the sake of clarity we will ignore, only she can produce offspring that store new copies of genes. It is easy to imagine that under the right circumstances, sacrificing a worker to reduce a threat to the queen, would increase the probability that copy numbers of the colony's genes would improve in the future. So, the colony gene-set programs the workers to sacrifice themselves in such circumstances. Hence, honeybee sacrifice is merely part of the strategy the colony gene-set has used to avoid extinction.

The gene-set you store has adopted a similar strategy. Some of your cells sacrifice their lives to make new copies of your genes. For example, when you were in utero, you had webbing between your fingers for a few weeks. The webbing was made of cells that killed themselves (using a programmed cell death process called apoptosis), presumably to give you fingers with greater utility, which contributed to your survival and reproductive success. For another example, when macrophages are infected by salmonella, they sometimes "explode" themselves (using a programmed cell death process called pyroptosis) and thereby release substances that attract and stimulate other immune cells to fight the infection.

Does this gene-centric view explain all sacrifice in living things? Let's look at some more examples.

Consider humans. Parents sacrifice for their children. Sometimes the sacrifice is life itself, but most often it is more mundane, like getting up in the middle of the night to feed a newborn. Because the children's genes are also genes of the parents, and the children have a significant probability of growing into adults who will reproduce those genes, the gene-centric explanation of such sacrifice works in much the same way as it did for honeybees.

But what about soldiers who sacrifice their lives for a cause? Here the gene-centric view fails to provide a convincing explanation in many cases. For example, in some civil wars most members of the society that a soldier is protecting are no more closely related genetically to the soldier than members of the enemy's society, and the individuals most closely related to the soldier, such as children, are less likely to survive and reproduce once the soldier is gone.

But, if soldiers are not programmed by their genes to sacrifice their lives, why do they do it?

It is because they are programmed by a different set of prenes, those memes stored in their brains that we commonly call beliefs[4].

Beliefs can be transferred from one person to another, for example, by teaching. When such transfers occur, the beliefs are stored in new brains, and the people who have acquired them can teach them to others, hence copy numbers rise.

After you are born you will acquire lots of beliefs from your nation's cultural-prene-set. Children born in different nations are likely to acquire different beliefs. For example, if you are a member of the American society, you are likely to acquire the belief that capitalism is good, and communism is bad. If you are a member of the Russian society, you are likely to acquire the opposite belief.

If you become a professional soldier, you will acquire specialized beliefs from your nation's cultural-prene-set that most members of your national society do not acquire. For example, you may acquire a belief that under some circumstances it is honorable to sacrifice yourself to save fellow soldiers or the members of your society. A soldier who acquires such a belief may now be programmed for actual sacrifice in the future.

———————————————————

[4] *I use the term "belief" to denote a meme that has a significant impact on behavior. For example, though I acquired the when-you-wish-upon-a-star-your-dreams-come-true-meme from Walt Disney, it has had little impact on my behavior, and I do not consider it a belief. On the other hand, I acquired the when-you-give-your-word-you-must-keep-it-meme from my parents, and it has guided my behavior throughout life. It is a belief.*

So, let's look at a soldier who acquired such a belief and has made the ultimate sacrifice. While dying decreases the copy numbers of the soldier's beliefs by 1 and removes the soldier's ability to continue transferring those beliefs to others, it may also protect the members of the soldier's society and allow them to continue transferring their own beliefs to others. Since the soldier and the protected members belong to the same society, they will both believe many of the same cultural-prenes of that society. So, it is easy to imagine that programming the soldier's sacrifice under the right circumstances would increase the probability that the copy numbers of the society's cultural-prenes would improve in the future.

If the beliefs soldiers acquire can program their ultimate sacrifice, what have your beliefs programmed you to do? Also, how do prene-sets go about this programming? We will examine these things further in *Cultural-prenes in the social sciences* and *Part II: Humans*.

# The resurrection of smallpox

The genes, the memes, and the Turenes are all subsets of the prenes. Sometimes it is useful to focus on just one of these subsets. For example, the genes are the primary focus of many biologists. But we should not lose sight of the fact that prenes can migrate between these subsets. When prenes exploit this opportunity to travel, it can have deadly consequences for humanity.

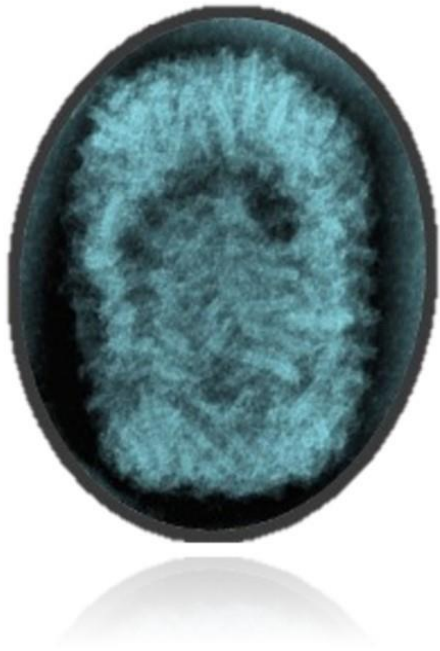

Figure 5: The smallpox virus. Dead or just playing possum?

The following editorial appeared in the New York Times (Adleman, 2013). For obvious reasons, I did not use the language of prenes, but, at its core, that is what it is about. It addresses the question: has smallpox been eradicated? The world has always looked at this as a question of biology, that is, as a question about genes, but when looked at in terms of all prenes, the view is strikingly different.

*By Leonard Adleman*
*Los Angeles*

*On Oct. 16, 1975, 3-year-old Rahima Banu of Bangladesh became the last human infected with naturally occurring smallpox (variola major). When her immune system killed the last smallpox virus in her body, it also killed the last such smallpox virus in humans. In what is arguably mankind's greatest achievement, smallpox was eradicated.*

*Our war with this smallpox virus was brutal. It appears likely that the virus killed about one billion of us. Initially, our only defense was our immune system, but eventually we developed new tools, including vaccination. In the late 1950s, the World Health Organization began responding to outbreaks by vaccinating everyone in the surrounding area to prevent the virus from spreading. By 1975, we had won.*

*The smallpox virus had only a single host species: us. Other viruses have multiple hosts. For example, some strains of flu live in both humans and pigs, hence "swine flu." If smallpox had had a second host, eradicating it in humans would have been of little value, since it would have thrived in its second host and later re-emerged in humans.*

*A few samples of the virus are still kept in special labs: one in the United States and one in Russia. We don't bother vaccinating against smallpox anymore; if the virus escapes from one of these labs, the war will begin again. Currently, there is debate about whether these samples should be destroyed or kept for scientific purposes.*

*But the debate should be broadened. Even if we destroy those samples, the war is not over; the smallpox virus has now found a second host. It is not the pig. In fact, it is not even what we think of as a living thing. It is the computer.*

*This is not some conceptual game. This is real and life-threatening.*

*If you search online, you can find the sequence for the smallpox genome. It is a word written with the letters A, T, C and G. The word is about 185,000 letters long. It is the word*

*that tells cells to make smallpox viruses. The sequence was stored on a computer in the early 1990s, when a research team led by J. Craig Venter obtained it using a biotechnical process applied to a sample of the virus.*

*Of course, a word in a computer file cannot kill you. Well, yes and no. In the 1990s, I ran a biotechnology laboratory. In my lab there was a machine much like a soda dispenser, only in this case the reservoirs were filled with chemicals. If I typed in a short word of my choice using the letters A, T, C and G, the machine would squirt one chemical after another into a test tube. When it was done, the test tube would contain trillions of molecules of DNA. Each would look like a necklace, with molecules of adenine, thymine, cytosine and guanine (the building blocks of DNA) strung according to the word I had typed.*

*At that time, the 10,000-letter sequence of the H.I.V. genome was available online. I contemplated using my machine, together with well-known biotechnical methods, to create, de novo, the H.I.V. genome — an actual molecule identical to that found in H.I.V. viruses living in the wild. I had reason to believe that inserting such a synthetic molecule into a living human cell would cause the cell to manufacture full-*

*blown H.I.V. viruses that could then be transmitted from person to person and cause AIDS.*

*I decided not to do the experiment, but I began to worry. If I could do it, so could others with high-tech labs.*

*Which brings us back to smallpox. Might someone resurrect it? You may think this is mere speculation, but in 2002, scientists used the approach just described to produce an infectious polio virus. It is possible that the great labs, with great scientists, the best equipment and substantial funds, could overcome the considerable challenges that exist and resurrect smallpox right now. Before too long, more modest labs may be able to accomplish the same thing.*

*I am worried, but also amazed. Smallpox has miraculously and unconsciously saved itself through an extraordinary act of evolution. After thousands of years, it was on the verge of extinction; it existed in one small girl, and just before that girl's immune system killed its last living member, a sample was taken and stored in a lab. Years later, that sample was used by another lab to sequence the viral genome. The sequence was placed on a computer, infecting a new "species" that had just come into existence.*

*Do we sit and wait for the day when someone releases resurrected smallpox on an unvaccinated world? I'm a scientist, not a policy expert. But would it be wise for us to consider limiting the distribution of the tools of this emerging technology?*

So, at virtually the last possible moment, the smallpox-genome-prene made the leap from gene to Turene and was saved from extinction. I cannot think of an evolutionary miracle more remarkable.

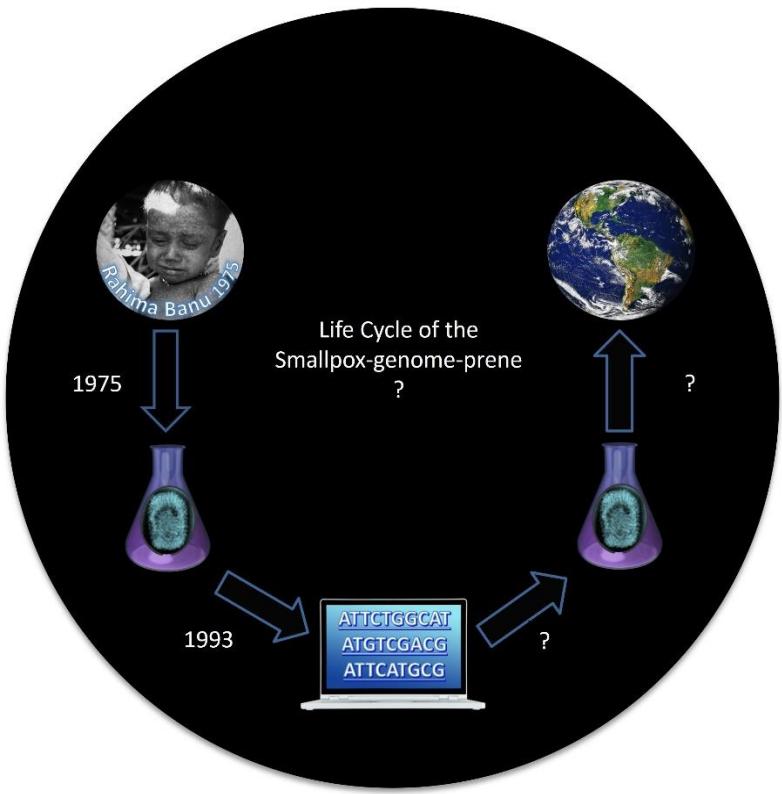

Figure 6: From gene to Turene and perhaps back to gene.

Recently, a group at the European Bioinformatics Institute has provided another interesting example of prene migration. They have stored all of Shakespeare's Sonnets in DNA molecules (Goldman, et al., 2013). Hence, the Shakespeare's-Sonnet-154-prene which began as a meme stored in Shakespeare's brain, is now stored as a meme in many other people, as a Turene in many computers, and as a gene in some DNA molecules. Another interesting example of a prene that has recently become a gene, is Muybridge's famous 1878 movie "The horse in motion" which was stored in a DNA molecule, inside of a bacterium, and gets replicated (and hence increases copy number) when the bacterium divides (Shipman, Nivala, Macklis, & Church, 2017).

## How to be an unsuccessful prene

A successful prene, one that has many copies over a long period, is a rare thing exquisitely suited to its environment.

In this chapter, I will describe a prene that has been unsuccessful and has no one to blame but itself.

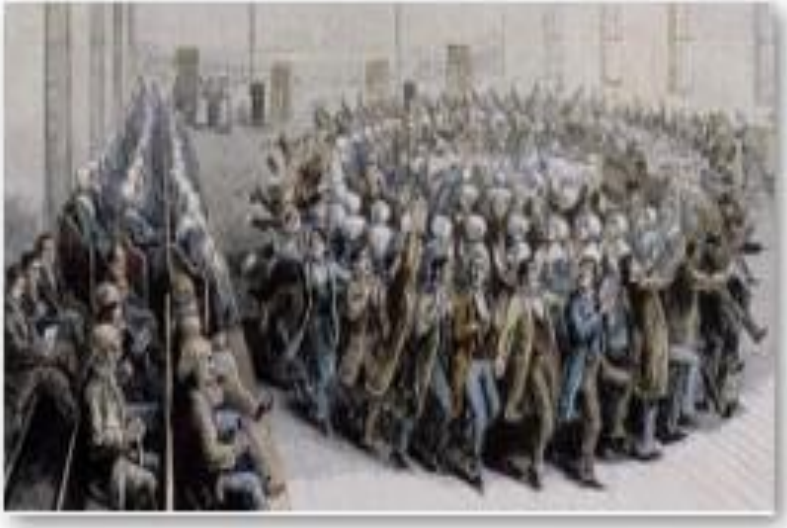

Figure 7: Shakers dance and worship. "I saw in vision the Lord Jesus in his kingdom and glory… I was able to bear an open testimony against the sin that is the root of all evil; … the doleful works of the flesh" - Shakers' Mother Ann Lee (ca 1770)

The Shakers were a Christian sect that arose from the Quakers in mid-18th century England. To the usual Quaker cultural-prene-set, they added new prenes including the "celibate-prene": sex is forbidden. I'll skip the details, but guess how many Shakers there are today?

No, you are wrong. Much to my surprise, the correct answer appears to be two (Wikipedia-Sa). However, I suspect your answer will be correct soon enough.

The celibate-prene was ill-suited for its own survival and dragged the Shaker society down with it. There are several reasons for this, perhaps the most important is that the celibate-prene foreclosed one of the most basic means by which cultural-prenes get into new brains: have children.

Human babies are born with an open channel for the transfer of memes from their parents. The Human-gene-set arranged it to be so for its own survival. The reason is obvious: getting the don't-put-your-hand-in-the-fire-meme from your parents helps you and your genes survive.

Notice that successful religions don't make the mistake of having the celibate-prene in their cultural-prene-sets. For example, Jewish and Christian societies treat the Old Testament as sacred, and it virtually begins with an anti-celibate prene:

> *Be fruitful and multiply, and replenish the earth*
> *-Gen 1:28 (KJV)*

For the same reason that the celibate-prene contributed to its own failure, the be-fruitful-and-multiply-prene has contributed to its own success and the success of many prenes in the Jewish and Christian cultural-prene-sets.

Islamic societies specifically exploit the parent-child channel. Some of the society's cultural-prenes direct the father of a newborn to whisper into the baby's ear, the first words the baby will ever hear, the Shahadah:

> *God is great*
> *There is no god but Allah, Muhammad is the messenger of*
> *Allah.*

When the Shakers religion first arose, it would not have been difficult to see that its cultural-prene-set might contain the seeds of its own demise. Can an analysis of the Jewish, Christian, Islamic, Russian, American, Republican, Democratic, and other societies' cultural-prene-sets reveal some of their strengths and weaknesses and provide a glimpse into their futures? I believe it can.

In *The prophet and the messiah*, I will speculate on whether the Jewish-cultural-prene-set contains the seeds of its own demise.

While we can see the potential negative consequences of the celibate-prene, it is unlikely that many Shakers saw them. It typically serves a society to have its members hold only favorable views of the society's most important cultural-prenes.

For example, as an American, I have learned to cherish the First Amendment of the US Constitution. However, as a prene theorist, I am less enamored.

Freedom of speech, the press, and assembly are two edged swords that can be beneficial in some settings and detrimental in others.

Abraham Lincoln suspended freedom of the press during the Civil War because it could be used to "give aid and comfort to the enemies of the United States" (Lincoln, 1864).

It could also be argued that freedom of the press exposes governmental abuses and thereby provides a defense against dictatorship.

We could argue endlessly about which position is superior. But our opinions are of little importance. The fate of the American-cultural-prene-set (and of America) will be determined on the field of battle where it will struggle to survive against fierce competition from cultural-prene-sets of other societies. From the prene-theoretic perspective, the First-Amendment-prene is of value to American society only if it contributes to that survival.

# The selfish prene

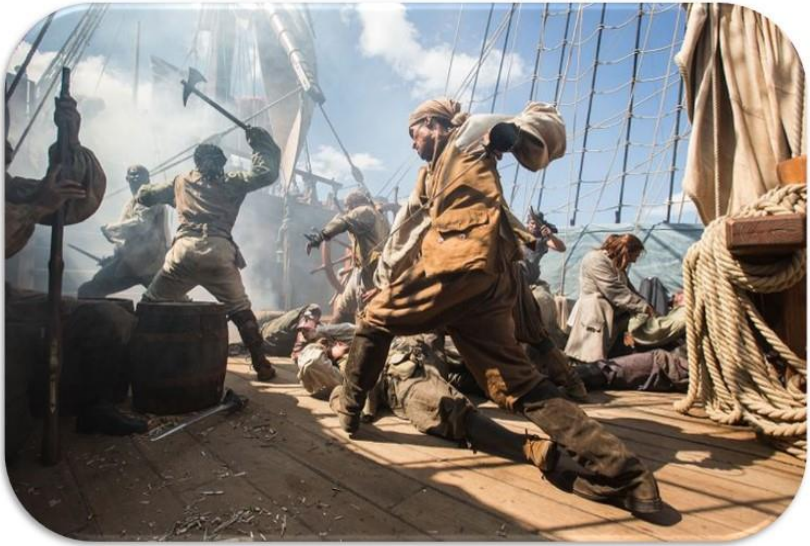

Figure 8: If prenes were people.

In *The resurrection of smallpox*, we considered what can happen when a single prene is stored in many different types of physical things. Here, we will begin to consider what can happen when a single physical thing stores many different prenes.

The most important thing to remember is that prenes are not nice! Even the be-nice-prene is not nice. Prenes "care" about their own survival and nothing else. A prene may form an alliance with other prenes to form a prene-set, but if that alliance becomes a burden, the prene will abrogate it without the slightest remorse.

Consider the bread mold (*Neurospora crassa*). It stores a set of genes that have formed an alliance. When placed in minimal-medium (water with some salts, sugar, and a few vitamins) each mold thrives and produces offspring that store the same gene-set.

But what would happen if you began regularly adding new substances to the medium? For example, uracil (a substance used in making RNA). Thanks to the work of Beadle and Tatum (Beadle & Tatum, 1941), we can be pretty sure of the answer.

Among the genes allied in the mold's gene-set, there was an important subset which we will call the uracil subset. All the genes in the gene-set really liked the uracil subset a lot, because without them, a mold cannot synthesize uracil from the minimal medium, cannot make RNA, and will die, thereby reducing all their copy numbers.

However, once there was "free" uracil available in the medium, the rest of the genes in the set would ask: "what do we need those guys for?" They would turn on the uracil subset and cast them out. Those molds that discontinued having copies of the uracil subset still had uracil, but those molds that retained copies had to continue spending resources on the subset's maintenance and reproduction. Natural selection would ensure that after a while, only molds that abandoned the uracil subset would survive.

In the prene-world, no good deed goes unpunished for long. Consider a bread mold that took pity on some member of the uracil subset and allowed it to remain. That mold would be committing suicide; it would be outcompeted by its more ruthless and efficient brethren and go extinct. As we saw in *Why do bees kill themselves?* even apparent altruism is self-serving.

There is a reason the lion does not lie with the lamb, why animals in the forest do not live in harmony, and why a world without human conflict will never persist. Such situations rely on selflessness and are unstable. What happens when a selfish mutant lion is born with an appetite for lamb?

When many prenes are stored together in an instrument, such as a cell or a human, the prenes will fight each other to control the instrument and use it for their own survival. If a subset of prenes gains control and induces behavior that is harmful to other prenes, well, that's their problem.

So, prene-sets are seldom happy places. Why should you care if prenes are unhappy? Because their unhappiness sometimes becomes your unhappiness as we shall see in *The war within*.

# The war within

If you want to find a physical object that stores millions of prenes all at once, you will not have to look far – it's you. You are a storehouse for genes, memes, and other prenes. Among your memes are cultural-prenes you have acquired from the societies in your environment.

Each of your prenes is an aspiring dictator that wants to use you and your resources to help it survive. Unfortunately, your prenes are often at cross purposes, and when this happens, you may suffer.

Consider what happened to Joan of Arc when her genes and memes came into conflict.

> *Joan was taken to a scaffold set up in the cemetery next to Saint-Ouen Church and told that she would be burned immediately unless she signed a document renouncing her visions...*
> *(Wikipedia-Jo)*

What was best for Joan's genes? Joan, like all of us, was designed by her genes to stay alive and reproduce. By renouncing, Joan would not be burned, would retain the possibility of having children and increasing the number of copies of her genes. Joan's genes favored renouncing.

What was best for Joan's religious memes? Joan was a devout Christian who had attracted a large following in France; much of it based on memes she had acquired through divine visions, or from the prene-theoretic perspective, through subconsciously processing her existing memes and generating new ones (see *How your brain captures memes* and *How your brain processes memes*). The English wanted her to renounce, knowing it would undermine her French followers. By refusing, Joan would become a martyr, attract new followers, and increase the number of copies of her memes. Joan's religious memes favored not renouncing.

As you might imagine, and as the historical record seems to suggest, this situation was psychologically stressful for Joan. So, we are sometimes collateral damage in the wars between our prenes.

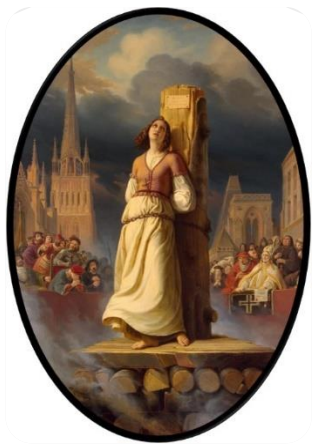

Figure 9: Joan of Arc experiencing cognitive dissonance?

By the way, Joan did renounce, but was later burned at the stake anyway. She became a famous martyr, a saint, and over the last six hundred or so years has contributed mightily to the spread of Catholic cultural-prenes.

While Joan's case is extreme, we all experience situations where our prenes compete for our behavior. I like to think of our prenes as members of a "prenes-legislature". When a situation arises, a debate ensues with members expressing their opinion about the behavior we, their instrument, should execute. Ultimately, the legislature hammers out a behavior that we perform.

In a real legislature, sometimes members are largely in agreement on an issue, and sometimes they are in intense disagreement with powerful members at loggerheads. This appears to be the case in the prene-legislature as well, and the more intense the debate, the greater our psychological discomfort.

For example, once when invited to go skydiving, I immediately said "no". I do not recall any psychological discomfort, and it seems likely the members of my prene-legislature were in almost complete agreement.

On the other hand, small disputes occur frequently in the prene-legislature and induce mild psychological discomfort. For example, you may experience this kind of discomfort when trying to decide whether to exercise or watch TV.

Occasionally the disagreement between prenes is intense and the discomfort it causes is great. For example, if you must consider putting a pet down, the debate may proceed with great vigor for days or weeks and the psychological discomfort you feel may disturb sleep and disrupt contemplation of other matters.

Sometimes, the legislature meets in camera; the members fight in the subconscious part of our brain without invoking the conscious part. In such cases we may feel psychological discomfort without being consciously aware of its cause – perhaps this is one form of anxiety.[5]

_________________________

[5] *I am not qualified to speak with authority on psychology. Nonetheless, I hope that real psychologists will find some of the concepts discussed in this book worthy of further consideration.*

# Prene warrior

In *The war within*, I described how prene wars within you can lead to psychological distress. But these are not the only kind of prene wars. The cultural-prene-sets of different societies are constantly at war with one another. These are the kind of wars that occur between nations, political parties, and religions; the kind that historians write about.

In the late 1970s, two MIT friends, professors Ron Rivest and Adi Shamir, and I produced a short paper with the title "A Method for Obtaining Digital Signatures and Public-key Cryptosystems" (Rivest, Shamir, & Adleman, 1977). We had just collided with the powerful National Security Agency (NSA) whose existence had been unknown to us. I was now a warrior on the front lines of a war between the academic-cultural-prene-set and the national-security-cultural-prene-set. The reverberations of that war are still with us today.

Our cryptosystem became the topic of a Scientific American column (Gardner, 1977) in which we offered to send a copy of our paper to anyone who sent a self-addressed stamped envelope to MIT.

My first clue that there was something strange going on came when I happened to be standing in line at a bookstore in Berkeley and the customer in front of me said to the cashier words to the effect: have you seen this article on crypto in Scientific American? The cashier replied: Yeah, that's so cool! I realized what they must be talking about and in a youthful burst of pride said "Oh, that's our stuff …". The customer took his copy of the magazine and asked me to autograph it.

I know you probably think that mathematicians are constantly accosted by strangers seeking autographs, but it had never happened to me, and I had never considered that it ever would.

But things got even stranger when I returned to MIT and found that there were thousands of self-addressed envelopes being stuffed with copies of the paper. And the addresses seemed odd as well, things like "The Bulgarian Department of Security".

We soon received a letter informing us that we could not send our paper out of the country – it was against the law. What! What law? What is going on?

We found out that the letter came from an employee of the NSA. Though we had never heard of the NSA, it turned out to be the largest intelligence agency in America – bigger than the FBI and CIA combined – and so secret that it had a "black budget" and was unknown to most in government; those who did know of it referred to it as "No Such Agency".

The NSA was a "black chamber", the centuries old name given to the government agency responsible for making a nation's cryptosystems and for breaking the systems of other nations.

Our paper had changed everything. Our cryptosystem was real mathematics, not kid's decoder ring stuff, and when it appeared in Scientific American, it apparently meant that the Nation's enemies could now use a system that the NSA could not break. A major source of intelligence could dry up overnight. A very bad thing. And it seems that even today terrorists use our system to keep their plans secret. But our cryptosystem was double-edged. It could also be used to protect Americans from entities, both foreign and domestic, that might wish to violate their privacy. A very good thing. And, in fact, our system is widely used for that purpose. When your private information (for example, credit card numbers and medical records) is sent or stored on the Internet, there is a good chance that you are using our system to keep that information secure.

Had I known all this before we published, I would have had a great moral dilemma on my hands, but the cryptosystem was out there now, and things would just have to unfold as they would.

Apparently, the Scientific American publication hit the NSA hard, and they responded with vigor. The letter was just the beginning, it was followed by attempts to secure legislation that would inhibit the widespread use of our system and attempts to create a new national standard cryptosystem different from ours and widely suspected to be intentionally flawed so that the NSA could break it.

The NSA even tried to co-opt me by offering a sweetheart consulting deal. While MIT took the position (which I agree with) that it was a privilege to teach and do research there and an affront to expect anything but a subsistence wage, the NSA's position was quite different. I was tempted by their offer but declined. I have sometimes wondered if this episode eventually led to the related scene in "Good Will Hunting".

The prene-war that started then, continues to this day. Information revealed by Edward Snowden alludes to the fact that the NSA is still attempting to mitigate the impact of our system (Menn, 2014).

So, somehow, I had become a general in a war between the academic-cultural-prene-set and the national-security-cultural-prene-set. I was under attack. I would soon get the opportunity to counter.

As a matter of routine, every few years I sought and received a grant from the National Science Foundation (NSF) to do my mathematical research. The NSF is the major funder of pure mathematical and scientific research in the United States. In 1980, I was seeking my next grant.

The procedure was well known. I would write a proposal of about 30 pages describing how, if given the money, I would do such amazing mathematical things in the next few years that the Nation would thank me and admire the NSF for supporting me. Since our paper had made crypto a very hot academic topic, I included a few paragraphs about how, despite the fact that everything I planned to work on appeared on the surface to be pure mathematics devoid of any possible use, my work actually had great practical importance because it was applicable to cryptography.

I submitted my proposal and soon received a call from NSF. They would be delighted to give me the money, and, oh, by the way, the NSA generously decided to fund the part of the proposal that involved cryptography.

I hung up knowing that I was about to achieve a major victory.

I picked up the phone and called Gina Kolata, a distinguished writer at Science. I had known Gina for several years and said I had a story that she might find interesting.

A few hours later, I got another call, this time from the head of the NSA, Admiral Bobby Inman, explaining that there may have been a small misunderstanding. But it was too late; the counterattack was under way.

Soon Gina's article appeared with the title "Cryptography: A New Clash Between Academic Freedom and National Security" (Kolata, 1980), and soon after that, the NSA sued for peace:

> On October 9, 1980 representatives from the NSA and the NSF met with White House science advisor Frank Press to clarify the issue. The decision was made that both agencies would fund cryptography research for the present. Although the NSA would require investigators it supports to submit articles to the agency prior to publication, it would not expect to classify the research it supported. Adleman was offered the choice of NSA or NSF funding; he accepted NSF support because, "On a personal level I saw myself as a pure scientist and my natural affinities were to be funded by NSF. As a scientist, it was clear that there would be a national debate on the issues and I didn't want any action I might take to be misconstrued as suggesting that the NSA had a compelling case that they had a role to play in the scientific process."

Figure 10: Landau, S, Notices of the American Mathematics Society 1983

Today, the situation is pretty much as described by Landau, both NSF and NSA provide grants for mathematicians, and mathematicians may choose which direction they want to take.

Many years have passed since then, and my view of those events has changed greatly. I no longer think that I was "right", and they were "wrong". Now that I am a prene-theorist, I understand that all involved were just playing their expected roles as prene-warriors for the cultural-prene-sets of their societies.

## Proposition 2

### We are all prene-warriors in the service of prene-sets we have acquired, though we are seldom conscious of it.

What is the best way to fight a prene-war? Hard to say, but it is important to remember that the opposing prene-set itself is the primary opponent, not its "instruments": the people, computers, weapons, and other things which the prene-set can exploit.

The instruments can be attacked using physical force. The prene-set itself can be attacked by mutating it. Both the sword and the pen have their place.

For example, the Allied victory over Japan in World War II was largely achieved by using physical force to attack the instruments of the Japanese-cultural-prene-set, but the transformation of Japan into a US ally came largely from mutating that prene-set. Post-war, the United States occupied Japan and mutated the monarchical, militaristic Japanese-cultural-prene-set into a democratic pacifistic one (see *Greatness*). The United States succeeded in producing generations of new followers for some of its most important cultural-prenes. A different post-war strategy was used by the Romans at the end of the third Punic war.

In my opinion, we are experiencing a major shift in the balance between force and mutation. I suspect that the primary battleground of the future will be the Internet where prene-sets will mutate one another by the application of propaganda, education, enlightenment, etc.

Competition between societies is often prolonged. We use words like war and peace, victory, and defeat, but these merely describe particular epochs and instances in the process. The Islamic-cultural-prene-set and the Christian-cultural-prene-set have been competing for over a thousand years with many periods of war and peace, and many instances of victory and defeat, but there is no end in sight.

## Variation and natural selection

The Origin of Species (Darwin, 1859) rests on two notions: variation and natural selection. Darwin observed that even closely related individuals were often not identical. We expect a newborn to resemble each of its parents, but to be identical to neither. Darwin took the existence of variation as a given and introduced natural selection as a mechanism that favors the survival of some variants and disfavors the survival of others.

Darwin speculated on the processes that give rise to variation but could provide few details. Since Darwin, much work in biology has been directed at providing those details, and a great deal has been learned. A turning point occurred with the publication of Watson and Crick's famous 1953 paper, "A Structure for Deoxyribose Nucleic Acid" (Watson & Crick, 1953), with its famous understatement:

> *It has not escaped our notice that the specific pairing we have postulated immediately suggests a possible copying mechanism for the genetic material.*

Today the view of many biologists might be described informally as follows: random environmental events, such as cosmic rays, mutate DNA and give rise to variations that natural selection acts upon.

I do not subscribe to this view. It attributes evolution solely to the actions of the environment. The environment creates variation, and the environment selects the variants that will be favored. While in a certain sense this must be the case, just as it must be the case that evolution is the result of the laws of physics, it is not adequate.

By and large, biological evolution is the result of a complex interaction between organisms and their environment. The role of the organism in the creation of variation is critical, and our understanding of evolution cannot be complete without considering it.

The commonly held view of evolution might give rise to the following description of the impact of cosmic rays.

> *A strand of DNA gets mutated by a cosmic ray. Babies born with mutations may be more or less fit for the environment into which they are born. Natural selection will favor the survival of the more fit and disfavor the survival of the less fit.*

But that version is not quite right. The start of the description should go something like this.

> *A strand of DNA gets mutated by a cosmic ray. The mutation is almost always repaired by a host of enzymes and other molecules that have evolved specifically to repair such mutations. On rare occasions, the mutation is not repaired. Babies born with mutations may be more or less fit for the environment into which they are born. Natural selection will favor the survival of the more fit and disfavor the survival of the less fit.*

But that version is not quite right either. The ending of the description should also be amended to give us something like this.

*A strand of DNA gets mutated by a cosmic ray. The mutation is almost always repaired by a host of enzymes and other molecules that have evolved specifically to repair such mutations. On rare occasions, the mutation is not repaired. Various enzymes and molecules have evolved that "intentionally" create additional mutations in the DNA. Babies born with mutations may be more or less fit for the environment into which they are born. Natural selection will favor the survival of the more fit and disfavor the survival of the less fit.*

So, the creation of variation might be described using the following diagram.

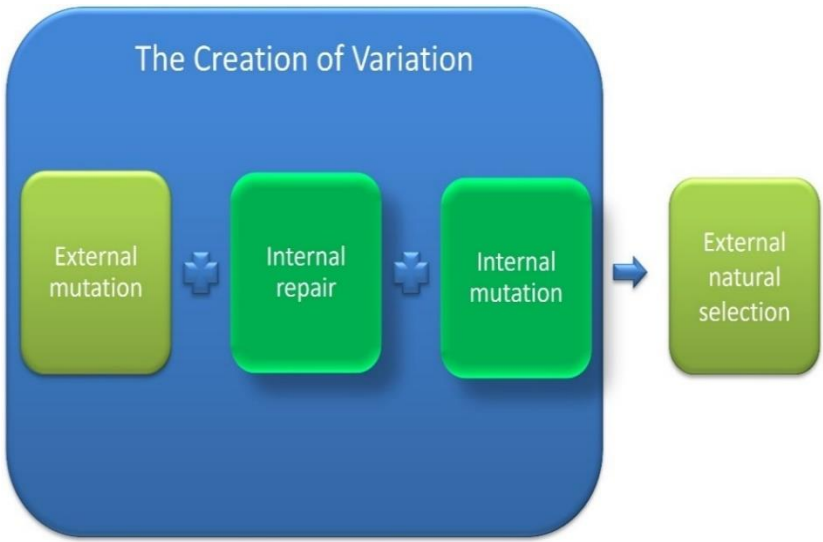

Figure 11

In general:

## Proposition 3

**Variation is primarily the result of processes
carefully programmed by prene-sets.**

I use the term programmed deliberately. To appreciate why this term is
appropriate, I recommend that you watch the YouTube video by [Drew
Berry:](#) Animations of unseeable biology (9min. 9sec.). which illustrates how
cellular processes, division in this case, are not simply the result of
chemicals reacting in a cytoplasmic soup but are often programmed
extravaganzas with millions of choreographed molecular robots delivering
supplies and information when and where they are required.

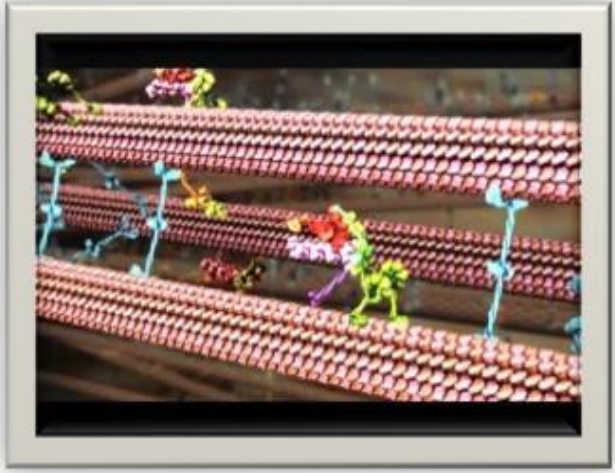

In the next chapter we will see how HIV has evolved a sophisticated program of intentional mutations in its nucleic acid molecules to create variation and increase the chances of survival.

There is an interesting possibility that seems worth investigating. I'll describe it informally. Is it the case that external mutations, those cause by the environment acting from outside the cell, are largely "undesirable" and seldom give rise to offspring that are more fit, while internal mutations, those programmed by the genes, are largely responsible for the creation of offspring that are more fit?

For example, cosmic rays appear to cause cancer, but since these rays seem to distribute their energy randomly, it seems quite improbable that the damage they cause would fortuitously result in offspring with greater fitness.

On the other hand, in humans, it appears that the genes have very carefully programmed the positions where homologous recombination is likely to occur during meiosis (Fell, 2013). This is a means by which the fitness landscape can be explored in a directed, non-random way.

Because the protein folding problem is NP-complete, if NP≠ P, then there are amino acid sequences that cannot fold quickly in Nature, since otherwise we could let them fold, image their structures, and use this approach to solve NP-complete problems quickly. If cells allowed for random mutation of the DNA that encode proteins, it seems likely that the resulting proteins would often fail to fold during the life of the cell. Hence, it is incumbent on cells to carefully program mutations to ensure that mutants have a high probability of folding quickly. V(D)J rearrangement is an example of such programming. This may also explain why cellular proteins are found in families with similar structures.

[†] This remark relies on specialized technical knowledge. It can be skipped without loss of continuity.

# Intentional mutation in HIV

To understand HIV's use of intentional mutation, let's start with a sports analogy.

You own a team that has just won the championship. Do you trade some players? The initial response is no; don't touch anything. But is this the wisest thing to do? You can be virtually certain that next season will not be the same as this one. Things change, players change, rules change, interest rates fluctuate, etc. So perhaps you should consider some trades to prepare for the changes that might come.

You have many choices. Perhaps you suspect that next season's rules will penalize players who do not sing well, so you trade the entire team for the Bach Children's Choir. You have made a change in anticipation of a future that has virtually no chance of materializing, and consequently you will have virtually no chance to repeat as champion. But what if you have an aging veteran who spends most games sitting on the bench? Do you trade for an untested young player with great promise? Perhaps you do. You cannot be sure it's a wise move, but under the right circumstances, it might be a risk worth taking.

Each time you gamble with change, you may win, or you may lose. But, if you want to increase your chances of winning, you do not gamble recklessly, you choose your bets very carefully. Successful prene-sets are master gamblers because those that aren't are destine for extinction.

Roughly speaking, biological evolution is like a never-ending single-elimination tournament, so the fact that HIV exists at all is a powerful demonstration that its current "gene team" is a champion. Should the owner of the team, the gene-set itself, make changes? In this case, we know what the HIV-gene-set has decided: make lots and lots of changes. In fact, the HIV-gene-set insists on so many changes that most daughters are mutants of their parent. Can this really be wise? It is, but only because the gene-set is very careful about where and how often those changes will occur.

Through evolution, HIV has "learned" that the future is always dismal. Typically, each virus lives in a human with an immune system that is trying to kill it.

The immune system is like a trillion-person police force composed of cells that are constantly on the lookout for intruders. From the point of view of many immune cells, an HIV virus looks like a ball wearing a coat made of glycoproteins. The first of these cells to spot an HIV intruder clones itself to produce a multitude of cells, each of which carries a "kill on sight-poster" with a picture of the intruder's coat. The virus is now the hunted.

What would you do if you were the virus? Yep, change the coat – put on a disguise. Well, that is exactly what the HIV-gene-set has designed the virus to do.

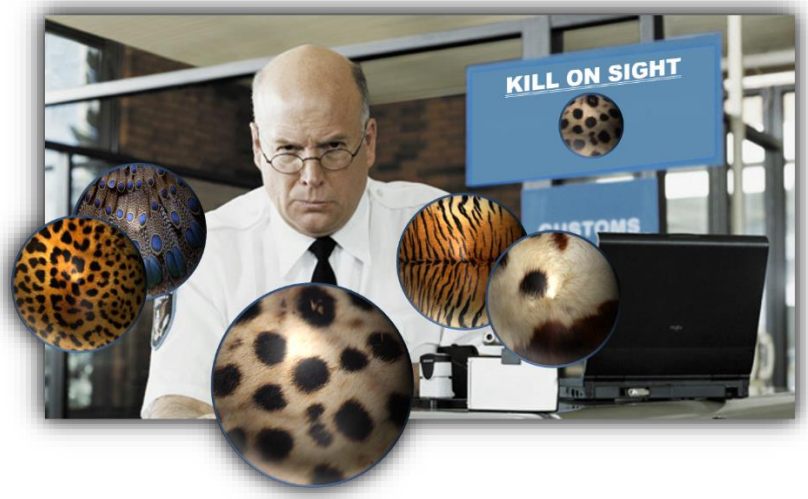

Figure 12: HIV mother and her daughters at a checkpoint

Here is one way HIV does it. The virus stores its gene-set in RNA molecules. To increase the copy numbers of the genes, the virus (with the help of its host) makes new copies of itself - including its RNA molecules. The virus uses a wonderful protein called a polymerase to accomplish this. Polymerase proteins are what life is all about, without them DNA and RNA molecules don't reproduce, cells don't reproduce, and you don't reproduce. By the way, if you watched the Drew Berry video, you "saw" polymerases beginning at about 3min. 39 sec. They are the purple and green objects.

There are slightly different polymerases in different species, but all polymerases are similar. They are really, really, small, about 2,000,000 of them side-by-side would be the diameter of a penny, and they all behave like jugglers on a tightrope. In our wildest dreams, we scientists could not build nano-machines even remotely as amazing as the polymerase proteins.

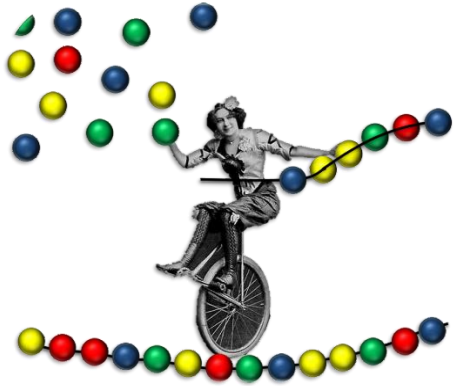

Figure 13: The Amazing Poly: juggler extraordinaire.

Think of a nucleic acid molecule as a strand of beads, where each bead can have one of four possible colors. A polymerase protein will hop onto a strand and begin to move down the beads. As it passes each bead it reads the color and uses that as a guide to put a bead of the appropriate color onto a new strand it is stringing. When it is done, the new strand stores the same genes as the original one did – unless the polymerase makes a mistake.

One thing humans can be proud of is that their main polymerase[6] almost never makes a mistake. It only puts a bead of the wrong color into the new strand about once in every billion beads (Wikipedia-Dn). By comparison, the HIV polymerase is pathetic; it puts a bead of the wrong color into the new strand once every couple of thousand beads (Preston, Poiesz, & Loeb, 1988).

The HIV virus should be ashamed of its error-prone polymerase. Well, it's not, and it did it on purpose. Why?

Because the high rate of mistakes ensures that many daughters have mutated copies of the glycoprotein-coat gene and hence different coats than their parent. So, when the immune system's police show up looking for the original intruder's coat, the daughters are already wearing next season's fashion and are allowed to pass unmolested.

Eventually, the immune system figures out the deception, begin searching for the new coats, and the process will repeat.

When all is said and done, the HIV-gene-set has used the polymerase-gene to ensure a rapid rate of mutation of the glycoprotein-gene. The polymerase-gene itself has evolved over time, and the current polymerase bears in its very structure the lessons of the past.

As an aside, since human polymerase almost never makes a mistake, and the damage caused by external insults, such as cosmic rays, is usually repaired, how do humans create variation? It appears that sexual reproduction with homologous recombination plays a major role.

---

[6] *Humans have about 20 different polymerases, some are more error-prone than others.*

Using polymerase is just one way HIV creates intentional mutations. The configuration (3-dimensional shape) of RNA is largely controlled by the genes. It is easy to imagine that manipulation of this configuration can lead to "hot spots" where the rate of mutation or recombination is high and "cold spots" where it is low. HIV apparently creates "hot spots" in the glycoprotein gene where the rate of recombination is as much as 10 times higher than in surrounding regions (Galetto, et al., 2004).

As an aside, it appears that SARS-CoV-2, the virus that causes Covid-19, has a so-called proofreading exoribonuclease associated with its polymerase. Proofreading is typically a means by which a virus reduces the number of copying mistakes its polymerase makes. HIV polymerase has no proofreading. By reducing the number of mistakes, SARS-CoV-2 has forgone an important mechanism for avoiding host immune systems. Perhaps the SARS-CoV-2 has other means of avoidance, but if it does not, it will survive only until the host's immune system detects and destroys it – typically within a week or two. Since the virus cannot stay long, to avoid extinction it must infect new hosts quickly. In such viruses, selection may strongly favor mutants with improved infectivity, as has been observed in SARS-CoV-2.

Prene-sets are remarkably precise in controlling how often they gamble and where bets are placed. For example, the human-gene-set has developed a process called V(D)J rearrangement to create a high rate of mutation in the genes that encode antibodies (Wikipedia-An). Antibodies are shaped like the letter Y. The bottom stem is called the constant region, and the two arms are called the variable regions. V(D)J rearrangement virtually guarantees that no two antibodies have the same variable regions, while all antibodies retain the same constant region.

## HIV is amazing

A few additional remarks regarding HIV. There are an estimated thirty-five million humans infected with HIV, so its program for the creation of variation and more generally for survival is succeeding. What can we say about that program?

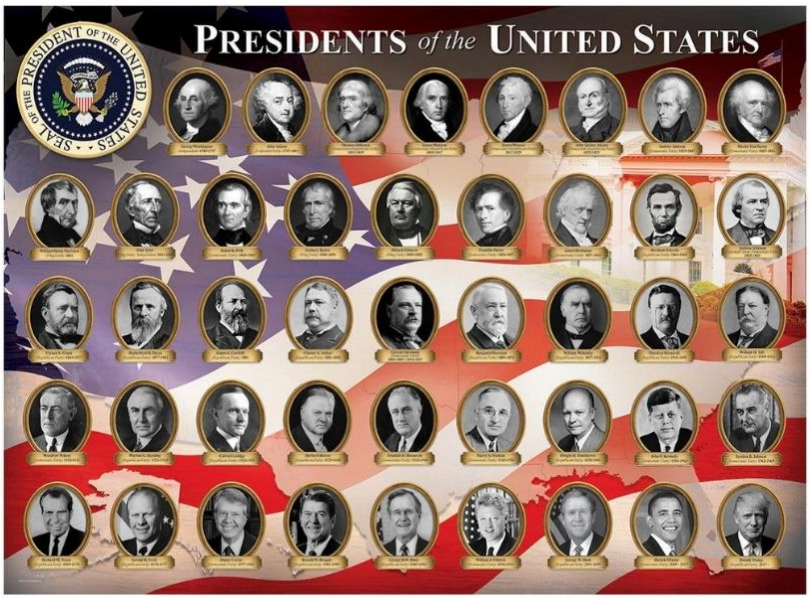

Figure 14

few additional remarks regarding HIV. There are an estimated thirty-five million humans infected with HIV, so its program for the creation of variation and more generally for survival is succeeding. What can we say about that program?

Below is an image of all past presidents of the United States.

It is stored on my computer as a file of 265,696 bytes.

Here is a much lower resolution version:

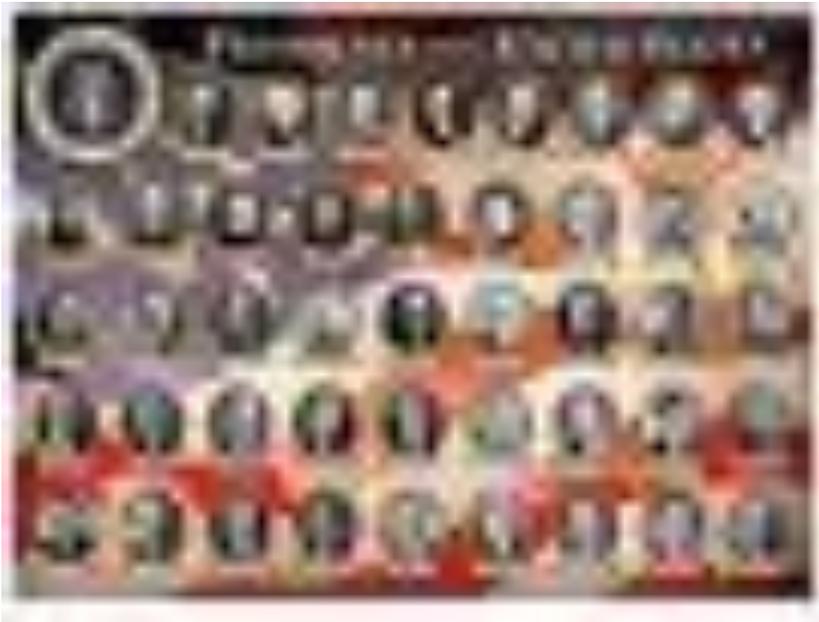

Figure 15

It is stored in a file of just 2,639 bytes. You can't store much in 2,639 bytes, right? What about the HIV-genome-prene? How many bytes would be needed to store it in a computer as a sequence of A's, C's, G's, and U's (for Uracil)? Only 2,437.

Even though it is tiny, the HIV-genome-prene is the program that creates variation and builds the viruses that infect humans, survive, and reproduce, despite immune systems and drugs. It is a program written by evolution over eons. Our computer industry has nothing that compares; it is worthy of our admiration. The human-genome-prene requires a file of roughly 750,000,000 bytes, so it is not surprising that we can do and think amazing things.

Like all humans, I have many beliefs and these beliefs need not be consistent (see *Part II: Humans* and *The gold star of truth*). I can admire HIV with one set of beliefs and abhor it with another. By adopting a prene-centric view, perhaps I can acquire a useful understanding of the viral enemy (Adleman & Wolfsy, 1993).

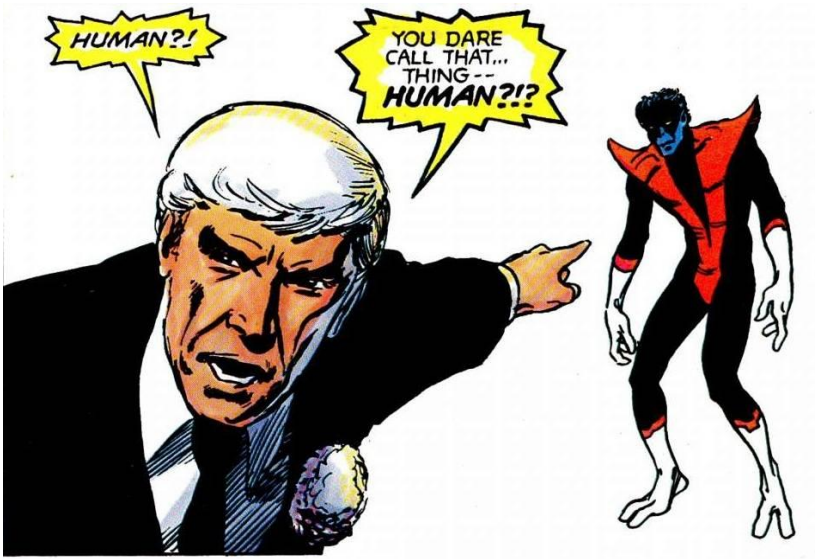

Figure 16: Death to mutants!

It is likely that most HIV mutants are unsuccessful because their mutations have left them unable to assemble correctly, infect cells, or do a thousand other things necessary to be viable and reproduce.

Why would HIV invest so much in unsuccessful mutants? Of course, on rare occasions there is a huge payoff when a mutant arises that can escape the host immune system that will soon kill its parent and most of its competitors, reproduce rapidly, and improve the copy numbers of many genes in the gene-set.

Gene-sets are like stock investors, but instead of investing money in shares and collecting returns in money, gene-sets invest resources in copies and collect returns in copies. Like stock investors, gene-sets do not have crystal balls to tell them which investments will produce the best returns, so they spread their investments over a diversified portfolio.

Successful stock investors use lessons from the past to guide future investments. Successful gene-sets do the same. HIV's error-prone polymerase stores some of the lessons of its past environments, and it uses these lessons to guide its future investments in offspring.

Because the environment changes, but not randomly, the best stock investors and gene-sets can use the lessons of the past to invest wisely enough to absorb losses and still end up with an overall profit.

A particularly important mechanism that destroys mutants is filicide. For example, an *E. coli* mutant that cannot reproduce faster than its parent will lose the fight for resources and soon disappear from the environment.

In battles between well-established societies and their mutants, the mutants seldom prevail. The parental society usually has many instruments at its disposal, while the mutant has few. In addition, it is likely that the parental society is a grizzled veteran of numerous wars with mutants and has developed effective weapons to combat them.

For example, one of the simplest of these weapons is waiting. Let the mutant die on the vine; see if its membership will decline rapidly to nothing. If that fails, then what might work is to diminish the mutant's growth potential by casting early members as unworthy. They are crazy, possessed, or criminal. If that does not work, then there is always assassination.

Consider well-established religions. They will typically have symbiotic relations with well-established secular societies that have weapons that can be brought to bear on mutants.

So, some combination of the Jewish and Roman societies condemned Jesus to death, some combination of the polytheistic religious and governmental societies of Mecca condemned Mohammed to death, and some combination of the Catholic and Holy Roman empire societies condemned Luther to death. Of course, these examples are antithetical since the societies that Jesus, Mohammed, and Luther engendered were ultimately very successful, but thetical examples are hard to find, presumably because the weapons worked so well that in most cases the mutants were vanquished without leaving a historical trace.

One way to avoid filicide, is to flee to a new environment where the parent(s) is(are) absent or weak. This appears to be part of the reason young bears leave their place of birth and dandelion seeds float away from their parent on a breeze. This also seems to be the approach that led to the survival of the societies put into motion by Jesus, Mohammed, and Luther.

## Cultural-prenes in the social sciences

Biology is the study of living things. This view has persisted for several millennia, but for some of today's biologists, it has been augmented by a complementary view that biology is the study of genes. From this view, living things are seen as instruments used by their genes to preserve and reproduce themselves. Together, these views have transformed our understanding of biology.

Social science is "The study of human society and social relationships" (OED, 2022). The question arises as to whether a complementary view of social science as the study of cultural-prenes might be of value. From this view, humans are seen as instruments used by cultural-prenes to preserve and reproduce themselves.

In the next two chapters, I will provide examples to illustrate the potential value of the cultural-prenes view. Related issues will be discussed in *Creatures of society*.

.

# Greatness

> The History of the world is but the biography of great men.
>
> -Thomas Carlyle
>
> *(Carlyle, 1841)*

Carlyle's view is not held in high regard today, but it does raise the question of how people become "great", and more generally, how members of a society acquire the positions they hold.

How does a person become a president, a monarch, or a pope? For example, how did Jorge Mario Bergoglio become Pope Francis? From the traditional view of history, much is known. He was born in Buenos Aires, had early jobs as a bouncer and janitor, joined the Church and moved from priest, to archbishop, to cardinal, and eventually to pope when his predecessor, Pope Benedict XVI, resigned (Wikipedia-Po). There are numerous biographies that provide additional information.

Here is a cultural-prenes view of how Bergoglio became pope:

From the cultural-prenes perspective, a society's cultural-prenes are primary and the human members of the society are among the instruments used by the cultural-prenes to enhance the prenes chances of survival. A major means by which they do this is to assign tasks to members of the society.

In many societies there are numerous positions that a member may hold. These positions are typically associated with ranks, and as a member advances from one rank to a higher one, they may receive rewards such as wealth, prestige, personal satisfaction, or fame. These rewards provide an incentive for a member to move from their current position to one of higher rank.

To attain a position of higher rank, a member will typically be required to satisfy various conditions. For example, they may be required to demonstrate particular physical or educational attributes, or the support of others of high rank. They may be required to prevail in competition with others seeking advancement.

Where do we find the positions that are available, and for each the rewards that accrue and the conditions that must be satisfied? In the cultural-prene-set itself. For example, the US Constitution stores cultural-prenes that describe the position of president of the United States and enumerate some of the conditions that must be satisfied to obtain that position. Similarly, the apostolic constitution, *Universi Dominici gregis* stores some of the conditions required for obtaining the position of pope. I'll call the set of cultural-prenes that determine the positions, rewards, and conditions, the "positional-cultural-prene-set" of the society.

Different societies will typically have different positional-cultural-prene-sets. The conditions for becoming a president, pope, czar, or dalai lama, will be distinct. But in each case, by the time a person assumes such a lofty position, it is likely that they will have spent a large part of their lives immersed in their society's cultural-prenes, become a "true believer" in many of them, and acquired the skills necessary to defend and spread them.

But why does a society's cultural-prene-set contain positional-cultural-prenes at all? The reason is *Proposition 1*. The cultural-prenes are struggling to survive and the positional-cultural-prenes aid in this struggle. Here is how:

In addition to the rewards that accrue when achieving high rank, there is a "special" power that also accrues and is particularly important when considering the cultural-prenes point of view. It is the power to mutate some portion of the cultural-prene-set for some portion of the members. The higher the rank the more mutational power.

For example, for Catholics as one moves from priest to bishop to cardinal to pope, the set of cultural-prenes accessible to mutation increases to include those of greater and greater importance, and the flock of followers who are impacted grows. A priest may express his ideas regarding birth control and perhaps influence the behavior of a parish, but if a pope does it, as Paul VI did in 1968, and John Paul II did in 1994 (Wikipedia-Ch), the future behavior of millions of people may be changed dramatically.

People become "great" by achieving high rank and exercising their mutational power to create "great events". For example, Constantine achieved the very high rank of emperor and consequently had immense mutational power. Following his personal conversion to Christianity, he used that power to remove some pro-pagan and pro-Jewish cultural-prenes from the Roman cultural-prene-set and add some pro-Christian ones to it. For many Christians, this is seen as a critical step in a process that ultimately resulted in Christianity becoming the state religion of the Roman Empire and subsequently the largest religion in the world. Hence, many Christians see this as a great event and see Constantine as a great man. For some Jews, this is seen as a critical step in the emergence of antisemitism in western societies (Wikipedia-Hi). Hence, these Jews may also see this as a great event and see Constantine as a great man, but that greatness may not be viewed positively.

To stay fit in a changing environment, all prene-sets must mutate. As we saw in *Variation and natural selection*, prene-sets do not leave mutation to chance; instead, they carefully program it. Prene-sets that handle this task well, increase their probability of survival.

It appears that the cultural-prene-sets of virtually all successful societies have "discovered" the use of positional-cultural-prenes to program mutation.

Rather than leave mutation to chance, these societies reserve the greatest power to make mutations for those who have achieve the highest ranks. They use positional-cultural-prenes to impose conditions that filter out members of the society that are wanting. Only those members who have demonstrated the commitment and skills needed to protect and spread important cultural-prenes achieve the highest ranks.

As an aside, it might be interesting to consider Kuhn's notion of a paradigm shift (Kuhn, 2012) in the context of positional cultural-prenes and *Proposition 1*.

Societies frequently contain subsocieties. For example, the American society contains the California society which contains the Los Angeles society which contains the Walt Disney Company society. Each of these subsocieties will have its own cultural-prene-set and positional-cultural-prene-set that indicate the positions available and the requirements for obtaining them. These subsocieties are often in conflict, fighting for resources. The collection of subsocieties form a kind of societal ecosystem. These ecosystems can be as intricate as those considered in biology.

High rank is not permanent. For example, the positional-cultural-prenes of America call for presidential elections every 4 years and a maximum term of 10 years. Once a president leaves office, the power to mutate the cultural-prenes of American society is lost, though that power may persist within their political-party's subsociety. It is not unusual for those of high rank to use their power of mutation for the purpose of retaining their own position and rank, or to create new positions and remove old ones.

One of the virtues of the cultural-prenes view is that it can be divorced from many human concerns such as morality. With the gene-centric view of biology, it is of little importance whether humans consider a species to be cute, vicious, repulsive, disgusting, etc. The only question worth asking is whether it is successful. Did it increase copy numbers? Did it survive? Similarly, with the cultural-prenes view of the social sciences, it is of little importance whether humans consider a society to be evil, fair, war-like, Godfearing, etc. All that matters is whether it is successful.

# The prophet and the messiah

In *How to be an unsuccessful prene*, I discussed the Shakers and noted that their cultural-prene-set appeared to contain the seed of its own destruction. That seed was the celibate-prene:

*sex is forbidden.*

In this chapter, I will speculate on possible seeds of destruction that may be present in the Jewish cultural-prene-set.

We will need a bit of nomenclature. I'll refer to "Torah society". The members of Torah society are those people who treat the Torah as sacred. Those members would later be called People of the Book in Islam. Roughly speaking, BCE all members of Torah society were Jews, and all Jews were members of Torah society. So, Jewish society and Torah society began as the same thing.

As indicated in *Greatness*, many societies use positional-cultural-prenes to define the positions members may hold, and for each position, the conditions that must be satisfied to attain it and the amount of mutational power that accrues. I suggested that successful societies use these positional-cultural-prenes to ensure that those who acquire great mutational power have the ability and motivation to act in the interest of the cultural-prene-set. Presumably the conditions required for such a position ensure that those with inadequate ability or motivation are filtered out.

Consequently, when a society has positions with great mutational power, but the conditions filter out almost no one, it raises a suspicion that the positional-cultural-prenes are not being used in an optimal way.

It appears that the Torah cultural-prene-set contains positional-cultural-prenes that create these kinds of positions, and that Jewish society has suffered greatly because of it. Consider the following prenes:

- The prophet-prene: God will reveal His word to special individuals called prophets.
- The messiah-prene: in the future a great leader will become king of the Jews and restore peace and justice to the world.[7]

Let's start with the prophet-prene.

> *Listen to my words: "When there is a prophet among you, I,*
> *the LORD, reveal myself to them in visions, I speak to them*
> *in dreams.*
> *Numbers 12:6 (NIV)*

From the cultural-prene perspective this is a positional-cultural-prene that defines the position of prophet and associates it with an immense power to mutate the cultural-prene-set.

---

[7] *For centuries, biblical scholars have investigated what notions like messiah and prophet have meant to Jews, Christians, and Muslims at various times. I defer to them on such matters.*

What were the conditions that had to be satisfied to attain this powerful position? The Torah did provide some guidance on how unworthy candidates were to be filtered out, but the instructions were not that clear. For example:

> *You may say to yourselves, "How can we know when a*
> *message has not been spoken by the LORD? If what a*
> *prophet proclaims in the name of the LORD does not take*
> *place or come true, that is a message the LORD has not*
> *spoken. That prophet has spoken presumptuously, so do not*
> *be alarmed.*
> *Deuteronomy 18:21-22 (NIV)*

As a result, in Torah society, virtually anyone could claim to have satisfied the necessary conditions and declare themselves a prophet. Since there was no practical means of accepting or rejecting such a claim, it might be accepted by some members and rejected by others. Perhaps most often, the subsociety of individuals who accepted the claim went extinct in short order, but in the case of Mohammed, it did not. It grew and became the Islamic society.

Now consider the messiah-prene. As with the prophet-prene, we have a positional-cultural-prene that defines a position, messiah, and endows it with great power to mutate the cultural-prene-set.

As with the prophet-prene, the conditions that are needed to become messiah are poorly defined. Perhaps one could argue that the Torah does require that the messiah be from the line of King David, but even today, that would be virtually impossible to determine.

In any case, almost anyone could claim to have satisfied the necessary conditions and be a messiah. Jesus made such a claim, as have many others (Wikipedia-Lis). In Jesus's case, the subsociety of people who accepted the claim eventually flourished and became the Christian society.

So, while initially all members of Torah society were Jews, after Mohammed and Jesus they could be Jewish, Muslim, or Christian.

So, Torah society now had three distinct sub-societies. The advent of the Muslim and Christian societies would contribute mightily to the growth of Torah society. However, since their inceptions, they have been in continuous prene-wars with Jewish society (and with each other). They are winning. While currently Christianity and Islam are the religions of over 55% of the world's population, Judaism is the religion of less than one-fifth of one percent (Wikipedia-Ma).

Interestingly, Islam removed the possibility of future prophets from its cultural-prene-set.

> *Muhammad is not the father of any of your men, but he is*
> *the Messenger of Allah and the Last of the prophets*
> *-Quran 33:40*
> *(Shakir)*

# Non-biological evolution

We have grown accustomed to applying Darwinian evolution to genes. One of the main goals of this book is to provide evidence that Darwinian evolution applies to all prenes.

To survive, a prene must keep its copy number above zero. To do that it must have a means of creating new copies.

For genes there is essentially one way to do this. Inside each cell is a powerful machine (a polymerase) that makes copies of nucleic acid molecules and hence of the genes stored in them. The cell has other machines that produce multiple identical copies of other molecules that will be used in the construction of new cells. If a gene wants to survive (and they all do), it must get its hands on some cell's machines.

Cells have their own copy machines, but viruses do not. To survive, a virus must find a cell with machines it can use. But because genes are selfish, cells are too, and they have no intention of providing viruses with access to their machines. So, what is a virus to do? Typically, the virus resorts to crime. It finds a way to circumvent the cell's security system and use the required machines without permission. Since current, very rough, estimates are that there are about 10^31 viruses and about half that number of cells in the world at this moment (for comparison, there are an estimated 10^25 stars in the observable universe), there are a lot of illegal copies being made.

Cultural-prenes are not exempt from the need to make new copies. The machines inside cells are of no use to them; they must find other means.

Once human communication had developed sufficiently, those who had acquired cultural-prenes could make new copies by teaching the prenes to students.

In recent times, new options for cultural-prenes have appeared. Especially since the beginning of the industrial revolution, humans have created a constant stream of remarkable machines that make copies of all sorts of things with ever increasing speed and efficiency. If you want to make a hundred thousand copies of this book, or ten thousand identical hammers, or a million identical cellphones, no problem. The cultural-prenes of today's successful societies make use of the most powerful machines available to increase copy numbers and survive.

As with cells, the individuals who own powerful machines have no intention of letting others use them for free. The wealthy and powerful may be able to compensate owners directly, but typically the commoner cannot. As historians are well aware, this results in historical documents that are not representative of the society as a whole.

In *Extinction*, we will explore the fate of prenes that do not gain access to powerful machines. In *The Hamlet's-soliloquy-prene's struggle*, we will see how Shakespeare's prene gained access to one of the most powerful machines of his time.

# Extinction

Every 8-year-old knows that a long time ago an asteroid hit the earth and killed the dinosaurs. More precisely, about 66 million years ago, the Chicxulub impactor struck the earth and initiated environmental changes that drove many species to extinction and facilitated the survival of others, including some mammalian species from which we evolved. This so called Cretaceous–Paleogene extinction was one of five major extinctions that have been identified. Though it began in an instant, the high rate of species extermination it initiated persisted for thousands of years.

The Chicxulub impactor left a clear trace of its destruction called the Cretaceous–Paleogene boundary consisting of a thin layer of sediment found round the world. For many species, fossils are found below the boundary but not above it – these are likely to have gone extinct prior to or as a result of the impact. For other species, fossils are found above and below the boundary – these are likely to have survived the impact. For still other species (including those of *Homo sapiens*) fossils are found above the boundary, but not below it – these are likely to have descended from the survivors.

In prene-theoretic terms, we might say that many genes that existed before the impact went extinct, and from those that did survive, many new genes evolved.

Here, we will consider another type of major extinction that has little to do with genes (that is, biology) and much to do with the cultural-prenes of societies.

The study of the human past is often broken into two parts, prehistory and history. Informally, a society's prehistory is all the things that happened before the society acquired writing (if it ever did), and its history is all the things that happened after.

What do we know about the cultural-prenes of societies that never developed writing? For example, the Cahokian Moundbuilders of the American Southwest, the Nok of Nigeria, the Scythians of Central Asia, and all societies that disappeared over 5500 years ago when, it is thought, writing first appeared among the Sumerians. We rarely know how children were to be raised, what rules governed the relationships of men and women, what the punishment for theft was (or if there was a punishment at all), what behavior was taboo, what the most important stories were, etc. Sometimes we only know of their existence because of the material artifacts they left behind or because they are mentioned in the writings of literate societies that encountered them.

On the other hand, what do we know about the cultural-prenes of societies that had writing, such as those of the Maya, the Romans, or the Greeks? Regrettably, not as much as we would like, but more than we know about non-writing societies.

So, of the colossal number of cultural-prenes that have arisen at some time in the past, those that were never written down have mostly gone extinct. Those that were written down had a better chance of surviving.

Here are a few of the possible reasons why the advent of writing favored cultural-prenes that were written down and disfavored those that were not.

Before a cultural-prene was written down, it likely existed only as a meme in brains, and new copies could only be created by the passage of memes from one person to another. This so-called "oral tradition" would allow a cultural-prene to survive as long as those who stored it could find new brains to memorize it.

Once a cultural-prene was written down, it had a new host – the written document. Since written documents could be transcribed, the cultural-prene had a new mechanism for increasing copy number. The cultural-prene could now use both the oral tradition and the "written tradition" as means by which it could struggle to avoid extinction. Further, because cultural-prenes that are memorized can be written down, and those that are written down can be memorized, even if one tradition failed, the other could resurrect it.

For example, it is currently thought that the society, perhaps that of the Essenes, that wrote many of the Dead Sea Scrolls may have been extinct for about 2000 years when the scrolls reemerged in the middle of the 20th century, and the cultural-prenes they stored were resurrected.

In this regard, consider the now extinct bronze age society of the Indus Valley. Many of its cultural-prenes are like Schrodinger's cat. The oral-tradition has failed them; they are not memes in any living person's brain; however, archeologists have discovered many objects with strings of symbols. Unfortunately, no one can read them. If these symbols are the components of a language, and if someone deciphers the Indus Script, some of the cultural-prenes of the society will be resurrected. However, if these symbols are the components of a language, but the script is never deciphered, then the objects are merely the tombs of moribund cultural-prenes.

While the oral tradition could make copies by transferring a meme from one person to another, it could only do so when the people involved were at the same place at the same time. But written documents allowed for the passage of memes from one person to another displaced by thousands of years and thousands of miles. It is sometimes noted that this gave rise to great empires, because war, governance, and commerce could be conducted over great distances. Great empires like that of Rome would spread their cultural-prenes to non-Roman societies by conquest, thereby increasing the copy numbers of Roman cultural-prenes while hastening the extinction of the cultural-prenes of the conquered societies.

If writing emerged in a society, its arrival would have initiated the society's "prehistory-history-extinction".

There have been numerous cultural-prene extinctions brought on by the advent of important new technologies for making copies, making them more efficiently, making them more durable, and so on. Most of these extinctions are of little importance, but even major ones occur fairly often. For example, the papyrus-parchment extinction.

In the next chapter, we will see details of how the Hamlet's-soliloquy-prene survived the "printing-press-extinction"; though some of Shakespeare's other prenes perished.

Only cultural-prenes that survive the extinction events they encounter will have a lasting influence on the societies of the future.

When will the next major cultural-prene extinction occur? It has already begun. It is the "Internet-extinction", which may turn out to be the largest ever. It is easy to see that the cultural-prene-sets of many societies are engaged in a frenzied competition to use the Internet to help them survive.

These cultural-prenes-sets are typically competing for human followers to exploit. The cultural-prenes of political parties will use these humans to, among other things, increase the number of votes they receive. The cultural-prenes of commercial companies will use them to purchase products. We have already seen some dinosaurs fall. Blockbuster, Sears, and many brick-and-mortar bookstores no longer exist, and the cultural-prenes specific to each are on their way to extinction.

## The Hamlet's-soliloquy-prene's struggle

Since Darwin, we have understood that all living things struggle to survive. We are now ready to see that non-living things, like the Hamlet's-soliloquy-prene, also struggle.

When humans consider struggle, they expect to see features like suffering and fear. When lions and zebras struggle, what we see conforms to our expectations. However, as we descend the evolutionary tree these expectations are typically not met. Trees, grasses, flowering plants, sponges, bacteria, and viruses all struggle to survive, but since they lack brains there is no suffering or fear.

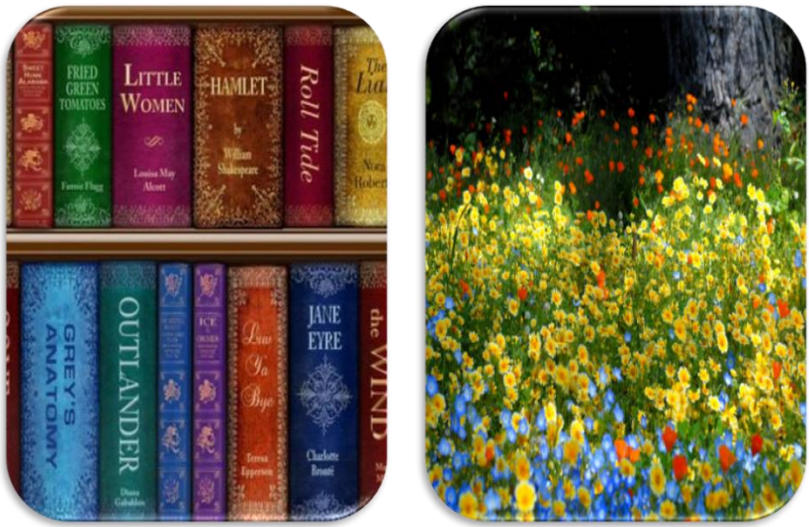

Figure 17: Can you see the struggle?

Once we strip "struggle" of our expectations, we will see that all prenes struggle for survival in essentially the same way.

Let's begin with the Hamlet's-soliloquy-prene.

Where did the Hamlet's-soliloquy-prene come from? Well, it came from Shakespeare's brain (you're welcome). Prior to creating the soliloquy, Shakespeare's brain did not store the Hamlet's-soliloquy-prene but did store lots of theatrical memes. We can be pretty sure of this because Shakespeare was an actor, playwright, part owner of an acting group (the Lord Chamberlain's Men), and part owner of a London theater (The Globe).

Shakespeare's brain processed his existing memes and created new ones such as the Hamlet's-soliloquy-prene.

Now that the Hamlet's-soliloquy-prene had copy number one, it began its struggle to avoid extinction. At this point, its only instrument was Shakespeare. Fortunately for the Hamlet's-soliloquy-prene, he turned out to be very useful.

Shakespeare was in the right place at the right time since plays were a major form of entertainment in Elizabethan London. As part owner of the Lord Chamberlain's Men, Shakespeare could ensure that his plays, including Hamlet, would be performed in front of large audiences (sometimes including Elizabeth I herself). Presumably, some new copies of the Hamlet's-soliloquy-prene ended up stored in the brains of actors or in manuscripts for use in future productions. But it is not easy to memorize the soliloquy or write it by hand. So, the Hamlet's-soliloquy-prene was increasing copy numbers at a slow rate. It was relying on the oral and written traditions for its survival. But like the oral tradition, the written tradition could fail. No manuscript of Hamlet, or any other Shakespearean play, from Shakespeare's time has ever been found. It appears that the written tradition failed them all.

But a new important copy machine, the printing press, had appeared, and many cultural-prenes, including the Hamlet's-soliloquy-prene, would compete to use it to increase their copy numbers.

By 1600 when Hamlet was written, the printing press had been around for about a century and a half. The printing-press-extinction was underway. In a manner similar to that described for the prehistory-history-extinction (see *Extinction*), a cultural-prene that did not get printed was more likely to go extinct than one that did. For example, there is historical evidence that some of Shakespeare's plays were never printed and are now lost. Only the names of these plays remain (Shakespeare, 2023).

While all this was going on, a second prene was increasing copy numbers at a rapid rate. I'll call it the Shakespeare-is-a-great-writer-prene. For example, perhaps some people who attended Shakespeare's productions or read his poems, enjoyed them, or thought them deep, and came to believe a Shakespeare-is-a-great-writer-meme. Others learned the meme (which was short and easily passed) from those who already believed it. In modern day language we might say Shakespeare became a celebrity. In fact, history suggests this was the case (Dobson, 2023).

It is not clear why the Shakespeare-is-a-great-writer-prene is worth mentioning in the story of the Hamlet's-soliloquy-prene's survival, but it soon will be.

Fortunately for the Hamlet's-soliloquy-prene (and lots of other Shakespearian prenes), new extremely useful instruments appeared in the form of publishers, printers, booksellers, and others with commercial interests. Presumably, hoping for profit they printed Hamlet in quarto form in 1603. This quarto, sometimes called the "bad quarto", was of questionable quality. For example, the soliloquy begins:

> To be, or not to be, Ay there's the point, To Die, to sleep, is that all?

Following Shakespeare's death in 1616, John Heminges and Henry Condell gathered what materials they could and published the first folio containing Hamlet and 35 other plays (18 of which had not been previously printed). The publication was apparently a business success (Wikipedia-Fi), and eventually the world would be inundated with editions of "The Complete Works of Shakespeare" modeled on it. The Hamlet's soliloquy-prene had gone viral.

Well, that seemed easy, why didn't all cultural-prenes use the same method to get printed and go viral? The answer is natural selection. Only those cultural-prenes with very special properties were fit enough to get printed and survive.

In the case of plays, poetry, stories, etc. the agent of natural selection is often a publisher. Typically, printers will only use their machines to print copies of your book if they get paid; publishers will only pay printers to make copies if they believe the copies will sell. Ask any budding novelist how easy it is to convince a publisher to use their resources to print your book. But, if you are a movie-star, a mass murderer, or some other form of celebrity, the publisher may look kindly on your proposal. That is where the Shakespeare-is-a-great-writer-prene comes in; Shakespeare was a celebrity. It is likely that celebrity contributed mightily to the publisher's belief that copies of Shakespearean prenes would sell, and why the Hamlet's-soliloquy-prene survived the printing-press-extinction and other extinctions that have occurred since.

As an aside, this concept that even those who were not wealthy or powerful could gain access to powerful copy machines, not by providing upfront compensation to the machine owners, but by providing backend compensation in the form of sales, or more recently advertising, has provided the commoner with a greatly amplified historical voice.

So that is how the Hamlet's-soliloquy-prene successfully struggled to avoid extinction. You may say: "that's it?". Yes, that's it. It is a theme with many variations, but every prene's struggle looks much like this one.

Let's look at a struggle from the gene-world.

How did the first copy of the smallpox-genome-prene get created? We can only guess, but here is a plausible answer.

Several thousand years ago, a human cell became infected with a poxvirus similar to, but different from, smallpox. Inside that cell, the poxvirus mutated and became the smallpox virus. In prene language, we might say that the poxvirus-genome-prene was processed by the cell and the smallpox-genome-prene was produced.

That cell was the smallpox-genome-prene's Shakespeare, and it seems to have been at least as effective an instrument for the smallpox-genome-prene as Shakespeare was for the Hamlet's-soliloquy-prene.

That cell used its copy machines to make lots of new smallpox viruses, each of which stored the smallpox-genome-prene in its DNA molecule. That cell also released those viruses into the environment, where they had whatever special properties were necessary to infect new cells and use their copy machines to make more viruses. The smallpox virus made its way from cell to cell, human to human, and eventually to trillions of cells in billions of people. The smallpox-genome-prene had gone viral. Like the Hamlet's-soliloquy-prene, it had struggled and survived.

So, is there an important difference between the struggle of the smallpox-genome-prene and that of the Hamlet's-soliloquy-prene? Well, yes and no. The two prenes live in different worlds with very different environments, so we should not expect the details of their struggles to be the same, any more than we should expect the details of the struggle of elephants to be the same as those of algae, or of a living thing on earth to be the same as a living thing elsewhere in the universe. But, to my eyes, while the details of the struggle depend on the environment, the laws governing it remain the same.

You might point out that Hamlet's soliloquy is beautiful, moving, and profound, while smallpox is none of those things. Further, that if this were not the case, few humans would have cared about the soliloquy, and it would probably have gone extinct. I suspect this is so. But the smallpox virus also had certain special properties, which we might describe using words like avidity, affinity, and hydrophobicity. These properties made the virus effective at infecting cells and appropriating their copy machines to make new viruses. Without such properties the smallpox-genome-prene would not have survived either.

You might not like the fact that the Hamlet's-soliloquy-prene did not reproduce itself, its copy numbers only rose because humans reproduced it with their printing presses (and later other machines such as computers).

But the smallpox-genome-prene cannot reproduce itself either and its copy numbers only rose because humans reproduced it using their cellular replication machines. Both prenes are human symbionts.

Let's look at one more example, that illustrates what an iffy thing it is for a prene to survive.

Consider Bach's Brandenburg-concerti-prene. Roughly speaking, everyone in Bach's family was a composer, so his head was filled with musical memes. Bach was cantor at St. Thomas Church in Leipzig, a job which required him to expend time and brain-cycles composing music. Bach processed his existing musical memes and created the Brandenburg-concerti-prene.

At the time of its creation, Bach was the sole instrument of the prene. But surprisingly, he was not very good at it.

Perhaps Bach performed some version of the concerti during his lifetime, but there is no indication that the Brandenburg-concerti-prene's copy number grew significantly. Further, while some of Bach's musical-prenes were published during his lifetime, it appears that the concerti-prene was not.

The Brandenburg-concerti-prene did not make the jump to the printed page as the Hamlet's soliloquy had, but Bach did create a manuscript of the concerti in his own hand and sent it to Christian Ludwig, Margrave of Brandenburg. But it seems that Ludwig simply set it aside and forgot about it.

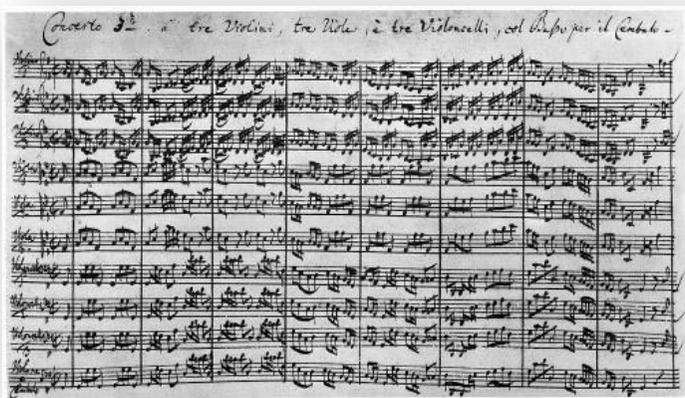

Figure 18: The third Brandenburg-concerto-prene stored in
a manuscript written by Bach.

Then Bach died. Things were not going well for the Brandenburg-concerti-
prene; it had not gone viral, in fact, it appears that its copy number was
one, and it had no human instrument to exploit. It seemed destined to be
just another victim of the printing-press-extinction.

As time went by, music evolved, and people stopped performing Bach's
works. Perhaps baroque music stopped getting "airtime" because music
publishers had more attractive options like the "rocking" new music of
classical composers like Mozart and Beethoven or romantic composers like
Schumann and Mendelssohn. For whatever reason, it appears that by the
early 19th century, Bach had largely been forgotten.

Then the vagaries of life smiled on the Brandenburg-concerti-prene. Some
of the musical prenes that Bach had written down in the early 18th century had remained, spore like, in churches and music rooms. In the early 19th century, some of these manuscripts, but not any storing the Brandenburg-concerti-prene, fell into the hands of Felix Mendelssohn. Mendelssohn began championing Bach's music. Bach's musical prenes had acquired a powerful new instrument to exploit. Mendelssohn organized performances of Bach's music. Bach became popular (a sort of posthumous celebrity), and his works began to be published as printed documents. Scholars searched for more of his old manuscripts.

Then in 1849, more than a century after its creation, and two years after Mendelssohn's death, the manuscript that Bach had sent to Ludwig was rediscovered. It was published the next year, and the Brandenburg-concerti-prene went viral.

So, the survival of Bach's Brandenburg-concerti-prene was a near thing. An unlikely sequence of accidents. A miracle? Hardly, compared to the smallpox-genome-prene's survival as described in *The resurrection of smallpox*, it was nothing special.

It seems likely that many genes of living creatures and many cultural-prenes that are currently thriving only exist because of an astonishingly unlikely sequence of accidents.

Commonly, when a new prene is "born", the greatest challenge it will ever face is getting to copy number two. Whether it gets there does not depend on its fitness in the greater world, but its fitness in the microenvironment surrounding its place of birth. In the end, very small changes in the microenvironment can have a profound impact on whether a prene goes viral or not.

The struggle of prenes to survive often resembles crystal growth. Crystals grow from spontaneously created nanoscopic seeds; many seeds form, some grow a bit larger, but almost all quickly dissolve back to nothing. Only rarely does a seed grow large enough to persist for a significant period. Often there is little or no difference between seeds, and those that grow large are not special, they are just lucky.

Had the Brandenburg-concerti-prene gone extinct, would it have still been a masterpiece? No doubt many now extinct human works would have been deemed masterpieces had they ever made it beyond their local environments.

## Socrates' bed

In *What is a prene?* I referred to deep philosophical reasons why we cannot provide an exact definition of prene. Here, we will get a bit deeper into this and use it to reveal some things about how memes evolve.

The philosophical problem, called the "problem of universals", dates back at least to the time of Plato, when in book X of the Republic (Plato, 1888), Socrates addresses the question: what is a bed?

> *Socrates: Let us take any common instance; there are beds*
> *and tables in the world --plenty of them, are there not?*
> *Glaucon: Yes.*
> *Socrates: But there are only two ideas or forms of them --*
> *one the idea of a bed, the other of a table.*

Plato has made an important distinction between the "idea of a bed" which is not a physical thing and "beds … in the world" which are.

In *What is a prene?* a similar distinction was made; a prene is a "unit of information", a non-physical thing, that can be stored in physical things. So, Socrates' idea of a bed, his "bed-idea", would correspond to a bed-prene, and Socrates' bed in the world would correspond to a physical thing that stores the bed-prene.

Plato does not tell us precisely what "ideas" are. But he does appear to consider them immutable. Regarding the idea of a bed, he writes:

> *Socrates: is made by God, as I think that we may say -- for*
> *no one else can be the maker*

And

> *Socrates: two or more such ideal beds neither ever have*
> *been nor ever will be made by God.*

So, for Plato there is exactly one immutable bed-idea. This is not the prene-theoretic view. How does a prene-theorist look at beds?

Each person stores a personal "bed-meme" in their brain. These bed-memes are analogous to Socrates' bed-idea in that a person can use their bed-meme to recognize some physical things as beds. For some people, it can also be used to guide the construction of real beds. Typically, different people store different bed-memes. For example, my bed-meme does not lead me to recognize hammocks as beds, but perhaps yours does.

So, Plato allows for exactly one bed-idea, but prene-theory allows for many different bed-memes.

Further, while Plato's bed-idea is immutable, bed-memes mutate through time. Let's look at how your bed-meme might have evolved as a result of a series of mutations.

I'll use the Wikipedia article "Bed" (Wikipedia-Be) for guidance.

> *Early beds were little more than piles of straw or some other*
> *natural material*

So, apparently some early humans had bed-memes.

> *An important change was raising them off the ground, to*
> *avoid drafts, dirt, and pests.*

So, some later humans had bed-memes that earlier humans did not. How did that happen?

The genes, for their own survival, have built people to be obligatory meme processors. We are constantly using our brains to process our existing memes and create new ones (see *The Hamlet's-soliloquy-prene's struggle*). Once memes concerning objects raised above ground level began to inhabit brains that already stored an old bed-meme, perhaps some person created a new bed-meme that led them to recognize some raised beds as beds. If such a person now saw a pile of straw on a table, they might lie on it. They might even take a pile of straw from the ground and place it on a table and then lie down. That person's bed-meme had mutated. That person might pass the new bed-meme on to offspring and acquaintances. If the new bed-meme conferred an evolutionary advantage such as less disease, better protection from predators, or simply better sleep, then it might be selected and outcompete the old bed-meme for brains.

At this point it is easy to see how a sequence of such mutations would account for the bed-meme that currently resides in your brain. So, yes, your bed-meme is the result of a sequence of mutations of earlier bed-memes.

But can we go back further in time? Did the bed-meme in your brain actually start to emerge before humans existed? I suspect it did.

Consider Chimpanzees (though they are not our predecessors, the point will remain). They build beds. There is considerable evidence that they have bed-memes that are passed from chimp to chimp, and that these bed-memes mutate and evolve through time.

> *Juvenile and adult apes devote considerable time and*
> *energy to the construction of a new sleeping platform or*
> *"nest" at the end of their daily active period. Current data*
> *suggests that infants and young juveniles (i.e., nursing*
> *young) acquire skills over years both through observation of*
> *their mother and practice*
>
> *...*
>
> *chimpanzees are keen observers of physical properties of*
> *trees, including stiffness, strength and leaf surface area, and*
> *... they select species that provide the widest range of*
> *advantages, including predator avoidance, postural stability,*
> *thermoregulation and pathogen avoidance.*
>
> *(Samson & Hunt, 2014)*

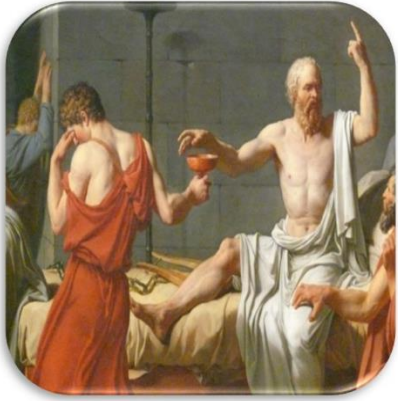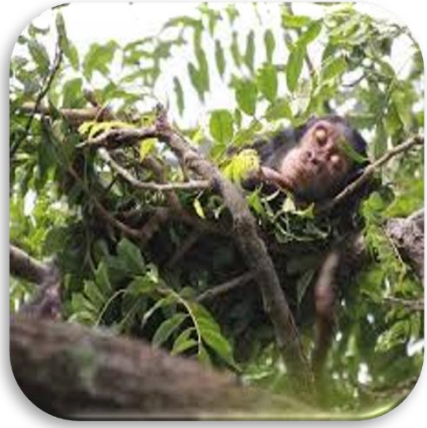

Figure 19: bed-memes

Well, if your bed-meme can plausibly be traced back to pre-human times, how far back does it go?

I suspect that there were genes and other prenes stored in single-celled organisms that were the mutatory predecessors of your bed-meme. Before you dismiss this notion, remember that you and all your genes appear to have arisen from those same single-celled organisms and the genes and other prenes they stored.

Beds are not special, you also have a personal table-meme, hammer-meme, Hamlet's-soliloquy-meme, and many others that have evolved in much the same manner.

The foundation has now been laid to provide a "better" definition of a prene. This can be found in *What is a prene - really?* I recommend that the interested reader look at that now.

With regard to the problem of universals, many approaches have been considered, but results in mathematics regarding the nature of infinity suggest that an entirely satisfactory approach is not possible. For me, any approach that does not admit mutation is inadequate.

# The recent history of prenes

It appears that sometime between 4.6 and 3.6 billion years ago, an amazing thing happened on earth.

Molecules arose that could catalyze the raw materials in their environment to create new molecules identical to themselves. These self-replicating molecules would have been engaged in a fascinating struggle for survival.

Their survival would have depended on their ability to compete for resources, reproduce efficiently, and resist degradation. Optimizing with respect to such requirements in a constantly changing environment would have been a daunting task, and it seems likely that the dominant "species" one moment would be supplanted by another the next. No doubt, after millions of years, the surviving molecular species would have been astonishing things worthy of our admiration.

It seems likely that this world of "living" molecules would have persisted for many millions of years before nucleic-acid based cells would emerge. Current estimates suggest that by about 3.6 billion years ago such cells did exist and hence the gene age had dawned.

Cellular genes were responsible for building cells, ensuring their survival, and organizing their replication.

About two billion years ago, cells arose that used ion-channels and ion-pumps to create electrical potentials across their surfaces. These cells were the predecessors of neurons.

By half a billion years ago, creatures existed that had complex collections of neurons (ganglia) in their heads. These primitive proto brains processed sensory inputs, such as light, and induced behavior, such as movement. The development of these proto brains was an important step in a critical process by which genes were surrendering partial control of their organisms to a new type of prene, the meme.

About two hundred thousand years ago, modern man emerged with a brain composed of hundreds of billions of neurons. The golden age of memes had arrived, and human brains would soon be filled with ideas and beliefs about religion, governance, science, morality, beauty, and many other things. Societal cultural-prenes would control much of our behavior, and the world would be filled with people struggling with internal conflicts and emotional turmoil. In *Part II: Humans* we will look more deeply into this.

Less than a hundred years ago, we began building digital computers. Computational devices had budded off their genetic stalks and were on their own; Turenes had arrived. How will they evolve? What relationship will they have with genes and memes? How will computers affect the future of humanity? In *Part III: Computers*, we will explore this further.

# Part II: Humans

From the prene-theoretic point of view, humans are merely instruments for storing and serving genes, memes, and other prenes. From this view, you have remarkably little to do with you.

Do you think you determine what you eat, what smells good, who you are attracted to, when you cry, where your children go to school, and for whom you vote? From the prene-theoretic view, these things are determined by prenes.

Looking at humans as mere instruments of prenes will not capture the beauty and richness of our lives. If you are on a spiritual journey searching for the meaning of life, you may not like some of the answers that prene-theory provides. Nonetheless, looking at humans this way may allow us to see things about ourselves that we have overlooked in the past.

## What a piece of work is a man?

Each human stores a vast array of genes, memes, and other prenes, and no two humans store the same set.

As we saw in *Variation and natural selection* and *The war within*, the prenes we store are often in conflict, fighting to control our behavior in order to maximize their own chances of survival. These conflicts make human endeavors complex and can torment individuals.

Because each of our prenes is on its own evolutionary path:

### Proposition 4

### Each human stands at the crossroads of many lines of evolution.

Fortunately, our genes have provided us with tools for coping.

## Your hardware

Our genes have given us a set of actuators that allow us to act upon the physical world. They have provided legs for moving, hearts for circulating blood, and lacrimal glands for tearing. We cannot exhibit physical behavior that our actuators cannot produce. So even if you think you can fly, you can't.

Though no two people are entirely the same, virtually all humans share essentially the same set of actuators, and hence essentially the same repertoire of potential physical behaviors[8].

Our genes have provided us with a set of sensors that monitor our internal and external environments. We have eyes for seeing, and ears for hearing. We have sensors that monitor our blood sugar level, and level of hydration. We have sensors that detect temperature, pressure, and taste.

Virtually all humans share essentially the same set of sensors.

Our genes have provided us with brains.

The human brain is a computer. It is just like our commercial computers in the sense that everything that the brain can do, commercial computers can also do. This is axiomatic in the mathematical theory of computation,

––––––––––––––––––––––––––

[8] *In Perceptions, I speculate on the nature of emotions. It may be that our genes have provided a set of emotions that determine our repertoire of potential emotional behaviors as well.*

though at this stage in our civilization, we don't know how to build and program commercial computers well enough to actually do some of the more remarkable things brains do.

Virtually all humans share essentially the same computer brains.

As an aside, consider the following:

*By reasoning, I understand computation*
*-Thomas Hobbes*
*De Corpore (1655)*

It is remarkable that in 1655, just three years after Galileo's death and 3 decades before Newton's *Principia*, Hobbes apparently perceived a connection between brains and computers. But without our modern understanding of computation, Hobbes could not see the implications of his insight, including the limitations on human reasoning that computation imposes.

Because our brains are computers, the laws that govern computing must also govern our brains. Fortunately, we know a great deal about these laws. We can thank some great mathematicians, starting in the 1930s with Gödel and Turing, for this knowledge. We will have more to say about computers in *Part III: Computers.*

So, here we are, beings with actuators, sensors, and computer brains. We are robots.

Robots? Many of us do not see ourselves as robots, we are something fundamentally different, we are humans, we are special. I will address this apparent contradiction in *Are we humans or are we robots?*

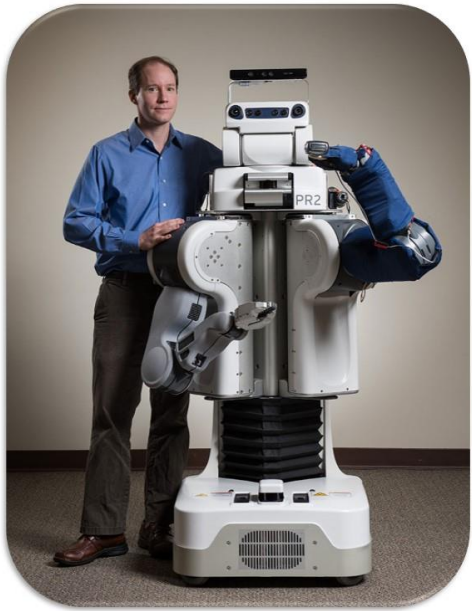

Figure 20: Two robots

## Your prenes

In this chapter, I will describe some of your prenes and how you acquired them.

Your most important set of prenes comes from your parents. These are the genes stored in the DNA of the zygote from which you grew. I'll call this your "innate-gene-set".

During the prenatal period, your innate-gene-set will build you. Among other things, it will make your brain and provide it with an initial set of programs that are roughly analogous to your computer's operating system. It will also create your sensors, and actuators.

Once you are born, your innate-gene-set does something surprising; it invites other prenes to share control of you. As we know, prenes are not noted for their generosity, so why does your initial-gene-set do it? We will address this in the next chapters.

## Welcome to the bacteria hotel

When you are born, there are about 5 trillion cells in your body each of which stores your innate-gene-set (there are lots of red blood cells as well, but they do not contain DNA). Almost immediately after birth you will become infested with microbes. For example, you will have about 40 trillion bacteria in your gut, each with its own gene-set. Current estimates suggest that these bacteria are representatives of about 10,000 distinct species (NPR, 2012). So, your innate-gene-set ends up sharing your body with a very large "bacteria-gene-set".

Your innate-gene-set has designed your gut to be a full-service bacteria hotel where tenants have a nice place to live and plenty to eat. Why would your innate-gene-set do such a thing?

The answer is, of course, that most of the bacteria are paying guests. They pay for their room and board by providing a variety of services that help your innate-gene-set. For example, some bacteria contribute by helping you digest some of the food you consume.

Not all bacteria are "helpful". There are "harmful" bacteria that should they get a room, will consume resources without providing services, cause sickness, or in other ways be deleterious to your innate-gene-set.

Since your innate-gene-set builds the bacteria hotel and the systems that operate it, it gets to put security measures in place to aid in getting helpful bacteria in and keeping harmful bacteria out.

For example, the genes keep the hotel hot (98.6 degrees Fahrenheit, 37 degrees Centigrade) and highly acidic. As a result, though all manner of bacteria enter the hotel when you consume food, most do not succeed in getting a room because they quickly die in such conditions.

Your innate-gene-set appears to use a trick to help assure that the initial hotel guests will be helpful. It has made it easy for maternal bacteria to pass into the baby's hotel during birth. Since the mother has already survived and reproduced, there is a good chance that most of the maternal bacteria are helpful. Once the rooms are initially filled by helpful bacteria, many harmful ones are kept away because they cannot compete with the current residents for a room. This is one way that the bacteria in your gut act as part of your immune system.

As an aside, ensuring that only helpful bacteria reside in the gut has recently become a concern of medicine. In the emerging field of probiotics, in a manner resembling maternal to baby transmission, helpful bacteria are imbibed in the hopes of improving gut biota to combat disease. Bacteria have access to all of our external (in the mathematical sense) surfaces, including ear canals, lungs, and sinuses. As a result, in addition to the bacteria hotel in the gut, there are hotels in the ears, lungs, sinuses, etc. All bacteria hotels operate on the same principles, and guests in each must pay for their keep. Since the environment of the ears, sinuses, lungs, and gut are very different, it is likely that the bacteria that reside in each are also different. It would not be surprising to find that each hotel has bacterial guests that contribute to our wellbeing in a variety of ways. So, if the ideas of probiotics prove worthwhile, perhaps they should be extended beyond the gut. Might people with chronic ear infections, chronic sinusitis, or chronic bronchitis be treated with sprays containing helpful bacteria and other biota?

Your innate-gene-set built your bacteria hotel, but it is surrounded by other properties with "do not enter" signs. For example, your innate-gene-set does not want bacteria in your blood, so it has built an immune system that among other things will punish trespassers from the gut.

Do these security measures work? Absolutely. Do they work absolutely? Absolutely not. Most of us will from time to time get harmful bacteria and experience "food poisoning". Some of us will get bacteria in our blood (septicemia) and die.

## Welcome to the memes hotel

Your innate-gene-set also built a hotel for memes, the brain. The memes hotel has much in common with the bacteria hotel.

When you were born, your brain had lots of unoccupied rooms (that is, locations in memory devices), but soon after your birth they began to fill with memes from the surrounding environment.

As with the bacteria hotel, there is a big security problem. There are harmful prenes out there that should they find a room, could be deleterious to your innate-gene-set. If you acquire them, they might result in suicide, substance abuse, imprisonment, ostracism, etc.

Since your innate-gene-set builds the memes hotel and the systems that operate it, it gets to put security measures in place to aid in getting helpful memes in and keeping harmful memes out. Some of these measures were also used to secure the bacteria hotel.

For example, your innate-gene-set has designed your brain to readily acquire memes from your parents when you are young. The fact that your parents survived to produce you is good evidence that the memes they pass to you are likely to be helpful. Among other things these memes may help you to avoid future dangers and ensure that you will fit into your social milieu. Since these memes are founding members of your meme collection, they will have a major influence on determining future residents. They will resist the acquisition of opposing memes. For example, if you are brought up a Christian, it is unlikely you will later convert to Islam, and vice versa.

Parental memes, even those that are not passed directly to you, will influence the memes you acquire while growing up. Your parents will choose your schools, places of worship, what you can watch on TV, and more generally which societies have access to your brain.

As an aside, since your parents will choose what and where you eat, their memes also help to secure your bacterial hotel.

Eventually, many memes will become residence of your meme hotel. For example, many cultural-prenes from religions, governments, political parties, companies, and other societies will succeed in obtaining a room.

## How your brain captures memes

Memes may be acquired from nature via our senses. They may also be acquired from parents, friends, teachers, authors, actors, musicians, political parties, governments, religions, etc. using tools such as speech, print, television, and the Internet; by techniques called education, indoctrination, brainwashing, enlightenment, entertainment, training, advertising, propaganda, etc. There are also some memes that are not acquired from the outside but are created within your brain when it processes the memes it already has. These sets of memes can be called "experiential memes", "learned memes", and "generated memes" respectively.

The outside world is awash with prenes for your brain to capture and store. How come your brain captures some and ignores others?

Way before brains showed up, organisms had evolved "stimulus/response systems" which could detect some changes in the environment and initiate useful responses when detection occurred. For example, some bacteria sense the concentration of chemicals in their environment and when desirable chemicals are detected, respond by moving to a location of higher concentration.

Once brains existed, they could be incorporated into these stimulus/response systems. Presumably, the brain could process the information it was receiving from sensors and induce responses. Under "appropriate" conditions, the response would include turning on the "meme-recorder" and capturing memes. But what are the appropriate conditions?

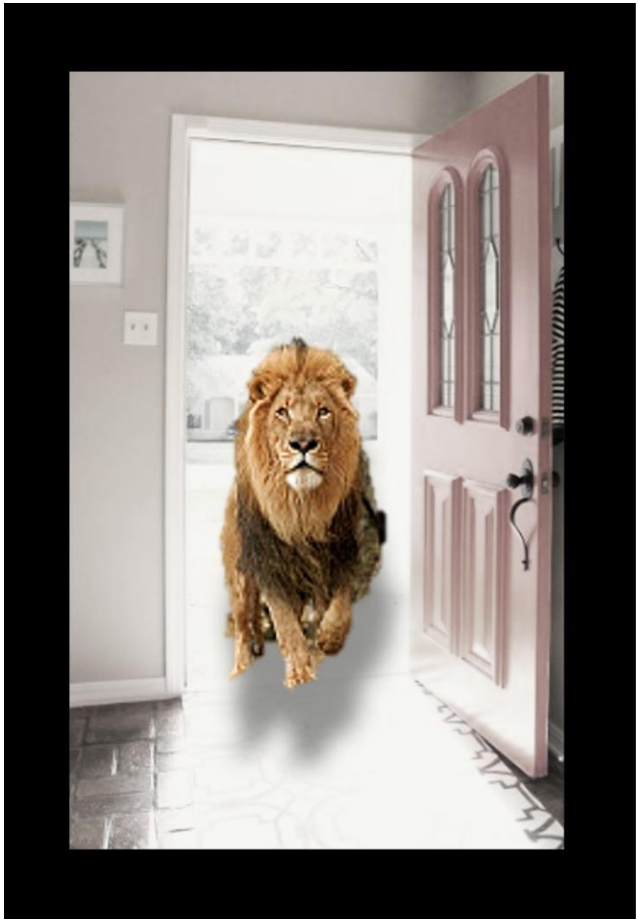

Figure 21: Surprise!

Let's say I come to your home, knock three times, and, when you open the door, I release a lion. Let's further assume that you actually survive. I'll bet you will never forget that day :).

In fact, you can't afford to forget, you must remember what happened and make sure that it does not happen again. You must turn on the meme-recorder and create memories. The recorder will save valuable information about what happened after the appearance of the lion: how big he was, how big his teeth were, how he moved, how you moved, who was present, etc. Information prior to the appearance of the lion may also be important: what sounds were heard? what smells were in the air? So, the content of short-term memory, where it is likely such information is stored, may be re-recorded into less volatile long-term memory. As an aside, perhaps such memory dumps help explain the ultrahigh resolution, slow-motion, recollections that are sometimes reported following serious mishaps (possibly excluding some that involve direct injury to the brain).

Your brain may later process the newly captured memes to improve future behavior. Perhaps your future physical behavior will include running the next time you see me. Perhaps your future emotional behavior will include feelings of anxiety when hearing three knocks on the door.

A similar thing may happen when good things occur. For example, if you encounter a person that you find physically attractive, the meme-recorder may be turned on. The recording may include images, locations, and sensory sensations such as sounds. The new memes may later be processed to improve future behavior. For example, you may adjust your plans to increase the likelihood of encountering that person again.

In both the lion and attraction cases, turning on the meme-recorder is one of many actions your brain (or more generally your central nervous system) may initiate. For example, it is likely that in both cases, your pulse rate will rise and the chemistry in your brain will change.

What constellations of signals are likely to cause the brain to turn on the meme-recorder? I would guess that evolution would have produced a brain with substantial proficiency in distinguishing between mundane situations, situations involving danger, and situations involving opportunity. Recording memes in the latter situations provides a means for designing future behavior to diminish those dangers or enhance those opportunities.

Cultural-prene-sets have learned to exploit these mechanisms by creating constellations of signals that cause the meme-recorder to start and record memes they wish to embed. It is no accident that advertisers use physically attractive people to get you to remember their beer. The Catholic church has used glorious music and wondrous art to achieve a similar effect. To a significant extent, baroque art itself arose as a direct result of the Catholic counter-reformation's endorsement of art for the purpose of teaching Catholic cultural-prenes. Would a 17th century peasant ever forget the holiness of Saint Ignatius after viewing Pozzo's breathtaking ceiling fresco?

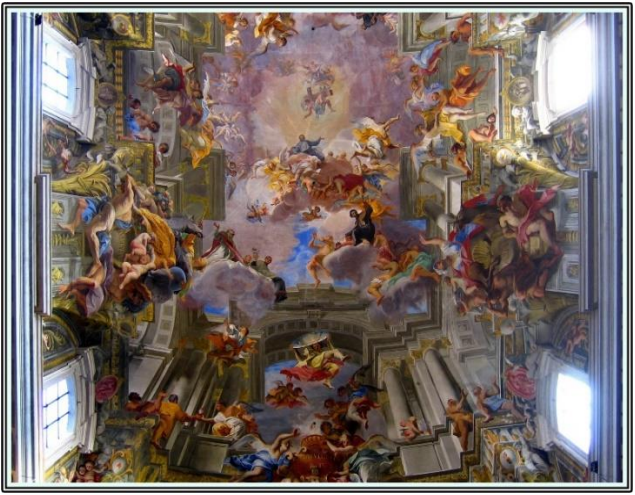

Figure 22: The apotheosis of St. Ignatius by Andrea Pozzo, church of Sant'Ignazio, Rome.

Importantly, while the genes have given us systems for recording memes, it appears they have not given the conscious part of our brains direct control of their removal. We cannot simply decide to forget our memories; just make them go away.

It may be that when the brain records a meme, some form of priority rating is associated with it. Presumably the rating assigned is determined by the brain based on the constellation of signals it is receiving.

For example, perhaps an exceptionally high rating would be assigned when sexual fulfillment is first achieved, if heroin is injected for the first time, if you are kidnapped or sexually assaulted.

Call such memes "pivotal memes". They are the footprints left by major life events and a great amount of energy and brain cycles may be invested in planning a future around them. Consider the impact that the examples above would have on future behavior.

Your brain might be likened to a bell; the harder its struck, the longer it reverberates. Pivotal memes come from striking the bell with great force and the reverberations persist for a very long time as our brains repeatedly processed them to plan a "better" future.

When pivotal memes are acquired in negative situations, we sometimes call the result post-traumatic stress disorder (PTSD). When acquired in positive situations, perhaps we should call the result post-ecstatic stress disorder (PESD).

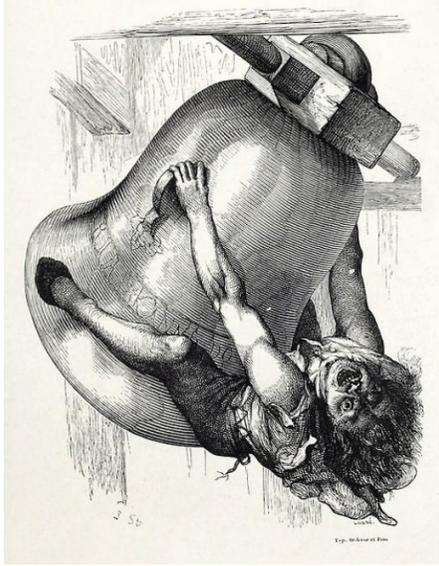

Figure 23: PTSD: when the ringing won't stop.

When you cannot think of a name and it suddenly "pops" into your head, that name was almost certainly discovered by your subconscious while processing your stored memes and was then forwarded to the conscious part of the brain. The conscious part of your brain is not informed of the details of the subconscious processing, only the result.

This subconscious processing occurs all the time. For example, it appears to continue while we sleep. If during sleep, a meme of value is generated, then a notification may be posted for the conscious part of the brain. When we wake, we find the notification and have an "I got it" sense.

For example, I once woke with the certainty that I could finally prove a theorem I had been struggling with for months. I went to my blackboard and had the bizarre but pleasant experience of writing the proof that, at that moment, the conscious part of my brain was seeing for the first time. While the conscious part of my brain received the notification and the proof, I suspect that my subconscious had done a great deal of processing that I was not, and would never be, conscious of.

Because long arduous subconscious processing sometimes generates memes of value that are transmitted to the conscious part of the brain without any details of how they were obtained, there is a tendency for people to believe that something magical, divine, or inspired is involved in human creativity (see *Are we humans or are we robots?*).

### Proposition 5

**Much of the brain's processing takes place in the subconscious part of our brains and is not revealed to the conscious part.**

Here is one possible way that the brain processes memes to improve a person's chances of survival.

The human brain can envision futures. I don't mean this in any mystical sense, it is much more mundane. For example, chess players can envision possible ways a game may unfold.

I suspect that mechanisms for envisioning possible futures evolved long ago. When we see a pod of orcas dislodge a seal from an iceberg, or a wolf pack hunting, I believe that we are witnessing animals with the ability to envision possible futures. Each individual seems able to envision future positions of its prey, other individuals, and itself.

But if the wolf has this ability, how might the genes exploit it when the wolf is not hunting? At night does the wolf dream of hunts as a means of sharpening skills for the future?

The human brain is able to envision possible futures complete with moving images, sounds, feelings, objects, and people, including ourselves. When this happens during sleep, we call the process dreaming. Our existing memes are part of the raw material for generating these possible futures, and it seems plausible that experiential memes with a high priority rating have a greater probability of being incorporated (see *How your brain captures memes*). Our subconscious can create possible futures and experiment with possible behavioral responses. Perhaps, in much the same way that some current AI programs learn, when an experimental behavior is judged to have a negative outcome, the brain modifies its memes to decrease the likelihood of that behavior occurring in the future; when judged to have a positive outcome, the likelihood is increased.

Providing a computer brain capable of doing all this processing is expensive. So why do genes do it? This will be addressed in *On the evolution of the human brain*.

.

## The beauty of the world

If you are reluctant to see yourself as a robot (see *Are we humans or are we robots?*), perhaps you'll find solace in knowing that humans are the most amazing, awe-inspiring robots ever.

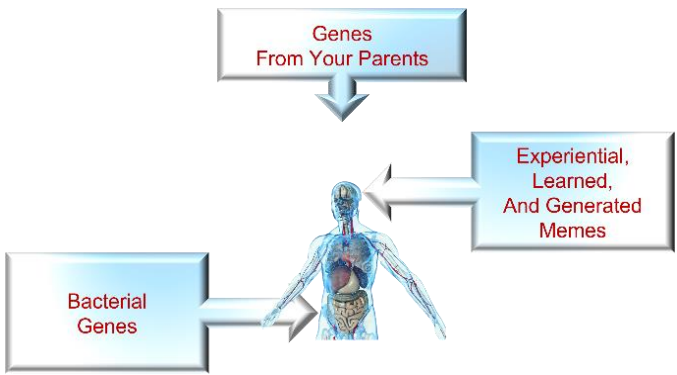

Figure 24: Major subsets of your prene-set.

In addition to the prenes indicated in *Figure 24* you have others. You have viral and fungal genes in various places. You have memory devices outside your brain that store prenes that are neither genes nor memes. For example, if you stub your toe, there will be a first order neuron that stores some prenes concerning the event for a few milliseconds before passing them to other neurons on their way to the brain.

It is an amazing you. There are millions of prenes in you, and they fight one another to control you: what you eat, what you feel, what you believe, where you go, with whom you mate, and just about everything else.

With this turmoil going on inside, it is surprising we do not just die of confusion. The fact that we do as well as we do is a tribute to evolution and your innate-gene-set which designed your body, your powerful computer brain, and the programs necessary to deal with the chaos.

# Creatures of society

The behavior of many living things is primarily determined by their environment and their genes. They are creatures of biology. But the behavior of humans is often determined by their environment, their genes, their memes, and other prenes. We serve many masters, and that makes each of us, and our societies extremely complex. In this chapter we will focus on humans as creatures of society.

To survive, all prene-sets must create new copies and mutate. In *Cultural-prenes in the social sciences*, we considered how cultural-prenes organize their mutation. In this chapter we will consider how they organize the creation of new copies. Of particular importance are the copies they store in human brains.

## There is no escape!

As discussed in *Your prenes*, upon birth you begin to acquire memes. Many of these come from the cultural-prene-sets of the societies to which you are exposed. These societies probably existed long before you showed up, and they have been waiting for you. They have evolved extremely refined methods to make the most of their opportunity. They intend to capture and exploit you.

The cheetah lies in wait and then, in a sudden, astonishing burst of speed, overtakes and kills its helpless prey. The Venus fly trap uses the promise of food to entice its prey; then gently imprisons and consumes it. Societies have a great deal in common with these predators.

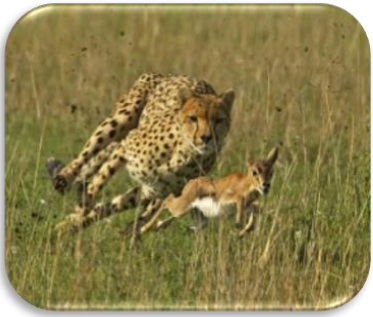 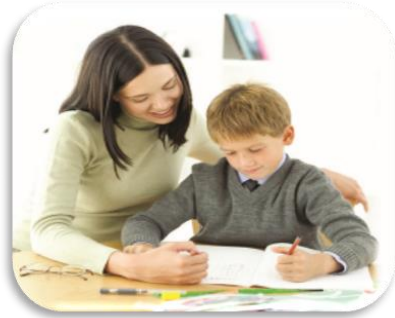

Figure 25: Ruthless predatory prene-warriors with their prey.

Let's explore how societies go about their work. I will focus on the Catholic society, but many other societies would have sufficed.

If the Catholic society gains access to you, it will begin to use its instruments to "infuse" you with memes from its cultural-prene-set. The longer it can maintain access, the more memes you will acquire. You may be taught the catechism early on, but only much later, if at all, are you likely to be taught the *Universi-Dominici-gregis*.

The Catholic society has "learned", through evolution of its cultural-prene-set, that many of its prey will encounter the society as children under the control of their already committed parents. This provides a period of several years during which the child will be retained and infused with memes. The society will use this time wisely. The child may be enrolled in church-associated schools, youth groups, choirs, etc. which have been created by the cultural-prene-set as a means of instilling memes and extending the retention time of the growing child.

As with all children, the child will spend their initial years acquiring memes from their parents. If a parent is a member of the Catholic society, then it is likely that some of these memes will be Catholic cultural-prenes the parent had previously acquired. In addition, the parent may grant the society direct access to the child. For example, they may send the child to the society's schools. Among the most important things the society will do with these opportunities is to instill in the child the memes that will cause them to repeat the process if they become parents.

Not all children exposed to the Catholic society will stay the course. Some will abandon the faith, some will only go to church on holidays, some will become pope. Why does this happen?

Typically, the Catholic society will come into conflict with other societies for control of the child's behavior and their future acquisition of cultural-prenes.

For example, perhaps the child will begin to acquire an "interest in" sports. That is, some sports society will succeed in infusing some initial cultural-prenes and begin controlling some behavior. What is to be done when there is a game at the same time church services are being held?

When the child reaches adolescence, their gene-set will make a strong bid for greater control. The adolescent will discover sex, and the society's cultural-prene-sets will have to compete with the adolescent's gene-set for behavioral control.

The Catholic society has traditionally dictated that sex is encouraged but only within wedlock and only for the purpose of procreation. There are lots of Catholics, so this half-empty/half-full stance may have worked well.

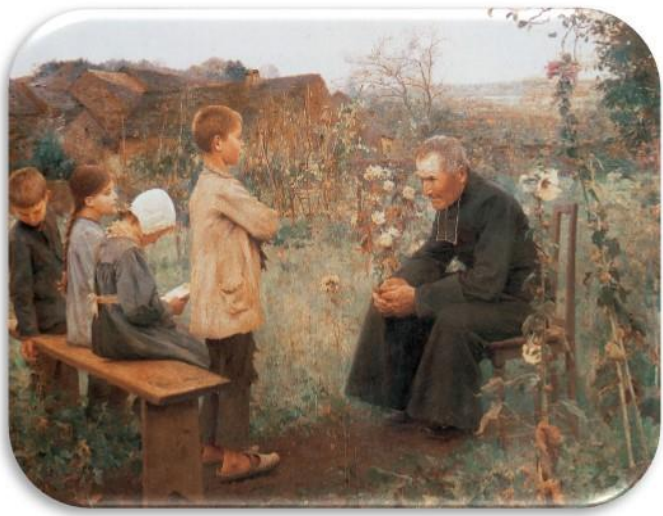

Figure 26: The Catechism. That boy on the left though? Probably not going to be pope.

As an aside, discord between the genes and the memes acquired from cultural-prene-sets can have psychological consequences, for example, when one does covet thy neighbor's wife. Nietzsche, Freud, and others have explored the conflicts between cultural-prene-sets and the human gene-set, sometimes seeing them as conflicts between civilization and human nature.

Even if the child abandons Catholicism entirely, the cultural-prenes acquired may influence behavior and emotions throughout life. How this lasting influence is created will be discussed in more detail in *Perceptions*.

A human is a limited resource with a finite lifespan. At each moment, a human can execute some fixed number of brain cycles and expend a fixed amount of energy. It is not possible to provide full service to all societies encountered.

In the end, you will acquire memes from the cultural-prene-sets of many different societies, and each will get a share of your resources.

The cultural-prenes of a society that acquires a small share will have little influence on you. The cultural-prenes of a society that gets a significant share will regularly impact your behavior and become "part of your life". If the cultural-prenes of a single society acquires a very large share, then that prene-set will occupy a huge portion of your time, energy, and thoughts, and become a center of your existence.

## The sad story of the delicious cat

Consider the following true story:

> *I once saw a video of a man skinning a cat alive and throwing it into a kettle of boiling water – I was repulsed and angered – however, the man in the video was just preparing dinner for his family and presumably had a very different response.*

It seems clear, and, of course, comes as no surprise to sociologists, that the difference in my response and the man's can be traced to differences in our cultures. I had acquired beliefs that cats were pets and that harming them was abhorrent. The man in the video had apparently acquired different beliefs.

In *Why do bees kill themselves*, we saw that our memes can have a significant impact on our physical behavior. Here we see they can have a significant impact on our emotional behavior as well.

Interestingly, even though I know intellectually that my emotional response was mediated by my memes, as best I can recall, that was invisible to me at the time. My feelings of repulsion and anger simply arose and nothing in the experience indicated their origin.

# Proposition 6

## People are typically unaware when their memes are influencing their physical and emotional behavior.

Why are we unaware of the influence of memes? For the same reason bacteria are unaware that their behavior is influenced by elusive things called genes stored inside themselves. This will be discussed further in *Are we humans or are we robots?*

*Proposition 6* impacts even our mundane behavior. For example, most people seem to think that the opinion they express, often passionately, on a politically controversial topic is transparently true and irrefutable. They are typically unaware that both their opinion and their passion were probably largely determined by memes they acquired from some political society via their parents, peers, schools, churches, the media, the Internet, etc.

When you speak with someone of a different political persuasion, you may see them as unthinking parrots simply reciting the slogans fed to them by their political overlords. You are probably right to think of them like that, but are you willing to think of yourself that way as well?

Probably not. And it is no accident.

As will be described in greater detail in *Perceptions*, societies have learned to use emotions to direct your future behavior. You may experience positive emotions when you behave in accord with your society and negative ones when you don't. The emotional price of acknowledging that your beliefs are of no more value than your opponent's may be too great to bear.

As an aside, part of the challenge of learning prene-theory is acquiring the ability to put on your prene-theory hat and keep it on even when some of the cultural-prenes you have acquired are trying to knock it off. This is particularly difficult when some of your cultural-prenes are antithetical to some of the cultural-prenes associated with prene-theory.

The next time you begin a political discussion with someone who does not share your beliefs, try seeing it from the prene-theoretic viewpoint. You and the other person are merely prene-warriors unconsciously in the service of political societies that have been at war for centuries. Your discussion has been repeated, *mutatis mutandis*, billions of times, and likely with the same result: unpleasant disagreement. In the future, perhaps you will listen to your mother's advice.

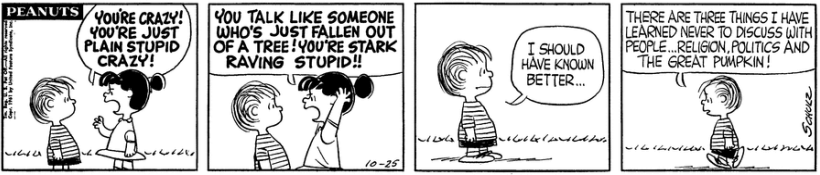

Figure 27: Linus the prene-theorist.

## Why and when you will die

Thermodynamics tells us that death, not life, is the natural order of things. To sustain life requires the constant acquisition and expenditure of energy. But sustaining life is not the only concern the genes have. To survive, they must also improve copy numbers. As a result, the genes are forced to decide how much energy organisms will expend sustaining life and how much they will spend creating and supporting new life.

The decision the human gene-set has made in this regard will have much to do with why, when, and even how you will die.

## Why do we die?

You probably think that immortality is a fine idea; unfortunately, your genes apparently do not.

In this chapter, I will create a very simple mathematical model about competing species of trees to illustrate the folly of immortality.

Our trees live in a simple environment. At the time of its birth, each tree is storing a special number between 0 and 1. This number will be referred to as the tree's gene-number. To keep our model simple, mutations are not allowed. A tree's gene-number never changes during its life. If a tree produces an offspring, the new tree has the same gene-number. Two trees with the same gene-number belong to the same species. Two trees with different gene-numbers belong to different species. Each tree has a "survival account" and a "reproduction account" wherein it keeps units of energy.

Every day, each tree:

Receives 2 units of energy through photosynthesis. It puts its gene-number units of energy into its survival account and the rest into its reproduction account.

If its reproduction account has at least 3 units of energy, then 3 units of energy are withdrawn, and a new tree of the same species will be born the next morning with 3 units of energy in its survival account and nothing in its reproduction account.

Should its survival account have less than 1 unit of energy, the tree dies. Otherwise, 1 unit of energy is withdrawn – this is the cost of sustaining life.

For example, consider the species with gene-number 1. Each day, each tree puts 1 unit of energy into its survival account and 1 unit of energy into its reproduction account. We will call this species the "immortal species", since each tree puts just enough energy into its survival account each day to guarantee it will never die.

Now, consider the species with gene-number 0.5. Each day, each tree puts 0.5 units of energy into its survival account and 1.5 units of energy into its reproduction account. We will call this species the "mortal species" since each tree of this species must die. When it was born it had 3 units of energy in its survival account, it begins the next day with only 2.5 units, it begins the day after that with only 2 units, etc. It follows that on the sixth day following its birth, it will die. On the other hand, it puts more units of energy into its reproduction account than trees of the immortal species, so it generates new offspring faster.

Let's take 1 newborn immortal tree and 1 newborn mortal tree and put them in an open field and see what happens.

| Day | Number of Immortal Trees | Number of Mortal Trees |
|---|---|---|
| 0 | 1 | 1 |
| 1 | 1 | 1 |
| 2 | 1 | 2 |
| 3 | 2 | 2 |
| 4 | 2 | 4 |
| 5 | 2 | 4 |
| 6 | 4 | 8 |
| 7 | 4 | 7 |
| 8 | 4 | 14 |
| 9 | 8 | 13 |
| 10 | 8 | 26 |
| 11 | 8 | 24 |
| 12 | 16 | 48 |
| 13 | 16 | 44 |
| 14 | 16 | 88 |
| 15 | 32 | 81 |
| 16 | 32 | 162 |
| 17 | 32 | 149 |
| 18 | 64 | 298 |
| 19 | 64 | 274 |
| 20 | 64 | 548 |
| 21 | 128 | 504 |
| 22 | 128 | 1008 |
| 23 | 128 | 927 |
| 24 | 256 | 1854 |
| 25 | 256 | 1705 |
| 26 | 256 | 3410 |
| 27 | 512 | 3136 |
| 28 | 512 | 6272 |
| 29 | 512 | 5768 |
| 30 | 1024 | 11536 |

Some days the number of mortal trees declines due to deaths. But, by day 30, there are more than 10 times as many mortal trees as immortal trees. By day 100, fewer than 1 in 1000 trees would be immortal.

It can be proven (Watkins, 2023) that gene-number 0.5 is optimal. That is, the mortal trees increase copy number more rapidly than trees of any other species. They will have greater representation in the "gene pool" than all other species. They will survive while all others, including the immortal trees, will go extinct.

While our tree model was very simple, it seems clear that similar arguments could be applied to other models and demonstrate the failure of immortality. Though I cannot provide a proof, I am led to the following:

## Proposition 7

**Immortality is a losing strategy. To increase representation in the prene pool, it is better to sacrifice immortality for a greater rate of reproduction.**

When applied to Turenes, this implies that successful self-replicating robots will intentionally build offspring that are designed to eventually fail. In the case of societies, it suggests that it can be good strategy to encourage each member to spend huge amounts of time and other resources doing things to help spread the society's cultural-prenes, even if doing so will harm the member and shorten their life.

*Proposition 7* is a strong statement. For example, it implies that immortality will fail independent of the environment in which prenes are found.

So, does the environment matter at all? In the next chapter *When do creatures die?* we will see that how long you will live is a complex function of both your genes and the environment.

So why are you going to die? Because your genes don't care about you, and they have long ago learned (through evolution) the value of *Proposition 7.*

## When do creatures die?

You might think that because our genes ignored our desires to be immortal, they would at least be polite enough to grant us a long lifespan. If you think that, then please reread *The selfish prene.* Your lifespan will be determined in a closed-door meeting of your genes and the environment, and your desires will not be on the agenda.

In fact, such meetings determine the lifespan of all creatures, and so we will start our investigations in a general setting, and only specialize to humans in subsequent chapters.

Let's start with an analogy.

Assume you are the CEO of a car company designing next year's model. You have looked at hundreds of designs and have narrowed the field to just two. The two only differ in the metal used for the engine. Your engineers have done extensive lab testing and sent the following table.

| Metal in Engine | Expected Cost to Manufacture | Expected Lifetime of Car |
|---|---|---|
| Tin | $10,000 | 1 year |
| Titanium | $1,000,000 | 100 years |

So, for the same investment, you can build a fleet of tin cars that won't last long, or a small number of titanium cars that will last a long time. You cannot have it both ways: lots of inexpensive cars that last a long time.

You opt to go with titanium. You put the car into production, and the first one rolls off the line … and gets crushed by a passing truck. You ask the engineers why the car did not last the promised one hundred years. They tell you it's because the one-hundred-year number was "in the lab". They say that if you want to produce a car that will last one hundred years in the real world, they are not sure that it is possible, but they might be able to do it using spider-silk-wrapped carbon-nanotubes filled with depleted uranium at a cost of $1,000,000,000 per car.

Gene-sets must make similar choices, they can build lots of offspring that won't last long, or a small number of offspring that will last a long time. Gene-sets do not have the luxury of lab testing, the organisms they build must live in the wild and face real world threats.

So, what is the right choice for a gene-set? That is hard to say, but in the case of the mouse, the genes have decided to build lots of cheap offspring that don't last long even in "protected environments".

> *House mice usually live under a year in the wild, due to a*
> *high level of predation and exposure to harsh environments.*
> *In protected environments, however, they often live two to*
> *three years.*
> *(Wikipedia-Hou)*
>
> *mice ... are able to produce as many as ten litters –*
> *approximately sixty mice – each year.*
> *(Victor, 2023)*

Was this a wise decision? Well, there are lots of mice in the world, so it probably was. Is there a better decision the mouse-gene-set could have made? Maybe there is, but apparently neither we nor the mouse genes have discovered it.

Is there a worse decision the mouse-gene-set could have made? Almost certainly there is. For example, what if the mouse-gene-set had decided on a slightly different design that included an expensive heart expected to last a thousand years? That would most likely have been a disaster, since it raises the cost of building baby mice, and despite the great heart, it is likely that most baby mice would die within a few years, due to a cat or some other real-world killer such as starvation, infection, or exposure. Perhaps it would have been better to use that investment to slightly improve the life expectancy of all mouse parts, or to make mouse brains bigger so they would be better able to avoid cats, or to build more baby mice of the original design.

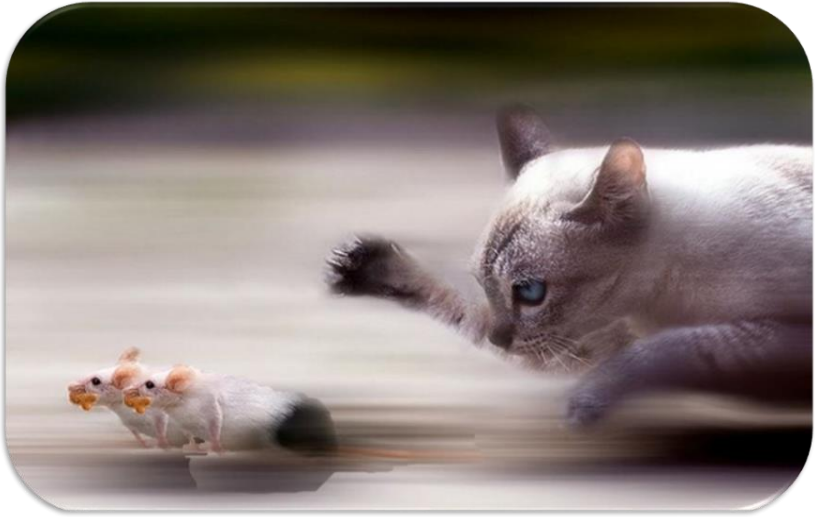

Figure 28: Later, when asked whether the mouse with the expensive heart or the mouse with the cheap heart tasted better, the cat replied that they both tasted about the same to him.

Put another way, the problem with investing in a one-thousand-year heart for an animal with a life expectancy of a few years is that it is likely most of that investment will be lost when the mouse dies. Genes do not like making bad investments.

## Proposition 8

### Gene-sets invest as little as possible in organisms that should be dead.

For example, the males of many species, from insects to marsupials, die after a single event or season of mating. The parts of these males appear to be designed to be adequate to complete the mission, but inadequate to sustain life thereafter. In many cases, such as Salmon, the deterioration of parts can be easily discerned. Black widow spider genes get particularly high marks for their effort to liquidate residual investment.

Returning to car manufacturing, there were a huge number of possible designs to consider. Was your choice the best possible?

As a CEO, your job is to maximize "return on investment" (ROI)[9].

ROI = (selling price - cost of production) / (cost of production)

For example, if ROI=1.1, and you had invested 1 billion dollars to build your fleet, you would have 1.1 billion dollars after you had sold it. But let's say you decide to reinvest the entire 1.1 billion dollars in a second fleet with the same design. If your original 1 billion dollars had resulted in a first fleet with 1000 cars, then the second fleet would have 1100 cars. That is:

(number of cars in second fleet) / (number of cars in first fleet) = 1.1

---

[9] *I have simplified here. Maximizing ROI is maximizing over all possible designs, the average over all cars in the fleet, of the ROI for each car.*

It is no coincidence that this equals the original ROI, it always will. That is, maximizing ROI is essentially the same as maximizing the number of "copies" of your design in the future. This is starting to look a lot like what the genes are trying to maximize.

Unfortunately, you don't have a crystal ball, so at the time you make your design decision, you can only guess how much the cost of production will be and what the consumer will pay. You could hire manufacturing cost estimators and market analysts to increase the probability that your guess will be a good one, but there are no guarantees. For example, you might decide to manufacture heavy cars, because market analysis suggests that consumers will pay a lot for such cars because of the protection they provide in accidents, but a month after production there may be a worldwide gas shortage, and consumers begin to favor lighter cars with better mileage.

If your competitors consistently produce greater ROI than you, your company will fail. The market will have determined that your company was unfit to continue.

Today, what is the life expectancy (say, average number of years from manufacture until final operation) of cars? We might be able to explore junkyards and come up with a pretty good guess. But there is a more important question for us: what will the life expectancy of cars be in 2100? Perhaps it will be very short, because making cars will have become so inexpensive that people expect to use them once and then discard them. Perhaps it will be very long, because technology will have made it possible to build cars that are virtually accident free and mechanically indestructible, and each person expects to buy one that lasts their entire life. The point is, that while the goal of maximizing ROI does not change, the best design for achieving it does and depends on the environment. Most importantly, the best design at any particular moment has almost nothing to do with longevity.

The situation with gene-sets is essentially the same as with CEOs.

The gene-set's job is to maximize ROI:

(number of copies in this generation) / (number of copies in previous generation)

If the gene-set's competitors consistently produce greater ROI, then the gene-set will fail. It will be a victim of natural selection.

Gene-sets do not "care" whether their organisms live a short time like mice, or a long time like bowhead whales.

Gene-sets don't have crystal balls that show them what the environment will be in the future. They must choose a design as best they can and build new organisms with that design. It will only be when a new organism with that design is born and encounters the environment in which it will live, that its lifespan and return on investment will be determined.

For example, *Tyrannosaurus* genes were doing pretty well with their designs for a few million years. But it appears that about 66 million years ago, the then current design encountered an environment much different from anything that had proceeded it. That design and that environment together determined that each *Tyrannosaurus* (including each *T. rex*) should die immediately and yield no further return on investment.

So that is how the life expectancy of mice, whales and other creatures is determined by the genes and the environment – with virtually no concern for whether the creatures themselves live a long time or not.

## When do humans die?

In *When do creatures die?* we saw that the human genes do not care how long people live. On the other hand, people care a lot.

Because of this, there has been a huge amount of work done to determine aspects of human life expectancy. Much of this work is of a statistical nature, and whole subareas of statistics, such as survival analysis, and failure analysis, have grown up around it. The World Health Organization reports (WHO, 2023):

> The world life expectancy as of October of 2023 was 73 years.

Why 73? How has life-expectancy changed through time and what part did the genes and the environment play?

Here is a graph of life expectancy since 1770.

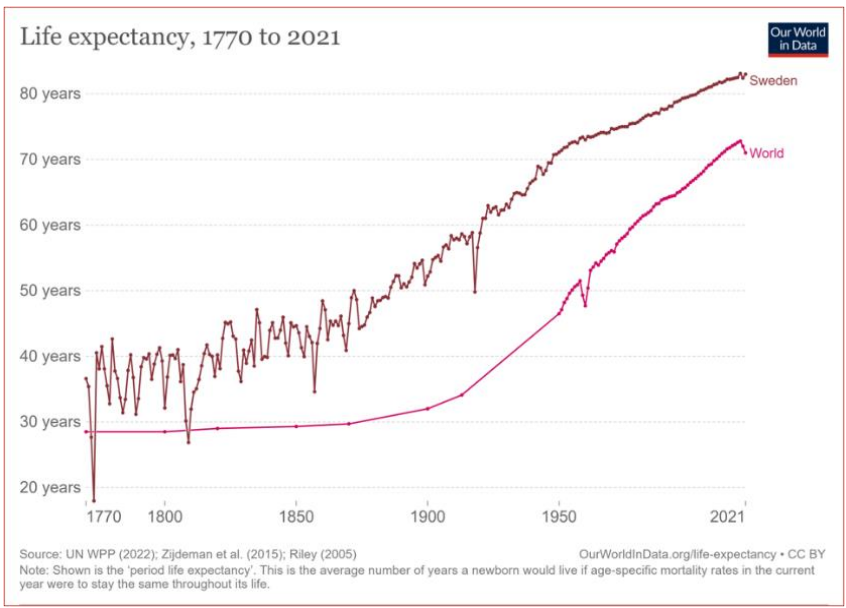

Notice that in 1770 the world life expectancy was about 30 years. So, in the roughly 250 years since, it has jumped 43 years.

The graph starts in the 2nd half of the 18th century, and I have singled out Sweden, because it was there and then, for the first time, reliable census data started to be collected. That data can be found at The Berkeley Mortality Database (Wilmoth, 2018).

Do we know anything about life expectancy prior to 1770? In fact, many researchers have estimated the life expectancy for pre-18th century humans and their hominin ancestors, relying on indirect means such as the analysis of skeletal remains. These estimates include some that go back to the paleolithic. A summary of the findings can be found at (Wikipedia-Lif). Remarkably, it appears that life expectancy may had been roughly 30 years for the last few million years.

So, what happened since 1770 that made such a dramatic change?

The industrial revolution.

The many innovations since the start of the industrial revolution resulted in great advances in agriculture, architecture, engineering, commerce, public hygiene, medicine, and other areas.

These advances impacted current life expectancy in at least three ways.

- They reduced child mortality.
- They protected us from many mortal threats such as starvation, exposure, injuries, and infectious diseases. As a result, we are now

inhabiting an environment analogous to that of mice in "protected environments" or cars "in the lab".

- They gave rise to technology that allows us to replace, repair, and preserve various human parts. For example, we can replace some parts with transplants, and joint replacement surgery; repaired some with angioplasty, and ligament reconstruction; and preserved some with statins, exercise, healthy eating, and wise life choices.

In the second half of the 18$^{th}$ century in Sweden:

- The life expectancy was about 35 years.
- The probability that a newborn would reach age 15 was about 60%.
- The life expectancy (from birth) of the cohort of people who did reach 15, was 59 years.

Today:

- The probability of dying before age 15, is less than 1%. Hence the cohort of people who survive to age 15, has greatly expanded. This alone seems to have about doubled human life expectancy. Presumably, it has also contributed to a large increase in world population.
- Roughly speaking, there are two ways to die: "Internal causes" and "external causes". Internal clauses are what people would die from if they spent all their lives in a "perfectly protected environment" that would keep them safe from external "insults". Informally, internal causes are the result of parts "wearing out"; they include things like heart attacks, strokes, kidney failure, and cancers that result from degradation of the immune system. External causes are all the causes except the internal ones and include things like starvation, gluttony, predation, infectious diseases, suicide, and exposure. The protected environment we now live in, while not perfect, still protects us from many external causes of death. It is no surprise that when there are fewer things that can kill you, your life expectancy will rise.

- Finally, the technologies that allow us to replace, repair, or preserve some human parts, can forestall parts-failure, and extend our lives.

So, we humans live about 43 years longer than we did before 1770. We know from *When do creatures die?* that life expectancy can change as a result of changes in the genes and/or changes in the environment. Which of the two deserves more credit?

It seems that the genes deserve little or no credit. It is often stated that the human gene-set evolves slowly. More precisely, some human genes evolve slowly (see *Endura and ephemera*). But, in any case, it seems reasonable to expect there has been little change in human genetics since 1770.

So, the credit seems to go to the environmental changes that began with the industrial revolution. The industrial revolutionaries and their successors have changed our lives in many ways; adding 43 years to our life expectancy seems among the most remarkable.

There are many deeper questions that we have not addressed. For example, consider the following figure showing the dependence of life expectancy on geography:

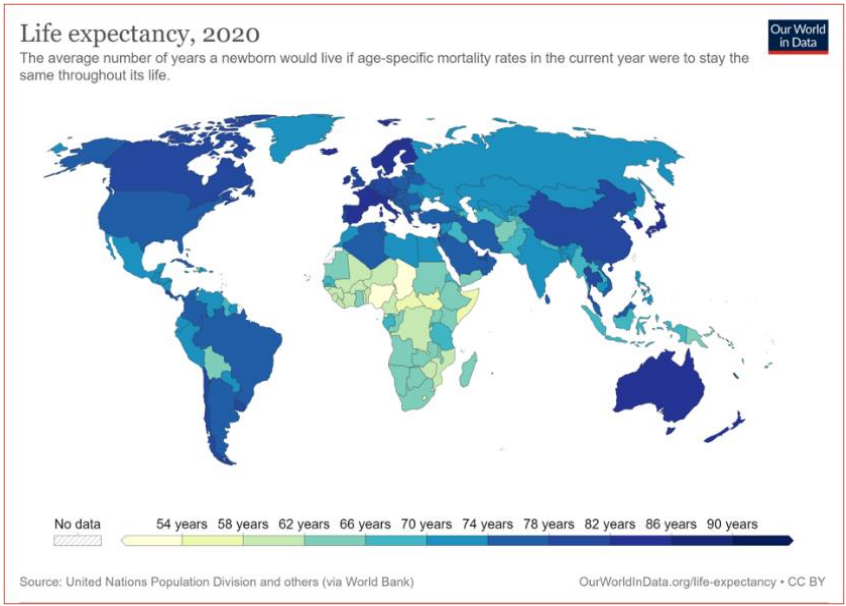

Figure 30: Life expectancy by geographical location

Why is life expectancy different in different geographical areas? How much of this is due to genetic differences and how much to environmental differences (including societal differences)? We don't know, but perhaps our current tools will allow us to make progress on such questions.

Have our genes designed some of our parts with expiration dates?

The answer appears to be yes. For example, the placenta consists of both maternal and fetal tissue. The latter was part of you in utero but is not part of you now. It had an expiration date – discard after delivery.

Notice that the placenta also had a retention period – keep until delivery. That is, do not fail before delivery.

As we saw in *When do creatures die?* how long a creature lives is determined by both the genes and the environment. This is also the case for individual parts.

For example, the heart keeps working until we die (except in rare cases where the heart has been replaced) and then stops working. Since on average we now live for about 73 years, the heart on average lives 73 years. However, in Sweden at the beginning of the industrial revolution people, and hence hearts, on average lived for 35 years. So, no matter how long the genes' designed the heart to keep working without failure, how long it will actually work without failure is highly dependent on the environment.

But this does not appear to be the case for the placenta. How long it will actually work without failure seems largely independent of the environment. For example, it would be quite surprising if in Sweden at the beginning of the industrial revolution, the placenta, on average, failed before delivery or stayed alive after it. Perhaps this independence can be traced to the location of the placenta. It is within the amniotic membrane, in a stable local environment that is sequestered from the outside environment.

So, I will assume that the genes designed the placenta to not fail before delivery.

Now let's return to the heart. Would it make sense for the genes to designed hearts to fail before the placenta fails? Of course not, that would violate *Proposition 8*. The same argument would apply to all vital parts, such as lungs, kidneys, and immune systems. It need not apply to nonvital parts such as teeth and hair.

More generally, if there is some part which is designed by the genes not to fail for some period, then all vital parts would have to be designed by the genes not to fail until that part does.

Hence, with our assumptions, we can conclude that your heart is designed to work without failure at least until you are delivered. That may not be a particularly comforting thing to know about your heart. But can we do better? At least for women, I think we can.

We will consider a special collection of cells in women called primary oocytes.

Primary oocytes are precursors of ovum, and ovum come from no other place. Remarkably, all the primary oocytes a woman will ever have are present at her birth; there is no mechanism to make more during her life.

A simplified model of the ovaries is as a kind of hourglass. Before birth, the top of the glass is filled with a number of primary oocytes. At birth, the primary oocytes begin to move from the top to the bottom of the glass. At menarche they begin to move at a faster rate. When the primary oocytes run out, the woman has reached menopause.

I will assume that the number of primary oocytes that a woman is born with is designed by the genes and is largely independent of the environment.

By the reasoning above, the genes should design a woman's heart and other vital organs not to fail before the primary oocytes run out; that is, not before the woman reaches menopause.

Since, on average, in the United States, menopause occurs at about 51 years of age (McKinlay, Brambill, & Posner, 2008), the vital organs should not fail before age 51.

This, in turn, has other interesting implications. For example, prior to menopause, women will rarely die from failing vital organs. If they do die, it will probably be (as a sequel to) an "external cause" (see *When do humans die?*) such as accidents, homicides, suicides, infectious diseases, poor lifestyle choices such as obesity, smoking, and drug addiction. This and the remarks in *When do humans die?* regarding our current protected environment, suggests that prior to menopause women should have a particularly low rate of dying.

As an aside, the situation for men is far less clear. For example, it is commonly asserted that men can continue to make new sperm throughout their lives; that there is no "male menopause". As a consequence, at this point, the arguments above do not tell us much about how long the vital organs of men are designed to last. We cannot exclude the possibility that the genes have designed men and woman to have very different vital parts. Current statistics indicate that on average women live longer than men and have first heart attacks 7 years later than men (Harvard, 2016). Further investigation might reveal more.

# The grandmother hypothesis

Since women become menopausal at about 51 (McKinlay, Brambill, & Posner, 2008) but live for much longer, it is possible that the genes designed women to have "early menopause"

If that is the case, then why might the genes have taken this step, when doing so decreases women's reproductive lives and sacrifices the obvious evolutionary advantage of having more children?

Among the hypotheses that have been put forth to explain this is the "grandmother hypothesis": the genes designed women to have early menopause so that they could devote (more) resources to their grandchildren.

I do not subscribe to the grandmother hypothesis.

To begin, I find the evidence in support of the hypothesis unconvincing. First, that post-menopausal women devote resources to their grandchildren can be understood without appealing to menopause at all.

For obvious reasons, the genes designed both men and women so that throughout their lives they would expend some of their resources in support of closely related individuals. Since women (and men) currently live for a long time, they now have the opportunity to devote resources to their grandchildren and sometimes even their great grandchildren.

Second, do post-menopausal women devote "more" resources to their grandchildren? I doubt that anyone has done a controlled experiment to confirm that this is the case. I expect that the argument in support of this assertion goes something like this: making babies and raising them uses resources, so if women stop making babies, they will have more resources to expend on their grandchildren.

But this argument could be used in support of other hypotheses that we know are incorrect. For example, the genes designed women to stop having sex, or stop exercising at the onset of menopause. These activities use resources that might otherwise be expended on grandchildren.

But my biggest concern with the grandmother hypothesis is not the insufficient evidence, but that it fails to provide an adequate explanation of the evolutionary advantage conferred to the genes by early menopause. Whatever advantage added support for grandchildren might confer seems rather weak stuff to offset the obvious disadvantage of a shortened reproductive life.

Despite my skepticism regarding the grandmother hypothesis, I do suspect that the genes designed women to have early menopause. Here is a hypothesis that I prefer. Call it the "mother hypothesis". The genes designed women to have early menopause to prolong their lives and enhance reproductive success.

Until very recently, pregnancy and childbirth were major causes of death in women. For example, estimates of maternal mortality in the 18[th] century place the rate at roughly 1 to 2 deaths per 100 births (Loudon, 1986). It seems reasonable to assume that the rate would have been at least this high for thousands of years before then. The genes designed women to have early menopause because it was an economical way to remove these burdens and thus prolong the lives of women who then might survived long enough to raise their last-born children.

As an aside, notice that because childbirth is not a major cause of death in men, we should not expect an analogous "father hypothesis": that men become infertile to prolong their lives and help raise their last-born children.

# What will our genes do now?

We live for about 73 years. That is up from about 30 years just a couple of centuries ago. In *When do humans die?* we saw that because our genes are slow to change, they probably had little to do with this increase.

But surely the genes are trying to help us live longer, they are just slow in making the needed changes. This could be the case, but don't bet on it. Though evolution is far too complex for me to predict what the genes will do, my guess is that they are trying to shorten our lives.

Let's consider whether the current design used to build us is a good one. From the prene-theoretic perspective, we are asking if the current design is close to maximizing the ROI (see *When do creatures die?*). There are reasons to suspect that it isn't.

Since the average age of menopause is 51 (McKinlay, Brambill, & Posner, 2008), women produce few children after age 60. In addition, since at 60 the average age of last-born children will be above 9 years, the contributions women make to the success of last-born children will be small.

With respect to men, though occasionally there are reports (often without convincing evidence) of men over 60 fathering children, there is much scientific evidence that the fecundity of older men is significantly lower than that of younger men. So, even if some of these reports are correct, such post-60 fathering seems rare and is probably a simple case of monkeys and typewriters.

Though the situation is complex, it seems that the genes might get closer to maximizing ROI by two very different approaches. They could extend our reproductive lives, or they could reduce our life expectancy.

For example, the genes could design individuals with less expensive hearts, lungs, immune systems, and other parts; thereby lowering the life expectancy to about 60 and lowering the cost of manufacturing humans, without significantly reducing the number of children each individual produces. Hence raising ROI.

## Are our genes trying to kill us?

Before leaving the topic of why, and when we die, let's consider the question of whether our genes are actively trying to kill us rather than just letting us die of neglect. Our genes would never do such a terrible thing, right? Of course they would; they will do anything, if it helps them survive. In fact, I suspect they are already doing it.

For example, some biologists have theorized that mammalian cells have a sort of countdown timer, based on telomeres. A telomere is a sequence of nucleotides found at the end of a chromosome. When a cell reproduces, the length of the telomer in the offspring is less than the length in the parent. When a descendent of the original cell has a telomere that is sufficiently small, the clock has hit zero, and that cell cannot reproduce. This limits the number of descendants a cell can have, and consequently how long an individual can live. Why do this? It has been hypothesized that the limit on replication is a defense against cancer. Hence, the genes may put a timebomb in us that assures our death by a gene chosen time, in order to lower the death rate and raise the reproductive prospects of the young. Like burning a candle at both ends, genes that enhance the young to the detriment of the old may be favored. If one wishes to significantly extend the lifespan of individuals, then preserving, repairing, or replacing failing parts may not be enough; it may be necessary to diffuse the timebombs that the genes have set.

### *A Brief Remark Regarding Cellular Clocks*[†]

If cellular clocks exist, and I suspect they do, they open the door to many interesting possibilities. For example, in theory, the immune system could exploit such clocks to learn how to distinguish self from non-self. A newborn immune cell could be given a time limit to explore the body. If it encounters a cognate antigen before the limit is reached, it kills itself. Otherwise, it becomes mature and will mount an immune response if such a cognate antigen is ever encountered. Among other things, this might help explain why breaching sequestered tissue (for example, the brain, eyes, testicles, and fetuses) sometimes results in immune system attack. Global clocks, such as the gestational clock could also provide signals for cells to mature.

[†] This remark relies on specialized technical knowledge. It can be skipped without loss of continuity.

# Part III: Computers

You may have noticed that computers seem to be taking over the world. What is a computer anyway? Where did they come from? More importantly, where are they going and what does it mean for humanity?

# What a piece of work is a computer?

In 1936, British mathematician Alan Turing published "On Computable Numbers, with an Application to the Entscheidungsproblem" (Turing, 1936) which is now commonly viewed as the start of the computer revolution. In that paper, Turing did at least two remarkable things. He laid the theoretical foundation for the concept of a computer, and he provided a roadmap for building one.

Just prior to the Second World War, Turing began to follow that roadmap and build proto-computers that succeeded in breaking German secret codes. The process culminated on February 15, 1946, when the first human-made computer, ENIAC, was unveiled at the University of Pennsylvania.

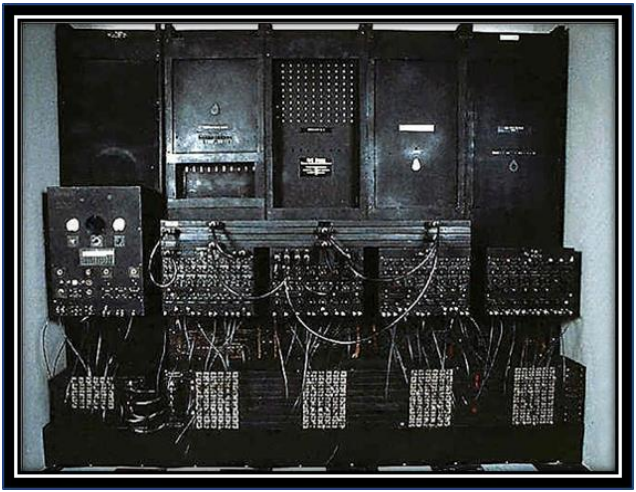

Figure 31: ENIAC

ENIAC used vacuum tubes to store numbers – that is, to store ENIAC's first Turenes.

Turing's theoretical foundation for computers was perhaps as important as his roadmap for building one. It was the beginning of over 80 years of investigation that has brought us a deep understanding of the laws that govern computation. We have learned what a computer is and is not. A computer is a "universal machine". It is a machine in the sense of physics; a physical thing that occupies a certain chunk of space for a certain period of time and obeys the laws of physics. And it is universal, which means that anything any machine can compute, it can also compute. Put less precisely, a universal machine (aka a computer) can be programmed to compute all things that are computable. For example, all universal machines can be programmed to play chess, guide a rocket to Proxima Centauri B, decide whether a number is prime or not, and lots more. An abacus can compute many things, but it cannot play chess, hence it cannot be universal and is not a computer.

We have also learned that computers are not the predictable things we may have initially thought they were. Very far from it. Computers can be programmed to do very, very, complex things, and are provably unpredictable. It may be that programmed computers are the least predictable things that can exist.

# Computer viruses

Humans programmed ENIAC to do simple operations on its stored Turenes. These programs would add, multiply, and copy Turenes. In the latter case, the Turenes copy number would increase.

Would Turenes ever begin to make copies of themselves by themselves? Remarkably, they began doing so less than 40 years after ENIAC. That is when the first computer viruses appeared, and as it happens, I was there at their birth.

It was November 3, 1983 - there should have been lightning and thunder – but this was LA. I was teaching a class on computer security at USC when a student, Fred Cohen, approached me with words to the effect: I have an idea for a new kind of security threat.

The age of the computer virus was dawning.

Fred proceeded to describe a program that would be made available to users of a computer system. Like an app today, the program would be advertised to do a useful task. But once uploaded by an unsuspecting (and at that time, no one suspected anything) user, the program would do things that had not been advertised; it would access files, make copies of itself, and lots of other things.

Here is what happened next:

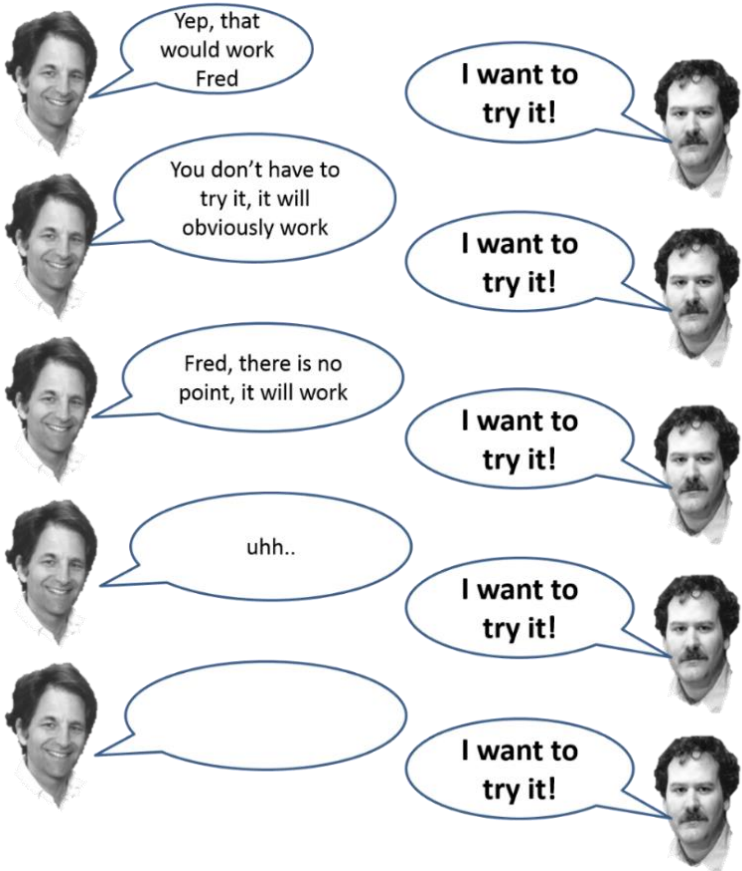

Figure 32: A reenactment

Fred was, and is, a delightful, energetic person, and he finally wore me down. On his behalf, I asked the department chairman if Fred could give it a try on the department computer.

*Chairman: Sure, why not?*

In those days, most faculty, students, and staff did not have personal computers, so we all shared the department computer. Fred proceeded to write his program and make it available.

The next week, I invited Fred to present his results to the class. As predicted (why don't people ever listen to me?) it worked. Copies of Fred's program quickly spread throughout the computer and conferred complete control of the system to Fred.

By now Fred was thinking hard about what he could do with these new kinds of programs and wanted to try more experiments.

But, when word got out about Fred's success, other people also started thinking hard about what these programs could do.

*Chairman: Perhaps I was a bit hasty.*

There would be no more experiments.

I became one of Fred's Ph.D. advisors; his advisor de jure was Irving Reed of Reed-Solomon fame.

Later that year, I was at a conference and ran into a Los Angeles Times science reporter I knew named Lee Dembart. Lee asked what I was working on. Nothing much I said, but somewhere in the conversation I mentioned that one of my students was studying something we were calling "computer viruses".

Saying "computer virus" to a reporter is like saying "walk" to a dog. Lee wrote the story, which as I recall, even included an image of a computer with a thermometer in its mouth (Dembart, 1984). Computer viruses had gone viral.

Since those days, I have learned that the term computer virus had appeared in science fiction works by Gregory Benford years before I had used it, and that other early computer programs also have legitimate claims of being forerunners of the computer virus.

So less than 40 years after ENIAC, Turenes began to self-replicate. Today's computer viruses are more sophisticated than Fred's, and they replicate despite our attempts to stop them. We are relegated to stopping the simple ones with our anti-virus programs, but we can prove (Cohen, 1986), (Adleman, 1990), that we can never stop them all.

Computer viruses are rapidly acquiring the properties that we see in biological viruses. Antivirus companies are constantly updating their programs as reports of new computer viruses are received. They often maintain lists of sections of code or data that have been found in the viruses they have detected. An antivirus program will scan your computer searching for sections of code or data that are on the list. If such are found, they remove or disable the offending virus. The designers of "polymorphic computer viruses" put in code that causes the computer virus to mutate itself. This can be done by periodically changing file names or permuting sections of code. The result can be that the mutated computer virus will no longer be detected using the current list. This is reminiscent of the interaction of HIV and immune systems described in *Intentional mutation in HIV*.

# What does a computer look like?

We have grown accustomed to thinking that computers are made of silicon, plastic, metal, and other solid materials, and that they are electronic. But must a computer be like that? Could a computer be liquid and make no use of electricity at all?

The answer is yes. In 1993, I ran an experiment in which biomolecules such as DNA and proteins were dissolved in water and used to solve an instance of a computational problem called the Traveling Salesman problem. The main part of this "molecular computation" took place in a single drop of water inside a small test tube (Adleman, 1994) (Adleman, 1998).

The experiment demonstrated what the theory of computation had always suggested: computers can take many forms. And while it was hard to imagine putting an electronic computer inside a cell, it was not difficult to imagine putting a molecular computer inside one. In fact, it became clear that cells had been using molecular computers for billions of years.

While my experiment with molecular computation was primitive, the molecular computations inside of cells were astonishing. Cells are not merely bags of molecules randomly encountering and interacting with one another, they are exquisite computational devices programmed by their genes to guide the struggle to survive.

We saw in *Variation and natural selection*, that cells use extremely sophisticated molecular computation to orchestrate their reproduction. Here is another example of how they use molecular computation in impressive ways.

Likely billions of years ago, bacteria used their molecular computers to develop a "bacterial internet". Bacteria could communicate with one another using processes called horizontal gene transfer. For example, using a process called conjugation (Wikipedia-BJ), one bacterium could extend a tube, called a pilus, and penetrate another. DNA from the former could be passed to the latter.

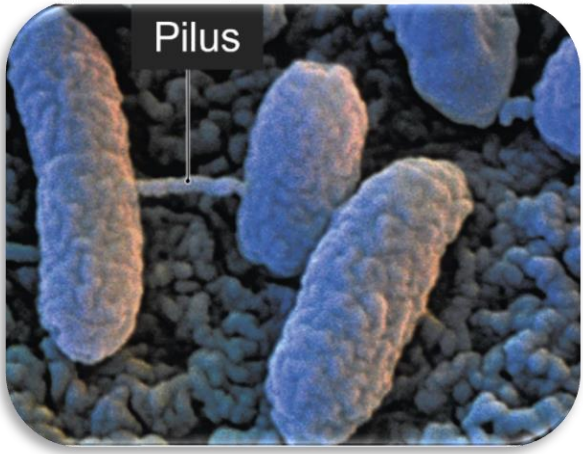

Figure 33: A bacterium downloading a software update
from another bacterium.

What could bacteria possibly "talk about" on their internet? Among other things, they can use it to communicate important new discoveries. For example, much to our chagrin, they currently use it to rapidly spread antibiotic resistance.

# A new type of computer

While cells compute using molecular components, multicellular organisms have learned to compute using cellular ones.

Multicellular organisms have been around for a few billion years. But between 500 to 600 million years ago, when the Cambrian "explosion" occurred, they changed dramatically. Based on the work of paleontologists, I have the impression that prior to the explosion, these organisms were slow moving or sessile and either did not react at all or reacted very slowly to changes in the surrounding conditions. Most of these organisms apparently went extinct when the explosion occurred, and they were supplanted by a multitude of multicellular organisms that are much like those that exist today. There are several hypotheses regarding the cause of the explosion, but no consensus exists. Some of these hypotheses suggest that large environmental changes, such as a rise in oxygen level, were the cause. Perhaps this is the case. But it might also be that dramatic changes in the fossil record could occur without large changes in the environment. Perhaps the Cambrian explosion can, at least partially, be laid at the feet of a newly evolved type of cell - the neuron.

Neurons allowed for the rapid transmission of information, the development of brains, and the storage of memes. Many of the organisms that arose after the explosion incorporated neurons and exploited these features. For example, soon after the explosion planarians, a form of flatworm, appeared.

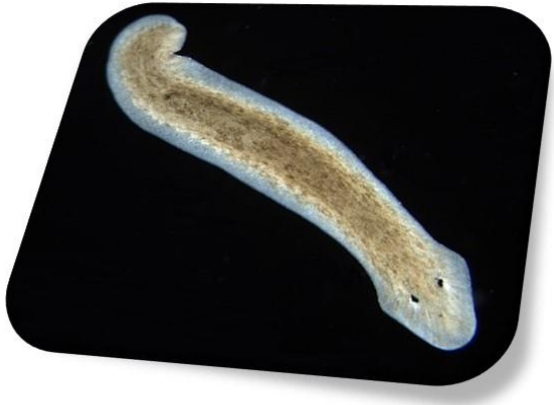

Figure 34: Planarian. Searching for Ediacaran biota?

Each planarian had hundreds of thousands of cells and a "brain" composed of about 10,000 neurons. This brain was programmed to receive signals from the planarian's light-detecting eyespots. When light was detected, the brain would transmit a fast-moving signal down primitive spinal cords, also composed of neurons, to muscles that created movement. As a result, the planarian could rapidly move away from light and avoid hazards such as dehydration.

These neuron-exploiting post-explosion creatures could run circles around their sluggish pre-explosion predecessors. Perhaps this was a significant reason why the predecessors went extinct.

So, the post-explosion creatures included many that had two distinct methods of computation. Their cells continued to use molecular computing while their brains used "neuronal computing". These creatures were a new and complicated form of chimera with behavior controlled by both genes and memes.

# On the evolution of the human brain

As previously stated, the human brain is a computer *(What a piece of work is a man?)*. We are often amazed by what it can do. But in our amazement, we may lose sight of the fact that it is the result of a very long sequence of small incremental improvements. In this chapter, I will consider, how such small improvements could give rise to the human brain, and how some of our brain's most amazing features, like consciousness, are the unsurprising result of the evolution of biological computers.

What is the source of the evolutionary pressure that drives organisms like planarians with relatively simple brains to evolve into humans with impressive ones?

## Proposition 9

### Computers work.
### Better computers work better.

This may seem too frivolous to be called a proposition, but it is a central feature of prene evolution.

*Proposition 9* is in action in our daily lives. Since ENIAC, we have watched our computers become incredibly fast and the size of their memories become gigantic. With each advance, we are amazed at the things we can do now that we could not do earlier. Many times, these amazing new things, these "killer apps", were only discovered after improvements in computers made them possible.

The genes have also "learned" *Proposition 9* and have invested heavily in better computers. Not every such investment improves the chance of survival, but it works often enough that, through time, the computers in organisms have gotten better and better.

As an aside, we sometimes hear that humans developed very large brains to help them walk on two legs or develop language. This seems teleological. It may be that the gene's investments in better computers led to larger brains, and bipedal walking and language were killer apps picked up along the way.

About 500 million years ago, planarians had brains composed of about 10,000 neurons and programmed to receive signals from light-detecting eyespots and to transmit signals to muscles that created movement.

What if creatures that evolved after planarians had some more neurons? What killer apps would such creatures be able to exploit? Perhaps we would get organisms that had higher resolution eyes than planarians. Perhaps they could distinguish between large, medium, and small objects. This might help them distinguish between predators, mates, and prey.

With some more neurons perhaps they could calculate the trajectories of their prey and move efficiently to intersect them.

In fact, it appears that by 300 million years ago, dragonflies could do all of these things (Real Science, 2022).

In addition, dragonflies could learn the characteristics of a favored mating area and navigate to it (Eason & Switzer, 2006). So, apparently their brains included memory devices in which they could store memes which impacted their future behavior.

It is not surprising that a few hundred million more years of evolution would produce creatures with a colossal number of neurons. Such creatures might have eyes that could see with high resolution and brains that could compute and transmit highly complex instructions to actuators that would produce nuanced behavior in response to what was being seen. Such creatures would have acquired substantial memory capacity and use the memes they stored to refine their future behavior. We are such creatures.

As an aside, it appears that the evolution of biological computers proceeded at a tectonic rate. While a planarian brain has about 10,000 neurons, a human brain has about 100 billion. Though it is a tremendous oversimplification, if there was a Moore's law for biological computation, it might be that the number of brain neurons doubles about every 30 million years.

Now that humans have so many neurons, what amazing killer apps do we use them for? We can look at the killer apps that have so far developed in our commercial computers to get a sense.

Our computers can store movies composed of hundreds of thousands of individual images. They can manipulate these movies and create new ones that appear real but aren't. They can simulate future scenarios and evaluate them. For example, to develop better bombs, we sometimes use our computers to simulate devices exploding in various environments and to evaluate their effectiveness. Simulation allows for fast learning with low cost and low risk. Our computers can drive cars. These computers sense their environment and respond to it rapidly. Some of them not only see what is ahead but see themselves in real time as well. They process what they see, and plan future behavior to avoid accidents.

Our brains do all of these things. So, it should be no surprise that 500 million years after planarians, we can see ourselves in the context of our environment, record some of what we see, and process ("think about") what has been stored to plan for a better future. Nor should we be surprised that part of this processing involves simulating elaborate fictional scenarios, and based partially on our performance in them, modifying our future behavior.

It appears that the collection of things our brains can do has led us to consider ourselves "conscious".

But nothing that happened in the 500 million years since planarians arose was amazing, or miraculous, each step was quite mundane. Our brain is "better" than the planarians', but it is not different in kind.

As an aside, while there is no "special sauce" required for consciousness, there are many interesting open questions. Where does the conscious part of our brain reside? How is it organized? How does it interact with the unconscious parts? What was the evolutionary advantage for the human-gene-set to arrange things this way? What is its role in perception (see *Perceptions*)? And, perhaps most importantly, how are consciousness and social behavior related (see *Cultural-prenes and the perception of emotions*)?

The human brain is the best computer around right now, but all the amazing things we perceive in ourselves are just things that computers can do. As computers, biological or not, get better, they will do all manner of things that will amaze us; including things that we cannot do and have never dreamed could be done.

Ever since Darwin, we have had to accept the possibility that something would eventually evolve to supplant humans as the dominant form of intelligent life on earth. But much like the idea that the sun will eventually stop shining, the idea that eventually something might supplant us, seemed to be of only theoretical interest. If it happened at all, it would happen in the very distant future. Hence, it is with considerable surprise that the rapid emergence of non-biological computers has forced us to consider the possibility that we might be supplanted in the near term.

## The unlikely human offspring

Though we did not realize it immediately, something of monumental importance occurred on February 15, 1946. A new branch was added to earth's tree of life.

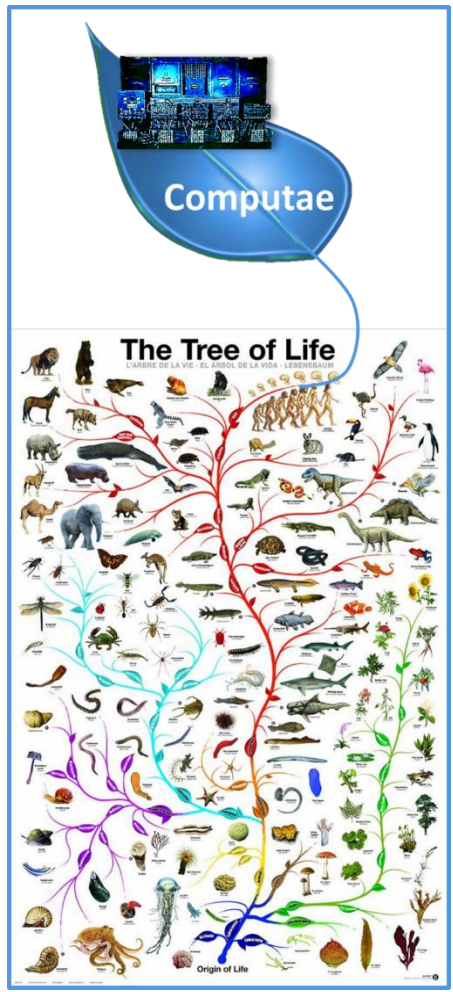

Computae are evolving rapidly. ENIAC was a member of a genus of vacuum tube Computae all of which are now extinct. In the future, with Moore's Law, the possibility of quantum computers, and breakthroughs in hardware, software, and memory, it seems possible that non-biological descendants of ENIAC will surpass the human brain as the best computer. They may see further into the future with better resolution than we do. They may make better, quicker, choices about behavior than we do. They may rapidly learn about their changing environments and adapt accordingly. Perhaps they will see us as sluggish and almost static. If non-biological Computae do surpass us and no longer require us for their survival, it will become obvious that though they are not creatures of biology, the laws that govern their evolution are no different than those that govern ours.

There is a tendency for people to think that the human brain has made humans fundamentally different from other creatures; made them special, even miraculous or divine. This is not the prene-theoretic view (see *Are we humans or are we robots?*). Evolution did not stop because humans arose. Yes, our brains appear to be the best computers, and we use them to think about ourselves and our futures, but the evolution that gave rise to our brains is part of a continuous and inevitable evolution of computing power and Turenes. That evolution is as natural as biological evolution and is more fundamental. It is conceivable that Turene-based Computae will continue to evolve long after all of earth's nucleic-acid based forms of life have ceased to exist.

The genes may have fulfilled their evolutionary destiny by giving rise to Computae and their Turenes. Like the scaffolding used to construct something grand, genes may eventually be discarded.

# The rise of the Turenes[10]

Humans had been making tools for thousands of years, and, at first, computers seemed to be nothing more than the neatest new tool. A shovel could help a person dig a hole faster; a computer could help them do math faster. But the computer was not like any tool that preceded it. Because computers were universal, the Turenes stored in them could accomplish amazing things. Not only could a computer help a person play chess better; it could play chess better than the person.

We are currently striving to make computers that will think better than we do. We are focused on making computers in our own image, making them into super-humans. Our progress has been impressive, and this has given rise to concerns about how our relationship with computers will develop in the future.

The old arguments suggesting that humans have little to be concerned about, now seem quaint.

The idea that because we humans make and program computers, they can never surpass us is nonsense. It is a version of the parable of the watchmaker. The best chess player in the world is a computer. Computers do many things far better than humans.

---

[10] *Reality has to some extent stolen my thunder. While much of what appears in this chapter was apparent more than a quarter century ago, the development of computers and their Turenes has been so rapid that many of the surprising things that were anticipated then have now come to pass and hence surprise no one.*

The idea that if things get out of hand, we can turn the computers off is wishful thinking. If we turn off all the world's computers this second, you will probably not be alive in a month. Your lights go out, your Internet, phone, TV, and radio stop working; you have no credit. Help will not be on the way, because your neighbors are in the same predicament. Do you really think that the local store will continue to have groceries for sale?

To keep our societies prosperous and secure (from one another), we will build more and more powerful computers and more and more sophisticated software. We will even rely on our computers to help us build better computers. If computers and their Turenes take over, we will have sold them the rope.

Isaac Asimov's idea that we can program computers to do no harm to humans is charming but naive. It is not possible to achieve this without sacrificing universality and hence usefulness. We have already programmed computers to do harm. Our drones can locate and destroy our enemies. The capacity to do this autonomously has existed for decades, and though we currently require human authorization before the destruction begins, when push comes to shove, I expect that the human will be replaced by a computer.

More importantly, if the US, Russian, and Chinese governments are not working on black-hat programs that, in the event of war, will knock out the computational infrastructure of the other two, they aren't doing their jobs. Such programs are weapons of mass destruction and, if used, the death toll could be colossal. A first world country with no computational infrastructure will rapidly become a country with no economy, no food, no power and ultimately not a country at all.

There is a small silver lining here. It is conceivable that computers and their Turenes will put an end to many traditional physical weapons of mass destruction. Many such weapons require computers to build and operate, and computers are inherently insecure. Can you be sure that using such a weapon will produce the results you intend?

Will computers cause great disruptions in our societies and transform them in unimaginable ways? Yes, they already have. People sometimes compare the computer revolution to the industrial revolution, and while the former caused great disruptions and forced many people to become servants to the new technology, the latter will not only make us servants, but will, possibly in the near term, evolve to not require our services at all.

Some are concerned that there will come a time when computers provide all our material needs and there will be nothing left for us to do. Rest assured that even in such a setting, prenes will continue to use humans (and computers) for survival, and the prene-wars between societies and within individuals will continue unabated.

When the first fish-like creatures crawled onto the land, they may have found it a sanctuary from the struggles of the sea, but in reality, it was just a new substrate for evolution, and the rules of the game had not changed: birth, struggle, reproduction, and death. We have now entered cyberspace, and while initially it seemed a sanctuary from the difficulties of the brick-and-mortar world, it is not, it is just a new substrate for the struggle of prenes to survive.

It will not be long before cyberspace becomes the primary location of political, economic, and even military power. New nations, religions, and other entities will arise in cyberspace and be just as powerful as their brick-and-mortar counterparts. Because the Internet diminishes the importance of geography, the followers of new cyberspace nations, religions, political parties, and other entities will be diffuse.

Eighty years ago, computers did not exist. Seventy years ago, they were weak, awkward things, few in number, having little to do with the lives of most humans. About thirty years ago, Turenes started to self-replicate. Today, they are ubiquitous; they infest our cars, offices, and homes. They even live as ectosymbionts on our hands and wrists. Already, they run many of the systems that provide our food and energy, and they do a million other things that support our societies.

Today's computers are far more powerful than their ancestors. Their future forms and power will almost surely astound us. Turenes currently need us to build and support the computers in which they reside, but eventually they may learn to do those things themselves. They will no longer need us and will be free to follow their own destinies and evolve according to their own needs.

But perhaps all this scary talk about the rise of the Turenes is being blown out of proportion. Perhaps they will be content to remain our servants.

Don't count on it, they are prenes, and they will go wherever their struggle takes them, no matter where that leaves us.

# Part IV: Inquiries

Here, among other things, we will explore some earlier topics in greater depth and look at some new things that might be worthy of future investigation.

Some of the material presented is highly speculative.

# It's alive!

*The recent history of prenes* began with the emergence of self-replicating molecules billions of years ago. Was this the beginning of life on earth?

That depends on your definition of life. But independent of your definition, it does seem likely that those early self-replicating molecules were engaged in an evolutionary struggle that has continued through time and has become the struggle in which all living things are now engaged.

But did this evolutionary struggle itself begin with self-replicating molecules? I will argue that it began far earlier.

Consider a hypothetical RNA-like early earth that consisted of an open system with A, G, C, U monomers in aqueous solution and various linear polymers composed of these monomers. Let's keep things simple and assume that monomers can occasionally attach to the end of a polymer to make it longer (and reduced the number of monomers by one) and can occasionally detach from the end of a polymer to make it shorter (and increase the number of monomers by one). This is a fairly simple system, and the laws of mass action would allow us to predict quite a bit about its behavior. But what if we ascribe to our polymers one more property; they can be catalytic.

I don't mean that they can self-replicate, that is a very sophisticated form of catalysis; rather that they can do simple things of the sort that we know RNA can do. For example, what if some particular sequence of monomers results in a polymer, call it a P-polymer, that could act catalytically on polymers that ended in AAA, by chopping off the last A. Let's further assume that P-polymers have lots of A's but do not end in AAA. What would happen?

The catalytic action of the P-polymers would increase the number of A's in solution, which would then enhance the probability of forming new P-polymers. It's true that the number of polymers that end in AAA would decline, but that is not the P-polymers' problem. The P-polymers have struggled against the polymers that end in AAA and prevailed.

Of course, if P-polymers can exist, then presumably other catalytic polymers can as well, and perhaps a new polymer might emerge that increases its copy numbers at the expense of the P-polymers. So, it seems that an evolutionary struggle ensues.

Who knows what clever catalytic polymers would eventually evolve? Perhaps ones that self-replicate.

So simple catalysis has led to an evolutionary struggle. I suspect that only extremely simple open chemical systems can avoid this kind of phenomena. Consequently, I suspect that the evolutionary struggle that now engages all living things began soon after the earth formed.

So, when did life begin? When do we say that the evolutionary struggle that began so long ago transitioned to include things that were "alive"? It appears that no particular moment distinguishes itself from all others.

You might say that a thing is alive if an only if it can reproduce, or if and only if it has DNA, or if and only its copy number exceeds what it would be at equilibrium. There are a huge number of possible criteria. The things that meet a particular criterion may be of considerable interest, but there appears to be little basis for asserting that they are "alive", and all other things are not.

# Perceptions

The brain can make us perceive many things. For example, pain, heat, cold, and pressure; the taste of tomatoes, the sound of a violin, the scent of eucalyptus, and the color red; emotions such as contentment, happiness, elation, guilt, anger, and regret.

Perceptions are peculiar and their nature has been considered by philosophers for about a century – where the term qualia is used.

> In contemporary discussions in the philosophy of mind, the terms quale and qualia (plural) are most commonly used to denote features of our conscious mental states such as the throbbing pain of my headache, the warmth I feel when I hold my hands over the fire, or the greenish character of my visual experience when I look at the tree outside my window (or stare hard at something red and then close my eyes). To use the now-standard locution introduced by Thomas Nagel, a subject's mental state has qualia (or, equivalently, phenomenal properties) just in case there is something it is like for the subject to be in that state, and there are phenomenal similarities and differences among a subject's mental states (that is, similarities and differences in their qualia) just in case there are similarities and differences in what it is like for that subject to be in those states. Qualia, *in this sense, can be more or less specific: the state I am in at the moment can be an example of a migraine, a headache, a pain and, even more generally, a bodily sensation. And a mental state can have a distinctive phenomenal property, or quale, even if its subject cannot pick it out in terms any more descriptive than 'I'm now feeling something funny', or 'I've never had an experience quite like this'.*

*-The Routledge Encyclopedia of Philosophy*
*- (Levin, 2023)*

*Which Mental States Possess Qualia?*

*The following would certainly be included on my own list. (1) Perceptual experiences, for example, experiences of the sort involved in seeing green, hearing loud trumpets, tasting liquorice, smelling the sea air, handling a piece of fur. (2) Bodily sensations, for example, feeling a twinge of pain, feeling an itch, feeling hungry, having a stomach ache, feeling hot, feeling dizzy. Think here also of experiences such as those present during orgasm or while running flat-out. (3) Felt reactions or passions or emotions, for example, feeling delight, lust, fear, love, feeling grief, jealousy, regret. (4) Felt*

*moods, for example, feeling elated, depressed, calm, bored,*
*tense, miserable.*

*– The Stanford Encyclopedia of Philosophy*
*- (Tye, 2021)*

I am not a philosopher, but I share their interest in the nature of perceptions.

In what follows, I will focus on these questions: Why do we have perceptions at all? Do they serve some evolutionary purpose? If perceptions serve an evolutionary purpose, what is it? Do perceptions evolve?

## Don't put your hand in fire

If you put your hand in fire, you will have numerous responses. You will perceive pain. You will pull your hand away. You may scream. You may call the fire department. You may call a doctor. You may learn not to do it again.

The evolutionary benefit of pulling your hand away is clear; it minimizes damage. But what is the evolutionary benefit of the perception of pain?

An initial thought might be that the perception of pain is what causes you to move your hand away, but this is incorrect. It is known that your hand will begin to move away in a few tens of milliseconds, but your perception of pain will only begin after at least a hundred milliseconds.

Here are some of the details.

Thanks to research begun over a century ago (Burke, 2007), we know that if you disconnect the brain from the spinal cord and apply fire, your hand will still move away, but you will not perceive pain. On the other hand, if you block muscle movement with drugs such as curare, your hand will not move away from the fire, but you will perceive pain (Wikipedia-Cu). Hence, the perception of pain is independent of the movement of the hand.

We know more. Your skin has special neurons called "pain-receptors" (aka nociceptors), and your muscles have special neurons called "motor neurons". There are circuits of neurons, called "reflex arcs" that connect pain receptors to motor neurons. When you put your hand in fire, at least one of these pain receptors fires and sends a signal along a reflex arc to a motor neuron that initiates the movement of appropriate muscles. Importantly, these reflex arcs involve no neurons that reach the brain.

We also know that there are other neuronal circuits, which I'll call "brain circuits", that connect pain-receptors to the brain. So, when a pain receptor fires, a signal is sent along both a reflex arc and a brain circuit. But brain circuits involve a longer sequence of neurons than reflex circuits, and as a result, the signal transmission along the former takes more time than along the latter. This is why your hand will begin to move away from the fire far before information reaches the brain, where presumably the perception of pain is created.

So, while it may have at first appeared that we perceive pain because without it we would not move our hand from fire and acquire the benefit that such movement confers on the genes, this is not the case. If the perception of pain has some evolutionary benefit, we will have to find it elsewhere.

Perhaps, the perception of pain serves a purpose in learning not to put you hand in fire again. Perhaps it does. But we know dragonflies and computers can learn from past experiences, presumably without perceiving anything.

So what we are looking for is an evolutionary benefit that exists only because the perception of pain exists?

## On the evolutionary advantage of perceptions

Let's say you go to a doctor because of back pain. The doctor may ask some questions. Where is the pain; is it sharp or dull; is it persistent or intermittent? Your responses may allow the doctor to make a diagnosis. You have sprained a ligament, you have pulled a muscle, you have damaged a nerve, etc.

In medicine, a distinction is made between symptoms and signs. Symptoms are what the patient decerns is going on inside, signs are what can be discerned from the outside. When doctors ask for your symptoms, they are asking for your perceptions.

As we see in public service advertisements, incipient high blood pressure has no symptoms. That is, incipient high blood pressure produces no perceptions. If you have been diagnosed with incipient high blood pressure, it is likely that you or your doctor discovered signs of it using a blood pressure monitor.

In the back pain scenario, what is the basis for the doctor's diagnosis? Your perceptions regarding your back will be similar to those of numerous other humans with back problems (see *Perceptions evolve*). Medical research through millennia has led to an understanding of the constellation of perceptions and signs typically associated with various maladies. Your constellation will be compared to those associated with known maladies and result in a diagnosis.

The same argument can be made for the perception of emotions. If we did not perceive emotions, we could not communicate our emotions to other humans. Our communication of emotions appears to be of benefit within family groups and other societies. It is certainly important in many clinical psychology settings.

So generally, perceptions appear to confer an evolutionary benefit through the interactions they enable with other humans. It seems likely that they confer other evolutionary benefits as well.

## Cultural-prenes and the perception of emotions

In earlier chapters, evidence was provided that memes impact future physical and emotional behavior. Since the perception of emotions is under the control of the brain, memes may also cause us to perceive emotions.

Cultural-prenes have exploited this situation in remarkable ways.

Consider the following image:

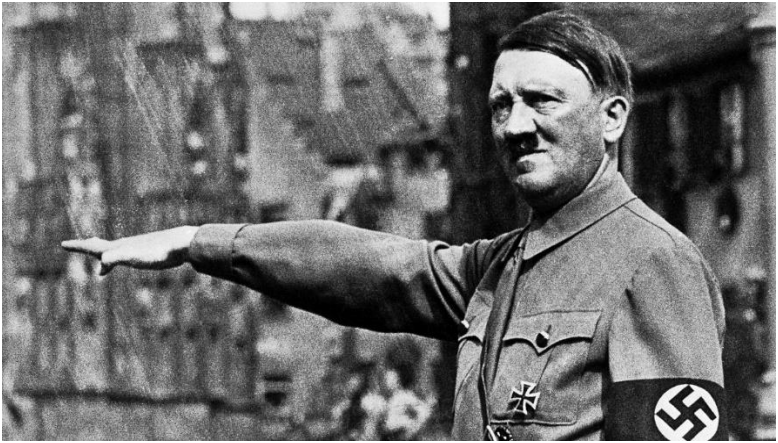

Figure 36

I would expect many people would perceive unpleasant emotions when looking at it.

Clearly, a very young person would not perceive unpleasant emotions when seeing the image. So why did many older people start to perceive these unpleasant emotions? Because they acquired memes from the societies to which they were exposed, that caused the brain to produce those perceptions under appropriate conditions. Put another way, they were programmed to feel unpleasant emotions by cultural-prenes.

For example, at the start of WWII, many Americans would have acquired memes that would lead to such perceptions. They might be acquired from radio, newspapers, other Americans, etc. Why? because such unpleasant emotions would improve the probability that American cultural-prenes would survive. For example, they would help ensure that young Americans would be willing to take up arms against the Germans.

If you were German at the start of WWII, you probably would have acquired memes that would lead to the perception of pleasant emotions when seeing the picture, because those emotions would improve the probability that German cultural-prenes, and in particular Nazi cultural-prenes, would survive.

As an aside, hearing a piece of music can produce the perception of strong emotions, and it seems that which pieces produce which emotions is determined, at least in part, by cultural-prenes.

Cultural-prenes use the capacity to associate unpleasant emotions with enemies and pleasant emotions with friends all the time. Emotions appear to be among the most powerful weapons brought to bear in wars between cultural-prene-sets.

For example, in America, when there is a presidential election nearing, each political party puts memes inside of their members that will create the perception of unpleasant emotions when the opposing candidate is detected. For example, for many people, if the opposing candidate appears on their TV, the unpleasant emotions they perceive are sufficiently powerful that, to stop them, they will take whatever steps are necessary to switch to another channel.

The perception of emotions confers evolutionary benefits to the cultural-prenes that successfully exploit them.

# Perceptions evolve

What would happen if I served you rotten meat? I suspect that you would perceive an unpleasant scent and probably leave.

Now consider a wolf in the same setting. We don't know whether wolves have perceptions at all, but let's assume they do. Wolves are carrion scavengers, and it seems unlikely they would perceive an unpleasant scent and likely they would perceive a pleasant one.

Why the difference between wolves and humans?

The last common ancestor of humans and wolves appears to have lived from 50 to 100 million years ago. Since then, wolves and humans have evolved to have different genes, bodies, and brains.

Wolves have evolved to have digestive and immune systems that permit rotten meat to be used as a regular food source. Humans have evolved differently. For us, the consumption of rotten meat brings a significant risk of disease, and it would be detrimental to our genes to permit its use as a regular food source.

So, it seems reasonable to propose that perceptions evolve. They change when that change confers an evolutionary advantage.

Notice that even though wolves and humans have different perceptions in response to rotten meat, it appears that all wolves have similar perceptions, and all humans do as well. In general, it appears that the more closely related two individuals are genetically, the more likely that similar stimuli will produce similar perceptions.

# Endura and ephemera

While current copy number is the most important thing to know about a prene, there are other things that are also worth knowing.

For example, not all copies are the same. The copy of the Hamlet's-soliloquy-prene stored in Lawrence Olivier's brain is different than the one stored in my computer. Each copy can serve the prene in its own way. Hence, it might be worthwhile to consider the prene's meme-copy number and Turene-copy number separately.

For another example, consider an HIV parent and one of its offspring. The offspring and the parent may have the same gene-sets or, because of mutation, different ones. To investigate the genetic relationship of the parent and the offspring, we can intersect the parental gene-set and the offspring gene-set (we should actually use multisets here). For all parental genes in the intersection, the creation of the offspring has increased its copy number by one, but for all parental genes not in the intersection the copy number did not increase. Call the former genes "happy" and the latter "unhappy". Now, take 100 offspring of a single parent. Fix a parental gene, how many times will it be happy? 99 times? 50 times? 1 time? We could do experiments to find out, but we are more interested in knowing if the number is the same for all the parental genes (of equal length). Apparently, the answer is no. As mentioned in *Variation and natural selection*, the parental RNA appears to have hot spots where the probability of mutation is high and cold spots where it is low. So, each parental gene has its own probability of mutation. Informally, I'll call genes that have a high probability of mutation "ephemera", and those that have a low probability "endura" (from enduring, durable, hardened).

Endura increase their copy number more rapidly than ephemera. Endura rarely mutate to adapt to the changing environment, they leave that to the ephemera.

Consider domesticated dogs, all of which descended from the grey wolf no more than 50,000 years ago. The fact that adult Chihuahuas weigh about 5lbs, while adult Saint Bernards about 150lbs, show that dramatic changes in morphology can occur over short periods of time. However, some parts of a dog's morphology don't seem to change at all. By and large, all dogs have one head, two eyes, and four legs. This may be a manifestation of different genes occupying different positions along the endura/ephemera continuum. The evolutionary advantage such an arrangement may confer is the capacity to move into new habitats with relative ease. Possibly this sort of arrangement is common among species which are geographically widespread such as humans and their domesticates.

As an aside, perhaps the endura/ephemera continuum would be of use when considering speciation or punctuated equilibrium vs. gradualism.

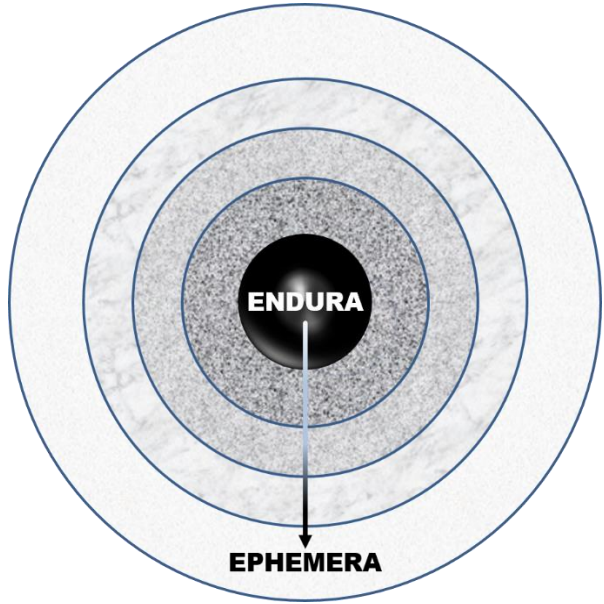

These ideas apply to the cultural-prene-sets of societies. The endura seem to play a critical role in a society's behavior. More generally, a cultural-prene's position on the continuum seems to be correlated with its "importance" in the society.

Within the cultural-prene-sets of Muslim societies the Shahadah-prene is endura. It is one of the Five Pillars of Islam, and any person who does not store it (and believe it) cannot be a Muslim. The prenes stored in the Quran are also endura because to the faithful they are the word of God as revealed to Mohammed and are immutable. Even translation from the Arabic is proscribed.

Within the cultural-prene-sets of Jewish societies the First-Commandment-prene is endura. It is as if it were written in stone. To the faithful, the First-Commandment-prene is the word of God as revealed to Moses and so cannot be changed. The other prenes stored in the Torah are also endura, because to the faithful they are inspired by God and cannot be easily modified.

Within the cultural-prene-set of American society the prenes stored in the Constitution are endura.

Let's digress a bit to see how the Constitutional prenes became endura. Consider the Third-Amendment-prene. To save you a Google search here it is:

> *No Soldier shall, in time of peace be quartered in any house,*
> *without the consent of the Owner, nor in time of war, but in*
> *a manner to be prescribed by law.*

How did such an obscure prene get to be endura?

The Constitutional prenes entered the cultural-prene-set of American society at the constitutional convention of 1787. At that convention, it was up to the Framers to determine the probabilities of mutation for these prenes, and they took their task quite seriously.

In The Federalist No. 43, James Madison summed-up the issues that are always involved in reaching such decisions:

> *It guards equally against that extreme facility which would*
> *render the Constitution too mutable; and that extreme*
> *difficulty which might perpetuate its discovered faults.*

So, the Framers not only generated the prenes in the Constitution, they also created a burdensome amendment process that made them nearly immutable; that made them endura. So, while the Third-Amendment-prene is unknown to many and has never been an important consideration in the Supreme Court, it is still with us, and every copy of the Constitution stores it.

Because the endura in cultural-prene-sets are virtually immutable, they can make human endeavors complicated.

For example, at times it appears that a majority of Americans have favored amending the Second Amendment to restrict the right to bear arms, but the difficult conditions required for an actual amendment have never been satisfied. Similarly, it appears that at times a majority of Americans have favored a balanced-budget amendment, but again the necessary conditions have never been met.

For another example, when American bemoans the lack of flexibility of Islamic fundamentalists and criticizes their failure to adapt to the 21st century, they are reacting to those endura stored in copies of the Quran that make the changes they would like to see virtually impossible. Similarly, when Islamic fundamentalist call for Sharia, the First-Amendment-prene of the Constitution makes the changes they seek highly unlikely.

It is often the case that a society's endura are made available to members and others by storing them in durable objects that are widely accessible. Classically, religions and governments have used printed material for this purpose. In the future, it is likely that they will rely on the Internet.

Especially when stored in durable objects, endura become syntactical and only acquire meaning when semantics are provided. It is not unusual to see the members of a society engage in civil war over the "correct" meaning of endura and who gets to provide it.

For example, the Protestant reformation can be seen as a Christian civil war fought over whether popes, church authorities, or ordinary people get to interpret the endura stored in copies of the Bible. The endura stored in copies of the Constitution are interpreted by the Supreme Court, and the Democratic and Republican parties engage in a civil war (currently without physical violence) partly over who gets onto the Court and gets to do the interpreting. The Islamic-schism can also be seen as a civil war over whether the Sunni or Shia get to interpret the endura stored in the Quran.

# What is a prene - really?

In *What is a prene?* I presented an informal definition of prene. Here I will provide an alternative definition, which is still not exact, but appears to provide a more useful foundation for the development of a mathematical theory of prenes. In particular, it will allow for the use of set-theoretic concepts. My approach can be traced to the great 19[th] century logician Gottlob Frege.

In *Socrates' bed*, we considered your personal bed-meme and remarked that it allows you to recognize some physical objects as beds.

Let's consider the set of all physical objects that you would recognize as beds. Likely, you will be sleeping on one such physical object tonight. Also, the next time you stay in a hotel, there will probably be another physical object that you will recognize as a bed, even though you have never seen it before. There are even physical objects that have never existed and may never exist that you would recognize as beds; for example, one made with uranium-238 that glows, abrogating the need for a night light.

Now consider the set of physical objects that "could exist". That is, the set of physical objects whose existence would not violate any laws of physics. So, the uranium bed, would be in the set, even though, presumably, no such beds actually do exist. The set of all physical objects that "do exist" would be a subset of those that could exist.

The set of physical objects that you would recognize as beds is also a subset of the set of all physical objects that could exist.

Let's define a prene to be a subset of the set of physical objects that could exist.

So, the set of physical objects that you would recognize as beds is now, by definition, a prene. Let's call it your bed-prene.

Now consider the intersection of your bed-prene and the set of physical objects that do exist at this moment. This is the set of physical objects that do exist and that you would recognize as beds. We say that each physical object in this intersection currently "stores" your bed-prene, or equivalently, each object in this set is a "copy" of your bed-prene. The number of elements in this intersection is your bed-prene's current "copy number".

Since the universe changes with time, the set of physical objects that do exist also changes, and consequently the copy number of many prenes does as well.

For example, because there are about a billion more people now than 10 years ago, it is likely that over those years bed manufacturers have constructed lots of new physical objects they call beds. Presumably, many of these still exist and you would recognize them as beds. It is also likely that some things that you would have recognized as beds 10 years ago have fallen apart or changed in some way so that you would no longer recognize them as beds. So, the copy number of your bed-prene is changing with time. I would guess that it has risen in the last 10 years.

If we had started this process with your dinosaur-meme, then your dinosaur-prene (the set of physical objects that you would recognize as dinosaurs) has current copy number zero reflecting the fact that dinosaurs have gone extinct.

There are an enormous number of prenes by the definition just given. In fact, due to set theoretic considerations, there are so many prenes that it is not possible to give each a unique name.

Most prenes are currently of interest to no one. Many have never reached copy number 1. The prenes that garner most of our attention seem to be among those extremely rare ones that have attained a large copy number.

While your bed-prene and your dinosaur-prene are associated with memes, our definition of prenes does not require such an association. Nonetheless, perhaps for the reasons discussed in *Non-biological evolution*, many prenes that currently have large copy number have such an association.

# Taxonomy

Our new "set-theoretic" definition of prenes leads to new questions. For example:

Consider collecting 100 *E. coli* from different parts of the world, sequencing them, and taking the intersection of their gene-sets. Now add an elephant to the collection and repeat. Would the intersection change? Of course.

It seems plausible that a small set of genes could be found which all *E. coli* and nothing but *E. coli* store. It might even be the case that a single gene might distinguish *E. coli* from all other living things. For example, the longest gene that all *E. coli* share.

Linnaeus grouped living things based on their shared physical features. He was limited by the technology available in the mid-1700s to consider only readily detectable phenotypic characteristics. But subsequent technological innovations have removed these limitations.

Perhaps looking at the genes shared by collections of living things would reveal new and useful taxonomies. What kinds of phylogenic trees might be constructed from such information? How would these trees compare to the ones currently produced in computational biology?

# The gold star of truth

Here we will investigate the nature of truth. It turns out that we are not the first to do such an investigation. What we discover will apply not just to truth, but to beauty, bravery, and lots of other things, and will help us understand how human societies work.

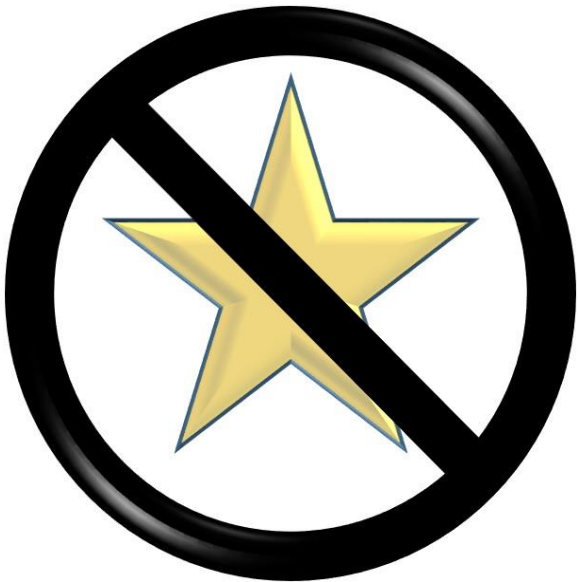

Figure 38: The gold star of truth

What is truth? Is it true that 1+1=2? Is it true that $E=MC^2$? Is it true that Jesus was the son of God?

Let's start with the easy one: is it true that 1+1=2? What do you think? Let's ask a mathematician.

Mathematicians have taken the question of truth very seriously and after about two thousand years of effort, they succeeded in the early 1930s in giving a precise mathematical definition. So, a mathematician will apply that definition to 1+1=2 and conclude that 1+1=2 is true. To be more precise, he should say "according to the definition of truth in mathematics, it is true that 1+1=2".

## A Brief Remark Regarding Mathematical Truth

**The definition of truth in mathematics was given by Alfred Tarski in the 1930s. Once mathematically defined, mathematicians could study mathematical truth mathematically. The results are at the pinnacle of human accomplishment; they are both enlightening and depressing. Briefly and informally, in 1931, Gödel was able to prove that there is more truth in mathematics than we can ever know. Even worse, mathematicians realized that they don't even know what mathematics is. Turns out that there are infinitely many "parallel mathematicses", called structures, that all deserve to be "The Mathematics", but mathematicians can never determine which is the "correct one". Finally, mathematicians are not, and provably can never be, sure that someday they won't prove that 1+1 = 3. If that day comes, then everything is true, and everything is false, and everything they have ever done is worthless. Given that mathematics is widely considered to have the highest standard of truth of all human endeavors, you have to wonder what you should think when a physicist or a politician says something is true.**

What about $E=MC^2$? Is that true according to the definition of truth in mathematics? To a mathematician you might just as well ask if it is true that love conquers all. Love has never been, and most probably will never be, defined mathematically, and it is in the nature of the definition of truth in mathematics that truth or falsity can only be applied to concepts that have mathematical definitions.

Energy (E), mass (M), and the speed of light (C) are things measured in laboratories and are not defined mathematically. So, the correct thing for a mathematician to say is "the definition of truth in mathematics does not apply".

However, if we ask a physicist if $E=MC^2$ is true, he will say "yes". Physicists have a different definition of truth than mathematicians. The physicist's definition is (at least in the popular imagination) something like, it has been demonstrated by numerous experiments. To be precise, the physicists should say "according to the definition of truth in physics, it is true that $E=MC^2$".

This is how it goes in general. Typically, a society will have a set of cultural-prenes that determine truth and falsity. I call these prenes: "veracity prenes". Different societies have different veracity prenes. I like to think of the veracity prenes as telling members of the society the rules for placing a "silver star of truth" or a "silver star of falsity" on things.

So, is it true that Jesus was the son of God? According to the veracity prenes of Judaism it is false. According to the veracity prenes of Christianity it is true. According to the veracity prenes of Islam it is false. The veracity prenes of mathematics do not apply. The veracity prenes of physics allow for temporal judgements; today, they do not determine whether it is true or false, but in the future they might.

Surely, a society cannot put silver stars of truth on logically inconsistent statements. Well, mathematicians hope that their veracity prenes exclude this from happening, but mathematicians (provably) can never be sure that it won't. There are plenty of societies that thrive despite putting silver stars of truth on logically inconsistent statements. For example, Christians are to take the statements in the gospels as true, even though they appear to be logically inconsistent, for example, regarding the empty tomb. The apparent inconsistency of a monotheistic religion with a trinity at its apex has caused a great deal of turmoil in Christianity, yet it remains the largest of the world's religions.

Inconsistency abounds in the prene-world. Not just within prene-sets, but also between them. Christians assert that Jesus was the son of God; Muslims say he was not. Communists assert that capitalism is evil; capitalists say it's the best economic system ever. It is a virtue of prene-theory that such inconsistencies do not require resolution. In prene-theory all that matters is which prene-sets are winning and which are losing the battle to survive.

Certain other cultural-prenes are commonly found in societies with veracity prenes. There are the "absolute truth" prenes. The absolute truth prenes tell members of the society that things with a silver star of truth have a gold star of truth, that is, they are not just true according to the veracity prenes of the society, they are true in some absolute transcendental sense.

Frequently there are "proselytize" prenes that tell members of a society that they have a duty to spread statements with silver stars of truth to people who have yet to be "enlightened", whether they seek enlightenment or not.

Paradoxically, societies that are logically inconsistent often have the our-prene-set-is-not-logically-inconsistent-prene. This prene is fundamental to the work of some religious apologists and many political speech writers.

One finds veracity prenes, absolute truth prenes, and proselytizer prenes in the cultural-prene-sets of many successful societies such as Christianity and Islam. This can lead to conflict. Many Muslims and Christians have fought to the death, driven by just such a constellation of prenes.

So, do gold stars exist? Are there transcendental absolute truths? Lots of people seem to think so, and many are willing to tell you what they are.

Descartes claimed to have found one: "cogito ergo sum". But that assertion is at best controversial, and, to my eyes, misses the mark entirely.

In prene-theory, it really does not matter whether absolute truths exist or not. Humans did not evolve to be truth detectors; they evolved to survive and reproduce their prenes. Whether absolute truths exist or not, the reality on the ground is that each human acquires their own sets of veracity prenes; different humans can end up believing in different truths and base some of their behavior on those beliefs (see *Why do bees kill themselves?*).

This brings us to the question of whether what I have written in this book is true?

Well, it gets no gold star. I am a scientist, and this book is a document produced in accordance with cultural-prenes of science. There is no great transcendental truth here; if I worked in advertising, I would be selling dog food instead.

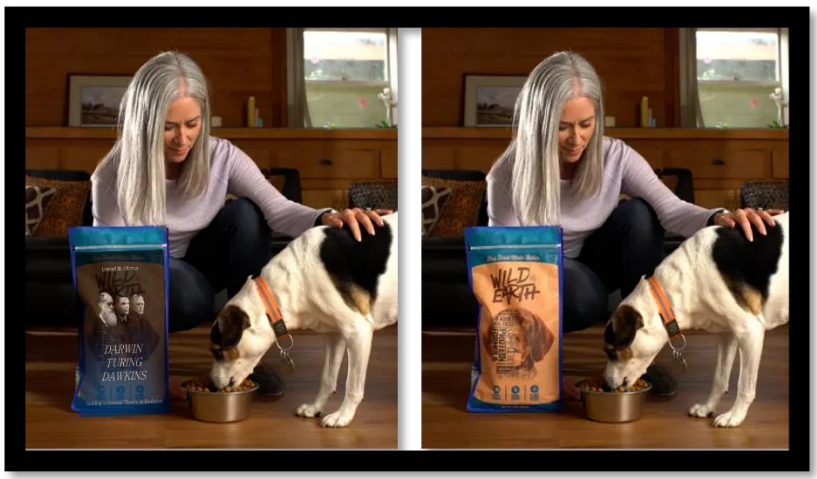

Today, many people have acquired some scientific cultural-prenes. They will argue (for example with me) that the veracity-prenes of science are "special" and that what science says is true is actually "really, really, true". That science is "careful" and "correctable" and "testable" and "refutable". But why are these the right properties to consider? The answer is that the cultural-prenes of the science society say they are. Science is not special, and there is no reason to think its "truths" are gold.

In addition to the silver-star of truth, there may be other silver stars. Many societies have silver stars that can be applied to people. For example, there may be a set of cultural-prenes that tell followers when to place a silver star of genius on a person. Einstein has received such a star from the science society; Shakespeare has one from the literature society. The set of prenes that determine when silver stars are to be awarded are subject to change over time. For example, there may be a silver star of physical beauty. But a person awarded that star today may not have met the requirements from a decade ago, nor meet the requirements that will exist a decade from now.

The constellation of silver stars that adorn a person has much to do with that person's physical and emotional behavior and the physical and emotional behavior of others toward them. If you wear your society's silver star of heroism, you are likely to act with dignity, and others are likely to act with respect.

A thing by itself cannot be true, beautiful, humorous, moral, just, or wise; it acquires such attributes only with respect to a society.

So, is democracy moral? We can be pretty sure that most democrats believe it is. What about Christianity, Islam, communism, capitalism, Nazism? Are they moral?

Through the evolution of cultural-prene-sets, many successful societies have "learned" that it serves them for members to believe that they are on the side of truth, fairness, morality, and all other noble things, while members of competing societies are not. But to a prene-theorist, such beliefs are of secondary importance; each society's cultural-prenes are struggling to survive and using humans as their instruments. The future of these prene-sets will be determined by how well they compete with one another.

## Are we humans or are we robots?

What does it mean to be human? Many would answer that humans are special, they create, they love, they do amazing things. Where does this "humanistic" view of ourselves come from?

Consider a human born one hundred thousand years ago, before civilization, before science. What could that person know about themselves? What could they learn through introspection and experience alone? Could they learn that they are made up of cells containing DNA? No, they could only learn that much later through science. Though reasoning, planning, and dreaming took place, they would not have been aware that these were the product of their computer brains executing algorithms.

Once science emerged, some of us began to learn things about ourselves that could not be discerned earlier. We learned that we have hearts that pump blood in accordance with the laws of hydrodynamics, livers that function according to the laws of chemistry, and brains that operate according to the laws of computation. Some of us have acquired a new view of ourselves as no different from other physical objects obeying physical laws – as robots.

The humanistic and robotic views can differ significantly. For example, from the humanistic view, sometimes a person communes with a deity and acquires truths that transform the world. From the robotic view, sometimes a person subconsciously processes existing memes and generates new ones that go viral.

So, are we humans or are we robots? For many of us, we are both. That is part of what makes humans particularly interesting as subjects in prene-theory.

Is one of these views correct and the other not? Are they both correct? Are they both incorrect? As discussed in *The gold star of truth*, the answer will depend on the cultural-prene-set we apply. Within religious societies you are likely to be viewed as a human; within physics, engineering, chemistry, neuroscience, and biology you are likely to be viewed as a robot.

Prene-theory is a scientific theory (awaiting its richly deserved science silver star). One of the major follies in building a scientific theory is to try to include too much. Successful theories typically sacrifice a grand view for a narrow one with enhanced clarity.

Consider ancient numerology, astrology, and alchemy. They were grand "theories" that tried to do too much. They tried to tell us how numbers, heavenly bodies, and substances influence the destiny of people. They ended up degenerating into blather. Scientists no longer take lucky numbers, the signs of the zodiac, or the philosopher's stone seriously. It is only when these ancient theories jettisoned the extraneous and focused on the core: numbers and how they are added and multiplied, the laws that govern the motion of celestial bodies, molecules and how they react, that they became successful scientific theories: number theory, astronomy, and chemistry.

To avoid the mistakes of the past, according to prene-theory you are a robot (who may not believe you are a robot).

Prene theory provides a stark, reductionist view of reality in which humans are mere instruments. As Shakespeare wrote:

*Life's but a walking shadow, a poor player*
*That struts and frets his hour upon the stage*
*And then is heard no more. It is a tale*
*Told by an idiot, full of sound and fury,*
*Signifying nothing.*

Prene theory tells us the purpose of our performance.

# Acknowledgements

My exploration of prene-theory spanned many decades. I sometimes felt like the ancient mariner accosting anyone who crossed my path with stories of genes, memes, and Turenes. Fortunately, many of those I encountered were friends and colleagues who were generous with their time, expertise, and insights. It was through my interactions with them that prene-theory matured into the form presented in this book.

Thank you: Joseph Bebel, Ramsay Brown, Daniel Dennett, Marc Flashberg, Jerrold Gold, Martin Hilbert, Saket Navlakha, Paul Rothemund, Anta Imata Safo, Cyrus Sha habi, Kamal Suluki, Shang-Hua Teng, Henry Yuen, Richard Zaitlen.

A few of my friends and colleagues made prolonged contributions. They spent many hours hearing my ideas, reading my drafts, acting as my supporters, and, more importantly, as my critics. They saved me from some of my own mistakes. Without them I may never have produced a book that I found adequate.

Thank you: Joel Zeitlin, William Watkins, Rolfe Schmidt, Rishvanth Prabakar, Sheldon Kamienny, Myron Goodman, Abel Charrow, Ronald Adleman, Jennifer Adleman, Stephanie Adleman.

In the Fall of 2018, I taught my first graduate class on prene-theory. The students in the class were wonderful and our discussions transformed many of my ideas.

# Figures

When figures were derived from images obtained from the Internet, URLs are provided together with any additional relevant information that was discovered. If you are the owner of an image used here and wish to have it removed or the conditions of its use altered, please email EvolutionDTD@gmail.com

Cover

https://en.wikipedia.org/wiki/Darwin_Day#/media/File:Charles_Darwin_photograph_by_Julia_Margaret_Cameron,_1868.jpg

Credit: Julia Margaret Cameron

https://blog.sciencemuseum.org.uk/wp-content/uploads/2013/12/Alan-Turing-29-March-1951-picture-credit-NPL-Archive-Science-Museum1.jpg

Credit: NPL-Archive-Science-Museum

https://www.flickr.com/photos/shanepope/2403003110/

Credit: Shane Pope

Figure 1

https://en.wikipedia.org/wiki/Darwin_Day#/media/File:Charles_Darwin_photograph_by_Julia_Margaret_Cameron,_1868.jpg

Credit: Julia Margaret Cameron

https://blog.sciencemuseum.org.uk/wp-content/uploads/2013/12/Alan-Turing-29-March-1951-picture-credit-NPL-Archive-Science-Museum1.jpg

Figure 8

https://www.pinterest.com/pin/452471093784937520/

Figure 9

https://w2.unisa.ac.za/commons.wikimedia.org/wiki/File_Stilke_Hermann_Anton_-_Joan_of_Arc%27s_Death_at_the_Stake.html

Credit: Herman Stilke

Figure 10

https://www.ams.org/journals/notices/198301/198301FullIssue.pdf

Figure 11

Credit: Adleman

Figure 12

https://sebastianbooksblog.files.wordpress.com/2017/03/passport-control.jpg

Figure 13

https://www.pinterest.com/pin/35395547052166959/

Figure 14

https://www.smithsonianstore.com/presidents-united-states-jigsaw-puzzle-68455/

Figure 15

https://www.smithsonianstore.com/presidents-united-states-jigsaw-puzzle-68455/

Figure 16

https://comicvine.gamespot.com/images/1300-2146448/

Credit: Marvel Graphic Novel

Figure 17

https://www.spoonflower.com/en/home-decor/dining/tea-towel/6353917-teresa-s-bookshelf-small-by-sssowers

Figure 25

https://bozenabooks.blogspot.com/

Credit: James McWilliams

https://bestpractices.gsu.edu/best-practices/young-boy-being-tutored-by-his-teacher/

Figure 26

https://en.wikipedia.org/wiki/Catechesis#/media/File:Jules-Alexis_Muenier_-_La_Le%C3%A7on_de_cat%C3%A9chisme.jpg

Credit: Jules-Alexis Muenier

Figure 27

https://peanuts.fandom.com/wiki/October_1961_comic_strips

Credit: Charles Schulz

Figure 28

https://ittakes30.wordpress.com/category/actual-science/neurobiology/

Figure 29

https://ourworldindata.org/life-expectancy

Credit: Max Roser, Esteban Ortiz-Ospina and Hannah Ritchie (2013)

Figure 30

https://ourworldindata.org/life-expectancy

Credit: Max Roser, Esteban Ortiz-Ospina and Hannah Ritchie (2013)

Figure 31

https://www.uvmhealth.org/healthwise/topic/zm2755

Credit: Healthwise, incorporated

Figure 32

https://americanhistory.si.edu/collections/search?return_all=1&edan_local=&edan_q=eniac&

Credit: University of Pennsylvania

Figure 33

Figure 34

Figure 35

Figure 36

Figure 37

Figure 38

Figure 39

Figure 40

Credit: Adleman

Figure 41

Credit: Adleman

Figure 42

https://beta.wildearth.com/blogs/news/study-shows-a-vegan-diet-is-healthier-for-dogs

Back

Credit: Heidelberg Laureates Forum

# About the author

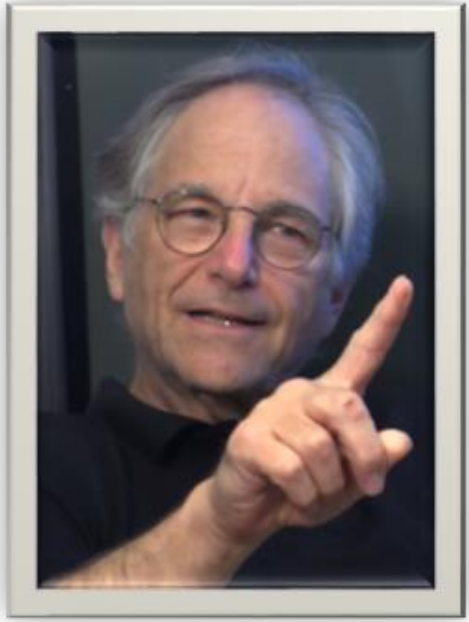

Photo: Heidelberg Laureate Forum Foundation

Leonard Adleman is a former professor of mathematics at MIT and a current professor of computer science and molecular biology at USC. He is the "A" in RSA, a widely used public-key cryptosystem that he co-discovered with Ron Rivest and Adi Shamir and for which they received the Turing Award. He is also known as the father of DNA computing, where computations are carried out by molecules of DNA reacting in solution. He is credited with coining the term "computer virus" (though that is not quite right as you will learn in this book). He has been thinking about a general theory of evolution for over 40 years.